\documentclass[10pt,a4paper]{article} % 10pt is ignored!
\pdfoutput=1
\usepackage{jcappub}

\allowdisplaybreaks

\usepackage{ifthen}
\usepackage{graphicx}% Include figure files
\usepackage{dcolumn}% Align table columns on decimal point
\usepackage{bm}% bold math
\usepackage{color}
\usepackage{amsmath}
\usepackage{amssymb}

\newcommand{\fnl}{f_\text{NL}}
\newcommand{\fnlsq}{f_\text{NL}^2}
\newcommand{\gnl}{g_\text{NL}}

\newcommand{\dqc}{\frac{\derivd^3q}{(2\pi)^3}}
\newcommand{\dqcp}{\frac{\derivd^3q'}{(2\pi)^3}}

\newcommand{\be}{\begin{equation}}
\newcommand{\ee}{\end{equation}}
\newcommand{\barr}{\begin{align}}
\newcommand{\earr}{\end{align}}
\newcommand{\la}{\left\langle}
\newcommand{\ra}{\right\rangle}
\newcommand{\partfrac}[2]{\frac{\partial{#1}}{\partial{#2}}}
\newcommand{\partsqfrac}[2]{\frac{\partial^2{#1}}{\partial{#2}^2}}

\newcommand{\cyc}{\ \text{cyc.}}

\newcommand{\hMpc}{\ h^{-1}\text{Mpc}}

\newcommand{\ihMpc}{\ h\text{Mpc}^{-1}}
\newcommand{\hGpcc}{\ h^{-3}\text{Gpc}^3}

\newcommand{\hMs}{\ h^{-1} M_\odot}

\newcommand{\elnn}{\nonumber\\}
\newcommand{\snr}{\text{SNR}}
\newcommand{\tim}[1]{\times 10^{#1}}%times 10 
\newcommand{\derivd}{\,\mathrm{d}} % upright d for derivatives and integrals
\newcommand{\eh}[1]{\exp{\left[#1\right]}}
\newcommand{\unit}[1]{\ \text{#1}}

\renewcommand{\vec}{\bm}

\title{Primordial non-Gaussianity in the Bispectrum of the Halo Density Field}

\author[a]{Tobias Baldauf,}
\emailAdd{baldauf@physik.uzh.ch}
\author[a,b,c]{Uro\v{s} Seljak}
\author[d,e]{and Leonardo Senatore}

\affiliation[a]{Institute for Theoretical Physics, University of Zurich, Zurich, Switzerland}
\affiliation[b]{Physics Department, Astronomy Department and Lawrence Berkeley National Laboratory, University of California, Berkeley, CA, USA}
\affiliation[c]{Institute for the Early Universe, EWHA Womans University, Seoul, South Korea}
\affiliation[d]{Stanford Institute for Theoretical Physics, Stanford University, Stanford, CA, USA}
\affiliation[e]{Kavli Institute for Particle Astrophysics and Cosmology, Menlo Park, CA, USA}

\abstract{
The bispectrum vanishes for linear Gaussian fields and is thus a sensitive probe
of non-linearities and non-Gaussianities in the cosmic density field. Hence, a
detection of the bispectrum in the halo density field would enable tight
constraints on non-Gaussian processes in the early Universe and allow inference
of the dynamics driving inflation. We present a tree level derivation of the halo
bispectrum arising from non-linear clustering, non-linear biasing and primordial
non-Gaussianity. A diagrammatic description is developed to provide an intuitive
understanding of the contributing terms and their dependence on scale, shape and
the non-Gaussianity parameter $\fnl$. We compute the terms based on a multivariate 
bias expansion and the peak-background split method and show that non-Gaussian
modifications to the bias parameters lead to amplifications of the
tree level bispectrum that were ignored in previous studies.
Our results are in a good agreement with published simulation
measurements of the halo bispectrum.
Finally, we estimate the expected signal to noise on $\fnl$ and show that
the constraint obtainable from the bispectrum analysis significantly exceeds the one obtainable
from the power spectrum analysis.}


\begin{document}
\maketitle
%===============================================================================
%				INTRODUCTION
%===============================================================================
\section{Introduction}\label{sec:intro}
The question whether the inhomogeneities in our Universe have been seeded by
a Gaussian initial distribution raised a lot of excitement recently 
(see \cite{Liguori2010,Desjacques2010} for reviews). Inflation \cite{Guth1981,Linde1982,Albrecht1982}
is a
theoretical paradigm that could generate the initial fluctuations, but has not
yet been
directly confirmed observationally. Standard slow-roll inflation predicts a
very low level of non-Gaussianity. However, there is no shortage of single and
multifield inflationary models with most of them predicting a fluctuation
distribution distinct from the simple Gaussian case. Thus detection of a
non-Gaussian signal would provide unprecedented information about the dynamics
driving inflation and the interactions of the inflaton field \cite{Cheung2008}.

The fluctuations in the inflaton field are imprinted in the distribution of
photons and matter in the Universe. This raises the question, which observable
is 
best suited to detect the tiny deviations from the fiducial Gaussian
distribution of field amplitudes. 
Certainly, it is promising to look at statistics that would vanish in the 
Gaussian case, such as the bispectrum of the Cosmic Microwave Background (CMB)
radiation that for a long time was believed to be the most promising probe of
primordial non-Gaussianity \cite{Komatsu2010}.
In Large Scale Structure (LSS) the detection of primordial non-Gaussianity is
hampered by the non-Gaussianity produced by late time non-linear clustering, a
caveat not present in the linear CMB physics. Only in recent years it was
realised that equally strong constraints can be obtained from the LSS
\cite{Sefusatti2005}. One of
the most promising features of primordial non-Gaussianity in the LSS is the
scale dependence of the halo bias. This scale dependent bias is most prominent 
for non-Gaussianities with non-vanishing squeezed limit of the bispectrum, 
such as the local shape~\cite{Salopek1990,Salopek1991} and the shapes with both 
equilateral and local limit recently found in the Effective Theory of Multifield 
Inflation~\cite{Senatore2010}. 
It was theoretically predicted for the local type non-Gaussianity by 
\cite{Dalal2008} and subsequently other derivations were presented by \cite{Matarrese2008,Slosar2008}. These were
confirmed in simulations by \cite{Dalal2008,Desjacques2009a,Desjacques2009b,Grossi2007,Grossi2008,Pillepich2010,Smith2010}.
First data analysis based on the scale dependent non-Gaussian bias lead to remarkably strong
constraints on local non-Gaussianity \cite{Slosar2008}.  

The two point function and its Fourier transform, the power spectrum, are 
the most important statistics that have been used to analyze LSS surveys so
far. 
Their drawback in terms of the detection of non-Gaussianities is that the 
signatures may be small and difficult to separate from non-Gaussianities 
generated by gravity, 
whereas the distinct features of
alternative 
inflationary models will be imprinted more clearly in higher order statistics 
such as the three point function or the bispectrum 
\be
\la\delta(\vec k_1)\delta(\vec k_2)\delta(\vec
k_3) \ra=(2\pi)^3\delta^\text{(D)}(\vec k_1+\vec k_2+\vec
k_3) B(\vec k_1,\vec k_2, \vec k_3),
\ee
which vanish for Gaussian fields.
Measuring the higher order statistics is more involved than the standard power 
spectrum analysis because these statistics show both shape and scale
dependence, 
and both of these dependencies need to be considered in order to constrain 
different forms of primordial non-Gaussianity. In this paper we try to
address whether the bispectrum analysis can improve on the power spectrum
analysis 
of LSS surveys in terms of detecting primordial non-Gaussianity of the local type. 
Previous studies of the halo bispectrum in presence of primordial
non-Gaussianity \cite{Jeong2009,Sefusatti2009} are based on the standard local
bias model. The non-Gaussian correction to the halo bispectrum in these
approaches arises mainly from loop terms and is thus dependent on the
smoothing scale. We will try to improve on these calculations and present an
independent and smoothing invariant approach to obtain the halo bispectrum.

In \cite{Creminelli2006gc} it was proven that the optimal estimator of 
non-Gaussianity is the bispectrum. This raises the question how the power
spectrum could possibly give tighter constraints than the bispectrum. 
In the analysis of LSS surveys it is extremely difficult to extract information 
from the very high-$k$ non-linear modes. This restricts the analysis of the 
bispectrum to relatively low-$k$ modes that are sufficiently linear. The 
situation is changed when one considers biased tracers of the density field. 
These can be related to bispectrum \cite{Matarrese1986}, hence they trace 
non-Gaussian information. The bias effectively allows us to extract some of the 
non-Gaussian information in the high-$k$ modes from the power spectrum analysis. 
This effectively increases the number of modes in the survey, and allows to 
tighten the limits on the non-Gaussian parameters. Still, one would expect that 
the bispectrum analysis contains further information and the question is how it 
compares to the power spectrum analysis of biased tracers. 

For simplicity, we will focus our attention on the local type of 
non-Gaussianity, where the potential shows a self coupling that is local in real space. The local
shape of non-Gaussianity is for instance predicted by multi-field inflation 
\cite{Lyth2003,Bartolo2003,Zaldarriaga2004} and in the bouncing cosmology model
\cite{Creminelli2007}. Recently \cite{Senatore2010} found new shapes with 
none vanishing squeezed limit whose LSS phenomenology is yet to be derived.

This paper breaks down as follows. We first review the basics of
non-Gaussianity in Section \ref{sec:basics} and then describe the multivariate
biasing scheme and peak-background split approach previously introduced by \cite{Slosar2008,Giannantonio2010} in Section
\ref{sec:bias}. Section \ref{sec:pt} introduces the perturbative solutions for
the distribution of matter and biased tracers as well as our new diagrammatic
prescription for the calculation of their $n$-spectra. Section \ref{sec:bispect}
is devoted to the calculation of the halo bispectrum whose constraining power is
compared to the power spectrum in Section \ref{sec:snr}.
We conclude our findings in Section \ref{sec:discussion}.
%===============================================================================
%				BASICS
%===============================================================================
\section{Basics}\label{sec:basics}
We consider the local type of non-Gaussianity
\cite{Salopek1990,Salopek1991,Maldacena2003,Wands2010}
\be
\Phi_\text{nG}(\vec x)=\varphi(\vec x)+\fnl\left(\varphi^2(\vec
x)-\left\langle\varphi^2\right\rangle\right)+\gnl \varphi^3(\vec
x),\label{eq:ngpot}
\ee
where $\varphi$ is an auxiliary primordial Gaussian potential.\footnote{
The coupling of the potentials in Eq.~\eqref{eq:ngpot} is naturally imposed in 
the early Universe during Inflation. This approach, which is followed by our study,
is denoted the CMB convention.  However, some authors impose the same equation in 
the late time evolved Universe (the LSS convention). Therefore one has to be careful 
when comparing quoted constraints on the non-Gaussianity parameters. Namely, the potential evolves as 
\be
\varphi(\vec x,a)=\frac{D(a)}{a} \varphi(\vec x, a_0=1)=g(a)\varphi(\vec
x,a_0=1),
\ee
where $D(a)$ is the linear growth factor normalised to unity at $a_0=1$ and thus
$g(a_0=1)=1$. In an Einstein-de-Sitter Universe the potential is constant in
time, whereas it decays as $g(a=0)/g(a_0=1)=1.34\approx 4/3$ in the
currently favoured $\Lambda\text{CDM}$ model. }
Following the peak-background split approach \cite{Slosar2008} we consider the potential as a 
superposition of small and large scale modes
$\varphi=\varphi_\text{s}+\varphi_\text{l}$ 
separated by a cut-off wavenumber $\Lambda$. Thus from Eq.~\eqref{eq:ngpot}
one
obtains
\be
\Phi_\text{nG}=\varphi_\text{l} + \fnl \varphi_\text{l}^2 +  \gnl
\varphi_\text{l}^3 + (1 + 2
\fnl \varphi_\text{l} + 3 \gnl \varphi_\text{l}^2) \varphi_\text{s} + (\fnl + 3 \gnl
\varphi_\text{l})
\varphi_\text{s}^2 + \gnl \varphi_\text{s}^3,\label{eq:ngpotsplit}
\ee
where all the fields are evaluated at the same spatial position $\vec x$.
Short modes in the above expression can be easily identified since only terms
containing at least one Gaussian short mode can contribute to the short
wavelength power. These short wavelength modes dominate the collapse of dark
matter haloes, whereas the long wavelength modes raise or lower the actual density in large patches of the sky, effectively lowering and raising the collapse threshold. In  the presence of 
non-Gaussianity the long wavelength modes furthermore affect 
the variance of the short modes and thus lead to an additional dependence of
the number density of collapsed objects on the amplitude of the long wavelength
modes. As we will see, in the presence of local non-Gaussianities, this effect is proportional to the value of the long wavelength Newtonian potential, leading to a distinct scale dependent effect. 

The actual effect of the long mode on the variance of the non-Gaussian
short modes can be estimated as follows.
Squaring the short part of the non-Gaussian potential we obtain
\be
\Phi_\text{nG,s}^2= \left(1+4 \fnl \varphi_\text{l}+6 \gnl
\varphi_\text{l}^2+4 \fnlsq \varphi_\text{l}^2+12 \fnl \gnl \varphi_\text{l}^3+9
\gnl^2 \varphi_\text{l}^4\right)\varphi_\text{s}^2,
\ee
of which we can easily compute the expectation over the short modes
\be
\sigma_\text{nG,s}^2=\la\Phi^2_\text{nG,s}\ra_\text{s}=\left(1+4 \fnl \varphi_\text{l}+6 \gnl
\varphi_\text{l}^2+4 \fnlsq \varphi_\text{l}^2+12 \fnl \gnl
\varphi_\text{l}^3+9 \gnl^2
\varphi_\text{l}^4\right)\sigma_\text{G,s}^2,
\label{eq:ngvariance}
\ee
where we identified $\sigma_\text{G,s}^2=\la\varphi_\text{s}^2\ra_\text{s}$.
Here we neglect all correlators of odd number of $\varphi_\text{s}$ as well
as the $\sigma_\text{G,s}^4=\la\varphi_\text{s}^4\ra$ term.
The resulting expression agrees with the expressions previously derived by
\cite{Dalal2008,Slosar2008}.
In contrast to the variance, the three point function or skewness
\be
\la\Phi_\text{nG,s}^3\ra_\text{s}=6\fnl \la\varphi_s^2\ra_\text{s}^2\left(1+4\fnl\varphi_\text{l}\right)+6\gnl \la\varphi_s^2\ra_\text{s}^2\varphi_\text{l}
\label{eq:phibispect}
\ee
vanishes in the Gaussian case. Similar to the variance, the skewness is rescaled
by the long wavelength potential.
\par
\begin{figure}
	\centering
	\includegraphics[width=0.49\textwidth]{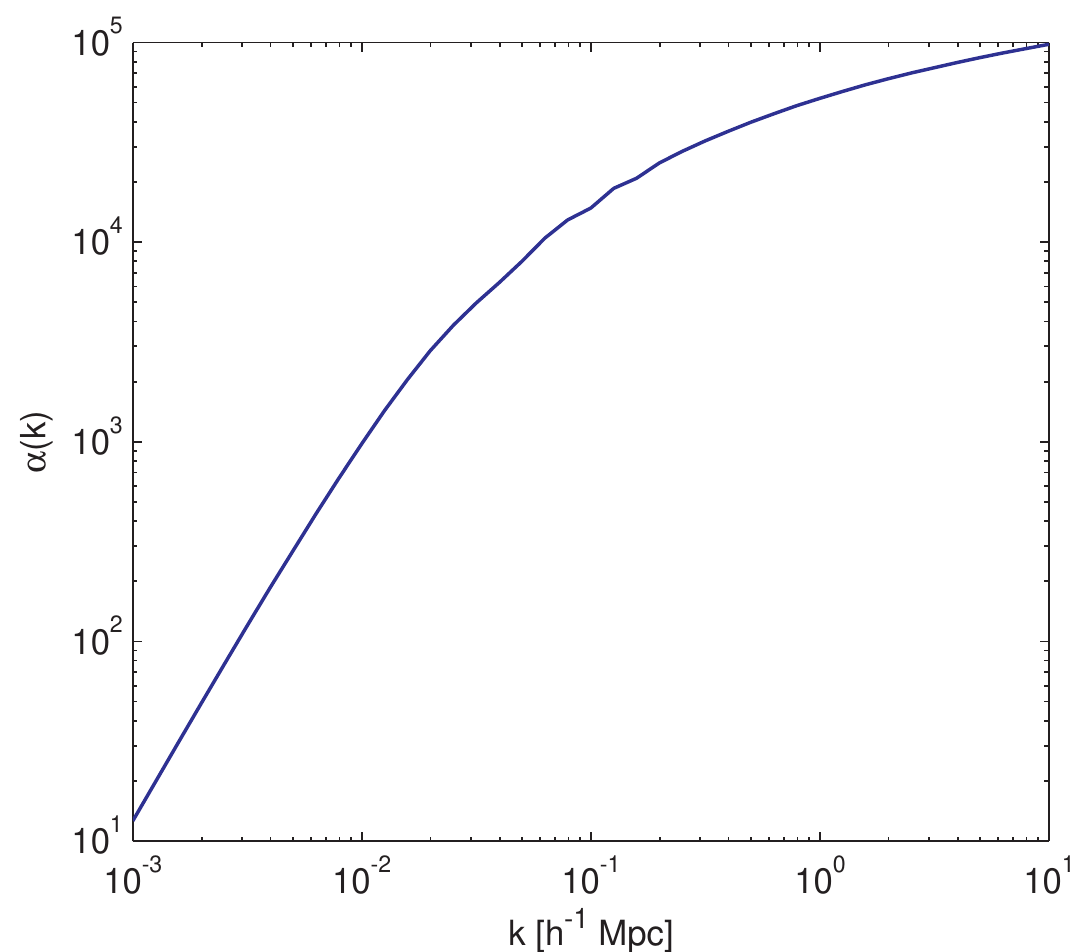}
	\includegraphics[width=0.49\textwidth]{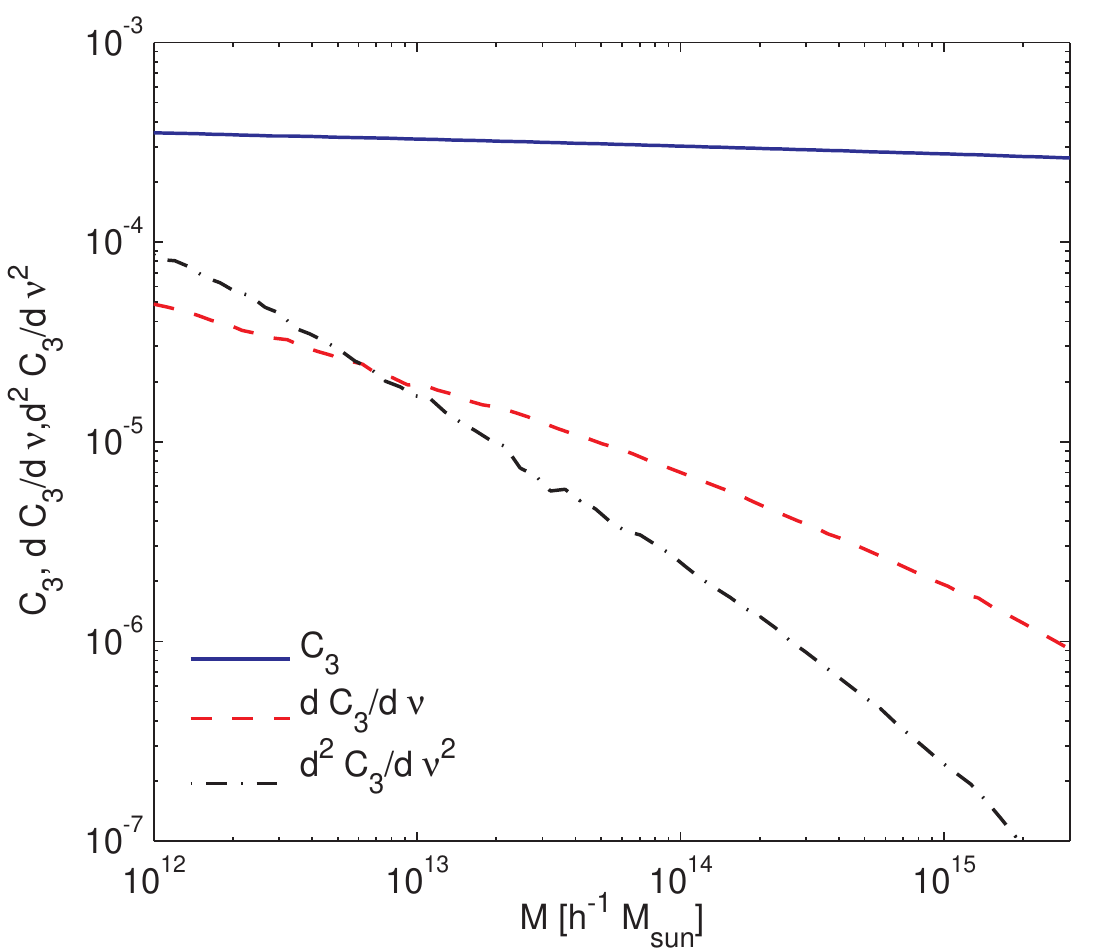}
	\caption{\emph{Left panel: }Poisson factor $\alpha(k)$ relating density
and potential via
	$\delta(k)=\alpha(k) \Phi(k)$. On scales of
	$k \approx 0.1 \ihMpc$ the potential is smaller than the density by a
factor of
	$10^5$, whereas they are equal on very 
	large scales of $k\approx 2\tim{-4} \ihMpc$. The non-Gaussian
corrections
	generally scale as $\fnl/\alpha(k)$ and are thus suppressed on high $k$'s.
	Note that the Poisson equation in Newtonian gauge receives general
	relativistic corrections as $\alpha(k)$ approaches unity.
	\emph{Right panel: }Skewness $C_{3}=\sigma S_3$ and its first and second 
	mass derivatives evaluated for $\fnl=1$ and $\varphi_\text{l}=0$.
	}
	\label{fig:poissfac}	
	\label{fig:skewness}
\end{figure}

Now it remains to connect the non-Gaussian effects on the gravitational
potential to the distribution of matter. In the Newtonian limit, valid well inside the 
horizon, the Poisson equation relates the long wavelength Gaussian potential 
to the density perturbation, 
\be
\Phi(\vec k)=\frac{\delta_\text{p}(\vec k,z)}{\alpha(k,z)},
\label{eq:linpoiss}
\ee
where we introduced the auxiliary function\footnote{Note that we
are not writing explicitly the norm of a vector but use the notation
$k=\left|\vec k\right|$}
\be
\alpha(k,z)=\frac{2 k^2 c^2 D(z) T(k)}{3H_0^2
\Omega_\text{m}}\frac{g(z=0)}{g(z_\infty)},\label{eq:alphaparam}
\ee
which scales as $k^2/H^2$ on large scales where the transfer function is unity. 
As discussed in \cite{Wands2009,Yoo2009,Yoo2010}, on horizon scales unphysical 
gauge modes and relativistic corrections to the Poisson equation require a more 
careful analysis.
Fig.~\ref{fig:poissfac} shows the Poisson factor as a function of 
$k$-mode. Note that the corrections to the Gaussian spectra are given as powers
of $\fnl/\alpha(k)$. The importance of potential and density terms is equal at 
$k\approx2\tim{-4}\ihMpc$ for $\fnl=\mathcal{O}(1)$ and at 
$k\approx2\tim{-3}\ihMpc$ for $\fnl=\mathcal{O}(100)$. Here and in the rest of the paper we use the transfer function for a cosmology with $\Omega_\text{m}=0.25,\ \Omega_\Lambda=0.75,\ \sigma_8=0.8,\ n_\text{s}=1.0$.
%>>>>>>>>>>>>>>>>>>>>>>>>>>>>>>>>>>>>>>>>>>>>>>>>>>>>>>>>>>>>>>>>>>>>>>>>>>>>>>>
%	 			Figure 1
%>>>>>>>>>>>>>>>>>>>>>>>>>>>>>>>>>>>>>>>>>>>>>>>>>>>>>>>>>>>>>>>>>>>>>>>>>>>>>>>
%<<<<<<<<<<<<<<<<<<<<<<<<<<<<<<<<<<<<<<<<<<<<<<<<<<<<<<<<<<<<<<<<<<<<<<<<<<<<<<<

%===============================================================================
%				BIAS
%===============================================================================
\section{Bias from the Universal Mass Function}\label{sec:bias}
Galaxies and their host haloes are believed to trace a smoothed version of the 
underlying distribution of dark matter. The relation between the local overdensity in the matter and
halo
fields is described by the bias function. This bias function can be related to
the
abundance of haloes of mass $M$ described by the halo mass function. In this
Section
we will review the basic Gaussian mass functions and their non-Gaussian
corrections.
Finally, we will present the peak-background split 
and multivariate biasing scheme previously introduced
by
\cite{Slosar2008,Giannantonio2010}.
%===============================================================================
\subsection{Mass Functions}
Haloes are assumed to form at the peaks of the underlying dark matter density
field. Numerical 
simulations and analytical calculations indicate that the abundance of 
collapsed objects can be inferred from the distribution of points that exceed the density threshold $\delta_\text{c}=1.686$. Following the first studies of
\cite{Press1974} (hereafter PS) it was found that the mass function, the number
density  of collapsed objects of mass $M$, reduces to an universal functional
form for 
different redshifts and cosmologies if it is expressed in terms of the
peak height
\be
\nu=\left(\frac{\delta_\text{c}}{\sigma_\text{nG}}\right)^2,
\label{eq:overdensityparam}
\ee
where $\sigma(M)$ is the variance of the density field smoothed on mass scale
$M$. 
This definition of $\nu$ follows e.\,g.\ \cite{Slosar2008}, while other
authors define $\nu=\delta_\text{c}/\sigma$. All the results presented
in this paper can be written in terms of either definition by replacing the
variables accordingly.
Then the number of collapsed objects of mass $M$ can be expressed as
\be
n(M)=\nu f(\nu)\frac{\bar \rho}{M^2}\frac{\derivd \ln{\nu}}{\derivd \ln{M}}.
\label{eq:universalmassfct}
\ee
Using a random walk in a Gaussian density field PS derived
\be
\nu f(\nu)=\sqrt{\frac{\nu}{2\pi}}\eh{-\frac{\nu}{2}},\label{eq:psmassfct}
\ee
which is however not in very good agreement with the mass function measured in
numerical simulations. To improve the agreement, \cite{Sheth1999} (hereafter
ST) proposed a modified version of the Press-Schechter mass function
\be
\nu f(\nu)=A(p)\left(1+\frac{1}{(q
\nu)^p}\right)\sqrt{\frac{q\nu}{2\pi}}\eh{-\frac{q\nu}{2}},\label{eq:stmassfct}
\ee
where the parameters $q=0.707$ and $p=0.3$ were obtained from a fit to numerical
simulations and $A$ is a normalisation factor.
\par
Performing the random walk using non-Gaussian statistics is more involved,
however using an Edgeworth expansion of the exponential \cite{LoVerde2008} (hereafter LV) obtained for the mass function
\be
n_\text{LV}(M)=\sqrt{\frac{2}{\pi}} \frac{\bar \rho}{M}\eh{-\frac{\delta_c^2}{2
\sigma^2}}
\left[\frac{d\ln{\sigma}}{dM}\left(\frac{\delta_c}{\sigma}+\frac{S_3\sigma}{6}
\left(\frac{\delta_c^4}{\sigma^4}-2\frac{\delta_c^2}{\sigma^2}
-1\right)\right)+\frac{1}{6}\frac{dS_3}{dM}\sigma\left(\frac{\delta_c^2}{
\sigma^2}-1\right)\right],\label{eq:loverdemassfct}
\ee
where the skewness is defined as
\be
S_3=\frac{\left\langle\delta_M^3\right\rangle_\text{c}}{
\left\langle\delta_M^2\right\rangle^2}.
\ee
The above derivation based on the Edgeworth expansion is expected to be
satisfactory for low peaks only. As LV show in their Appendix B, higher
cummulants than $S_3$ gain importance for high mass haloes
$M\gtrsim10^{15}\hMs$. As we will see in the next section, bias parameters are
basically derivatives of the mass function, and thus the bias parameters derived
from the LV mass function should be trusted for low and intermediate mass haloes
only. The full treatment of the excursion set theory using non-Markovian random
walks \cite{Maggiore2009} seems to be a promising alternative in the high mass
limit.
In the case of local type non-Gaussianity we obtain for the skewness
\be
C_{3}=\sigma_\text{G} S_3=\frac{6 \fnl}{\sigma_\text{G}^{3}} \int \dqc \int \dqcp \alpha_M(\vec q) \alpha_M(\vec
q')\alpha_M(\vec q+\vec q')P_{\varphi\varphi}(\vec q)P_{\varphi\varphi}(\vec q'),
\ee
where $\alpha_M(k)=W_M(k) \alpha(k)$ and $W_M(k)$ is the filter of mass scale
$M$.  The LV mass function in Eq.~\eqref{eq:loverdemassfct} is not in the form 
of an universal mass function. We can however rewrite it simply applying the chain
rule to $\sigma S_3$.
\begin{align} \nonumber
n_\text{LV}(M)=&\sqrt{\frac{1}{2\pi}} \frac{\bar
\rho}{M^2}\eh{-\frac{\delta_c^2}{2 \sigma^2}}\frac{d\ln{\nu}}{d\ln{M}}
\left[\frac{\delta_c}{\sigma}+\frac{S_3\sigma}{6}
\left(\frac{\delta_c^4}{\sigma^4}-3\frac{\delta_c^2}{\sigma^2}
\right)-\frac{1}{3}\frac{d S_3 \sigma}{d \nu}\left(\frac{\delta_c^4}{\sigma^4}-\frac{\delta_c^2}{\sigma^2}\right)\right]\\ 
=&n_\text{PS}(M)\left[1+\frac{S_3 \sigma}{6}\left(\nu^{3/2}-3\nu^{1/2}\right)-\frac{1}{3}\frac{d S_3 \sigma}{d \nu}\left(\nu^{3/2}-\nu^{1/2}\right)\right]\elnn
=&n_\text{PS}(M)R(\nu),
\label{eq:lvmassfct}
\end{align}
where we introduced the auxiliary function
\be
R(\nu)=1+\frac{S_3 \sigma}{6}\left(\nu^{3/2}-3\nu^{1/2}\right)-\frac{1}{3}\frac{d S_3 \sigma}{d \nu}\left(\nu^{3/2}-\nu^{1/2}\right).
\ee
The presence of the $\sigma S_3$ terms still spoils the universality because this
term carries a mass dependence via the smoothing scale. As shown in 
\cite{Desjacques2009b} and in Fig.~\ref{fig:skewness},
$\sigma S_3$ is only very weakly dependent on the smoothing radius, such that we can safely treat it as a constant in the mass function.

Thus we managed to write the non-Gaussian mass function in the form of an
universal mass function and can benefit from the known results for
universal mass functions.
Usually the LV mass function is multiplied by a correction factor
$\gamma(M)=n_\text{ST}(M)/n_\text{PS}(M)$ to improve the agreement with
simulations, leading to $n_\text{LV}(M)=n_\text{ST}(M)R(\nu)$. The underlying
assumption is that ST corrects PS for triaxial collapse and LV corrects PS for
non-Gaussianity.\footnote{It is not clear how well this statement is
theoretically justified, but it is enough for us for obtaining an estimate on
the non-Gaussian effects on the mass function and the induced biases.}
Alternative derivations of non-Gaussian mass functions that are valid in various, if not all, regimes have been
presented by \cite{Matarrese2000,Maggiore2009,Amico2011}. For our study, the
important
result following from these analyses is that to a good approximation, all of the
above mass functions can be treated as universal, that is as being function of
$\delta_\text{c}/\sigma$ only, and that the inferred values of the biases are
not very different. While we are aware of disadvantages of both the Gaussian
and non-Gaussian mass functions, it goes beyond the scope of this paper to
discuss their detailed validity or to develop an improved mass function.
%===============================================================================
\subsection{Multivariate Lagrangian Bias}
The spherical top hat collapse model states that a spherical overdensity
collapses to form a gravitationally bound object once it exceeds a certain
overdensity threshold $\delta_c$. Long wavelength density perturbations raise or
lower the mean density in a patch of the Universe and thus effectively raise or
lower the collapse threshold $\delta_\text{c}\to\delta_
\text{c}-\delta_\text{l}$. So far, the bias parameters were calculated by
expanding the local number density in the amplitude of the long wavelength
density fluctuation $\delta_\text{l}$ only. As we saw above in
Eq.~\eqref{eq:ngvariance}, coupling between long and short modes
leads to an enhancement of the variance of the short wavelength density
perturbations. This enhancement is proportional
to powers of the long wavelength Gaussian potential and thus suggests to
calculate the distribution of collapsed objects by expanding the local number
density in terms of the long wavelength modes of both density and
potential \cite{Slosar2008,McDonald2008,Giannantonio2010}
\begin{align}
n(\vec x)=&\bar{n}+\frac{\partial n}{\partial\delta_\text{l}}\delta_\text{l}(\vec
x)+\frac{\partial n}{\partial\varphi_\text{l}}\varphi_\text{l}(\vec
x)+\frac{1}{2}\frac{\partial^2n}{\partial\delta_\text{l}^2}\delta_\text{l}^2(\vec
x)+\frac{2}{2!}\frac{\partial^2n}{\partial\delta_\text{l} d\varphi_\text{l}}\delta_\text{l}(\vec
x)\varphi_\text{l}(\vec
x)+\frac{1}{2}\frac{\partial^2n}{\partial\varphi_\text{l}^2}\varphi_\text{l}
^2(\vec x).
\end{align}
Here we write down explicitly the spatial dependence to highlight the local
relation between bias, overdensity and potential. 
The multivariate bias expansion  is not stating that density and potential are
independent parameters, but rather that the different scale dependence of
density and potential requires one to expand in both of them in order to keep
the bias parameters scale independent. As we will stress at the end of this subsection, it is important that these fields are restricted to long-wavelengths. Notice that the fact that the overdensity depends locally only on $\delta$ and $\varphi$ is a consequence of the functional dependence of the mass function on $\delta$ and $\varphi$ even in the non-Gaussian case. 
Defining
$\delta_\text{h}(\vec x)= n(\vec x)/\bar{n}-1$ and identifying the partial
derivatives with the bias parameters we obtain in Lagrangian space
\be
\delta_\text{h}^\text{L}(\vec x)=b_{10}^\text{L}\delta_0(\vec x)+b_{01}^\text{L}
\varphi(\vec x)+\frac { b_
{ 20 } ^\text{L}}{2!}\delta_0^2(\vec x)+b_{11}
^\text{L}\delta_0(\vec
x)\varphi(\vec x)+\frac{b_{02}^\text{L}}{2!}\varphi^2(\vec
x)+\ldots,\label{eq:lagbiasexp}
\ee
where $\delta_0$ is the initial Lagrangian overdensity.
\par 
We can now calculate the Lagrangian bias parameters under the assumption that
the local number density can be expressed with an universal
mass function as in Eq.~\eqref{eq:universalmassfct}. 

The presence of a long wavelength mode can be accounted for by replacing
$\delta_\text{c}\to \delta_\text{c}-\delta_\text{l}$ and
$\sigma_\text{G}\to\sigma_\text{nG}$ in the peak height, such that
that the conditional peak height in presence of a long wavelength mode can be
written as
\be
\tilde{\nu}=\left(\frac{\delta_\text{c}-\delta_\text{l}}{\sigma_\text{G}
(1+2\fnl\varphi_\text{l}+3\gnl \varphi_\text{l}^2+2 f_\text{NL}^2
\varphi_\text{l}^2) }\right)^2
\ee
and the bias parameters are given by (see Appendix \ref{app:B} for details on
the calculation)

\begin{align}
b_{10}^\text{L}=&\frac{1}{\bar{n}}\partfrac{n}{\delta_\text{l}}=-\frac{1}{\bar
n}\frac{2\nu}{
\delta_\text{c} }
\partfrac{n}{\nu}\\
b_{01}^\text{L}=&\frac{1}{\bar{n}}\partfrac{n}{\varphi_\text{l}}
=-\frac{4\fnl\nu}{\bar n}\partfrac { n } { \nu}
=2\fnl\delta_\text{c} b_{10}^\text{L}\\
b_{20}^\text{L}=&\frac{1}{\bar{n}}\partsqfrac{n}{\delta_\text{l}}=\frac{4}{\bar
n}
\frac{\nu^2}{\delta_c^2}\partsqfrac{n}{\nu}+\frac{2}{\bar{n}}\frac{\nu}{
\delta_\text{c}^2}\partfrac{n}{
\nu}\\
b_{11}^\text{L}=&\frac{1}{\bar{n}}\frac{\partial^2 n}{\partial
\varphi_\text{l}\partial\delta_\text{l}}=\frac{8\fnl}{\bar
n}\left(\frac{\nu^2}{\delta_\text{c}}
\partsqfrac { n } { \nu }
+\frac{\nu}{\delta_\text{c}}\partfrac{n}{\nu}\right)\\
=&2\fnl\left(\delta_\text{c} b_{20}^\text{L}-b_{10}^\text{L}\right)\\
b_{02}^\text{L}=&\frac{1}{\bar{n}}\partsqfrac{n}{\varphi_\text{l}}
=\frac{8\fnlsq}{\bar n}\left(2\nu^2\partsqfrac{n}{\nu}+3\nu\partfrac{n}{\nu}
\right)-\frac{12\nu\gnl}{\bar n}\partfrac{n}{\nu}\\
=&4\fnlsq\delta_\text{c}\left(b_{20}^\text{L}
\delta_\text{c}-2b_{10}^\text{L}\right)+6\delta_\text{c}\gnl b_{10}^\text{L}
\end{align}
where all the derivatives are evaluated
for $\delta_\text{l}=0,\varphi_\text{l}=0$.
We left the derivatives of the mass function unevaluated and obtained
expressions that are sufficiently general to enable application to different
Gaussian and non-Gaussian mass functions.\footnote{Actually one only has to
calculate the derivative of $\nu f(\nu)$ since the proportionality factors
cancel out.} For instance, the derivatives of the
ST mass function Eq.~\eqref{eq:stmassfct} are given by
\begin{align}
\frac{1}{n_\text{ST}}\frac{\partial n_\text{ST}}{\partial
\nu}=&-\frac{q\nu-1}{2\nu}-\frac{p}{\nu\left(1+(q\nu)^p\right)}\\
\frac{1}{n_\text{ST}}\frac{\partial^2 n_\text{ST}}{\partial
\nu^2}=&\frac{p^2+\nu p
q}{\nu^2\left(1+(q\nu)^p\right)}+\frac{(q\nu)^2-2q\nu-1}{4\nu^2}
\end{align}
and for the LV mass function Eq.~\eqref{eq:lvmassfct} by
\begin{align}
\frac{1}{n_\text{LV}}\frac{\partial n_\text{LV}}{\partial
\nu}=&\frac{1}{n_\text{ST}}\frac{\partial n_\text{ST}}{\partial
\nu}+\frac{1}{R}\frac{\partial R(\nu)}{\partial
\nu}\label{eq:lvbiascorr1}\\
\frac{1}{n_\text{LV}}\frac{\partial^2 n_\text{LV}}{\partial \nu^2}=&
\frac{1}{n_\text{ST}}\frac{\partial^2 n_\text{ST}}{\partial
\nu^2}+2\frac{1}{n_\text{ST} R}\frac{\partial n_\text{ST}}{\partial
\nu}\frac{\partial R(\nu)}{\partial
\nu}+\frac{1}{R}\frac{\partial^2 R(\nu)}{\partial
\nu^2}\label{eq:lvbiascorr2}.
\end{align}
\par
The amplitude of the halo power spectrum at a given wavelength must not
depend on the smoothing scale, as it is an observable. Let us imagine to perform
the one loop computation of the halo power spectrum with two different smoothing
scales $\Lambda_1$ and $\Lambda_2$, with $\Lambda_2>\Lambda_1$. Since the final
answer must not change as we change $\Lambda$, we need to renormalize the bias
accordingly. This means that the renormalized bias at scale $\Lambda_1$ is
related to the renormalized bias at scale $\Lambda_2$ by a relation of the
following form \cite{McDonald2006}
\be
b_{10}^{\Lambda_2}=b_{10}^{\Lambda_1}+\left(b_{20}^{\Lambda_1}\frac{68}{21}+b_{
30}\right)\sigma^2_{\Lambda_2,\Lambda_1}
\ee 
where $\sigma^2_{\Lambda_2,\Lambda_1}$ is the real space variance computed
including only wavenumbers between $\Lambda_1$ and $\Lambda_2$. Notice that the
term proportional to 68/21 arises from having inserted a gravitational
interaction vertex $F_2$. This tells us how the renormalized bias parameters
change as we change the smoothing scale and include loops that renormalize the
bias parameters.

\par
In the PBS method we assume that the long mode is infinitely
long. This corresponds to having taken the smoothing scale $\Lambda$ to be zero
(or equivalently infinitely long smoothing length). In this regime, loop
corrections, that always include only modes that run from zero wavenumber to the
smoothing scale, are zero by construction, as we took $\Lambda=0$. This tells us
that the PBS method provides already the renormalized bias parameters at
infinite smoothing length. As we change the smoothing length and we make it
shorter and shorter, we should include loops and renormalize the bias parameters
accordingly following formulae similar to the one above. However, the above
formula is by construction such that the effective bias parameters do not change
as we change the smoothing scale. Therefore, as we change the smoothing scale,
we can simply avoid including the loops, like the ones above, that renormalize
the bias parameters and use directly the bias parameters that we obtain at
infinite smoothing length. This will lead to the same answer as if we had
included all the loop corrections and changed the bias parameters accordingly.
Although this statement has not been carefully verified from a theoretical
point of view, the corrections are anyways quite small on large scales and
the model seems to be in accord with what is inferred from simulations.
In fact in~\cite{Slosar2008,Dalal2008}, a good agreement
between peak background split predictions for the non-Gaussian bias $b_{01}$
and simulations has been found.
Although there is evidence for a weak dependence of the bias
parameters on the smoothing scale, it has been shown by  \cite{Padmanabhan2009,
Manera2009} that even the second order bias $b_{20}$ is quite well approximated
by the local bias model in Fourier space.
A more precise treatment of this point goes beyond the
scope of the present paper as it would require calculations at loop
level and comparison to numerical simulations.
\par
However, the fact that the peak background split method leads directly to the 
renormalized bias parameters is independent of the assumption of Gaussian initial 
conditions. Thus it is not unreasonable to assume that this fact also extends to
the non-Gaussian bias parameters used in our study. 
\par
The derivation presented so far assumes an universal mass function,
\emph{i.\,e.}~that
all the dependence on the long modes is implicitly encoded in the peak height $\nu$. 
However, intrinsically non-Gaussian properties of the distribution enter in the 
non-Gaussian mass function, for example $S_3\sigma$ in the case of the LV mass
function. Following an argument similar to the one we used to derive the
$\varphi_\text{l}$ 
dependence
of the variance in Eq.~\eqref{eq:ngvariance}, one can show based on Eq.~\eqref{eq:phibispect} 
that the three point function in the presence of a long fluctuation gets rescaled as 
$\la\delta_M^3\ra\to\la\delta_M^3\ra(1+4\fnl \varphi_\text{l})$.
This dependence on $\varphi_\text{l}$ can not be encoded in $\nu$ and thus an additional
explicit derivative with respect to the long wavelength potential arises, which leads to
the following corrections to the bias parameters, for example for the LV mass function:
\begin{align}
\Delta b_{01}=&\frac{2}{3}\frac{\fnl C_{3}}{R}\left(\nu^{3/2}-3\nu^{1/2}\right)\label{eq:db01}\\
\Delta b_{11}=&-\frac{2}{3}\frac{ \fnl C_{3} \nu}{R \delta_c}\left[\frac{1}{n_\text{ST}}\frac{\partial n_\text{ST}}{\partial \nu}\left(\nu^{3/2}-3\nu^{1/2}\right)+\frac{3}{2}\left(\nu^{1/2}-\nu^{-1/2}\right)\right]\label{eq:db11}\\
\Delta b_{02}=&-\frac{16}{3}\frac{f_\text{NL}^2\nu C_{3}}{R}\left[\frac{1}{n_\text{ST}}\frac{\partial n_\text{ST}}{\partial \nu}\left(\nu^{3/2}-3\nu^{1/2}\right)+\frac{3}{2}\left(\nu^{1/2}-\nu^{-1/2}\right)\right]\label{eq:db02}
\end{align}
These corrections are generally of the same order 
as the bias corrections arising from the non-Gaussian LV mass function in Eq.~\eqref{eq:lvbiascorr1}
and \eqref{eq:lvbiascorr2}. However, for high $\nu$ the latter dominate. For realistic values
of $\fnl$ all the bias corrections arising from the non-Gaussian mass function
are on the percent level. The mass function itself can be trusted at the $10 \%$
level only and thus these corrections can be safely neglected.

%===============================================================================
\subsection{Transformation to Eulerian Space}
%>>>>>>>>>>>>>>>>>>>>>>>>>>>>>>>>>>>>>>>>>>>>>>>>>>>>>>>>>>>>>>>>>>>>>>>>>>>>>>>
%	 			Figure 2
%>>>>>>>>>>>>>>>>>>>>>>>>>>>>>>>>>>>>>>>>>>>>>>>>>>>>>>>>>>>>>>>>>>>>>>>>>>>>>>>
\begin{figure}
	\centering
	\includegraphics[width=0.49\textwidth]{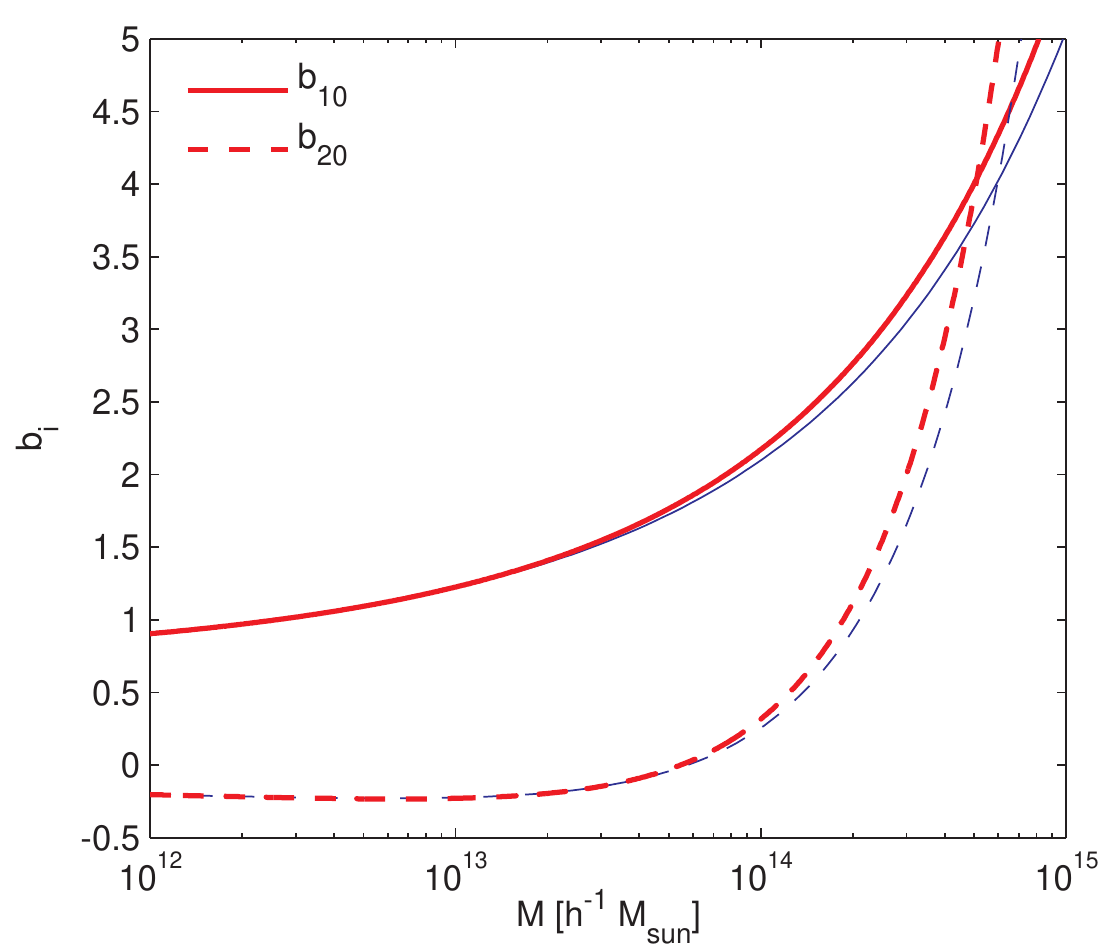}
	\includegraphics[width=0.49\textwidth]{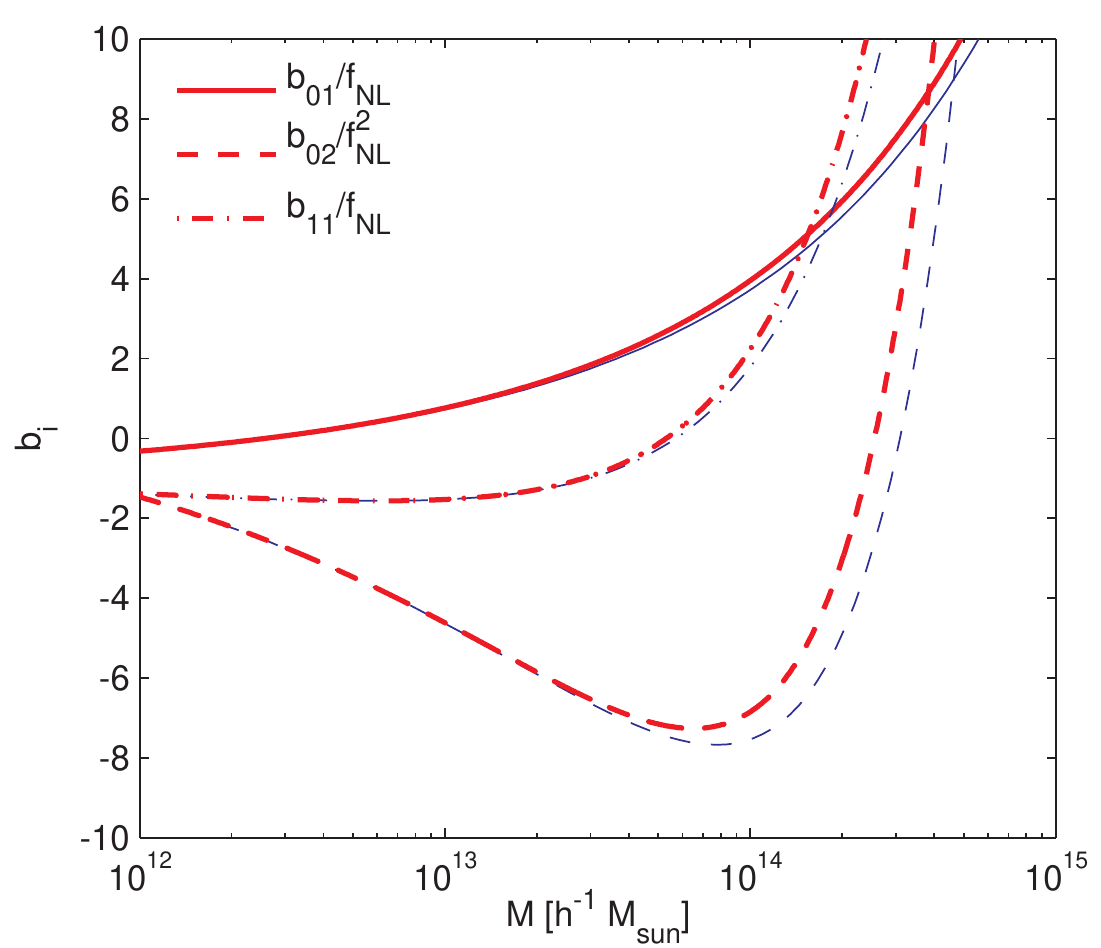}
	\caption{Eulerian multivariate bias parameters for $\fnl=100$, $\gnl=0$. 
	\emph{Left panel:} Bias parameters for $\delta$. 
	\emph{Right panel:} Bias parameters for $\varphi$.
	For $\gnl=0$ one has $b_{02}\propto \fnlsq$, $b_{01}\propto \fnl$, $b_{11}\propto
	\fnl$ and thus the rescaled bias parameters are independent of $\fnl$.
	$b_{02}$ shown by the dashed line shows a pronounced minimum for
	$M\approx 1\tim{14}\hMs$ that is multiplied by $\fnlsq$ and can thus
	lead to a large contribution. The red lines are derived from the ST
	mass function, whereas the blue thin lines are derived from the LV mass
	function including the explicit $\varphi_\text{l}$ correction of 
	Eqs.~\eqref{eq:db01}, \eqref{eq:db11} and \eqref{eq:db02}.}
	\label{fig:biases}
\end{figure}
%<<<<<<<<<<<<<<<<<<<<<<<<<<<<<<<<<<<<<<<<<<<<<<<<<<<<<<<<<<<<<<<<<<<<<<<<<<<<<<<
Observations are performed in the late time evolved Eulerian density field. It
thus remains to translate the above result to Eulerian space. The halo density
fields in Eulerian and Lagrangian space are related by \cite{Mo1996a}
\be
\left(1+\delta_\text{h}^\text{E}\right)=(1+\delta)(1+\delta_\text{h}^\text{L}).
\ee
Finally, one wants to write down the Eulerian analogue of
Eq.~\eqref{eq:lagbiasexp}
\be
\delta_h^\text{E}(\vec x)
=b_{10}^\text{E}\delta(\vec x)+b_{01}^\text{E}\varphi(\vec x)+\frac{b_{20 }
^\text{E}}{2!}\delta^2(\vec x)+b_{11}
^\text{E}\delta(\vec x)\varphi(\vec
x)+\frac{b_{02}^\text{E}}{2!}\varphi^2(\vec x)+\ldots \label{eq:eulbiasexp}
\ee
In what follows we will absorb the prefactors in the bias parameters, thus
$b_{20}/2!\to b_{20}$ and $b_{02}/2!\to b_{02}$.
The linearly evolved Lagrangian overdensity can be expanded in powers of the final
Eulerian overdensity 
\be
\delta_0=\sum_i a_i \delta^i=a_1 \delta +a_2\delta^2 +a_3\delta^3+\ldots
\ee
where the expansion parameters $a_i$ are given by the spherical collapse
dynamics $a_1=1,\ a_2=-17/21,\ a_3=341/567$.
We calculated the corresponding Eulerian bias parameters using the
relations in \cite{Mo1996a,Mo1996b,Giannantonio2010}
\begin{align}
b_{10}^\text{E}=&1+b_{10}^\text{L}\\
b_{20}^\text{E}=&2(a_1+a_2) b_{10}^\text{L}+a_1^2 b_{20}^\text{L}\\
b_{01}^\text{E}=&b_{01}^\text{L}\\
b_{02}^\text{E}=&b_{02}^\text{L}\\
b_{11}^\text{E}=&b_{11}^\text{L}a_1+b_{01}^\text{L}
\end{align}
\par
Calculations of the evolved density field are most conveniently done in Fourier
space. Thus we translate the above expansion Eq.~\eqref{eq:eulbiasexp} to
$k$-space.
\be
\delta_\text{h}(\vec k)=b_{10}^\text{E}\delta(\vec
k)+b_{01}^\text{E}\varphi(\vec k)+b_{20} ^\text { E } [ \delta*\delta](\vec
k)+b_{
11}^\text{E}[\delta*\varphi](\vec k)+b_{02}^\text{E}[\varphi*\varphi](\vec k),
\label{eq:biasexpk}
\ee
where we introduced the notation $[\delta*\delta](\vec k)$ as a shorthand for
the convolution integral. From now on we will omit the superscript E and
all bias factors should be understood as the Eulerian ones.

\par
In Fig.~\ref{fig:biases} we show the mass dependence of the bias parameters
arising from the Gaussian ST and the non-Gaussian LV mass function.
The bias factors of the potential and the density-potential cross term are
scaled by their $\fnl$ dependence. From this figure it is obvious that the
second order biases can be negative over large parts of the interesting mass
range. This can lead to subtle cancellations between terms in the resulting
galaxy or halo $n$-point functions.
Furthermore, the $\fnlsq$ dependence and the pronounced minimum in $b_{02}(M)$
around $M=10^{14} \hMs$ can boost the corresponding terms on large scales.
As noted already by \cite{Desjacques2009a} and apparent in
Fig.~\ref{fig:biases},
 the non-Gaussian mass function Eq.~\eqref{eq:lvmassfct} leads to an additional 
scale independent offset in the bias. 
From Fig.~\ref{fig:biases} we see that the difference in the bias parameters is
most apparent for high mass haloes, whereas there is only a small difference for
low mass haloes. The differences 
between the bias from the two mass functions are very small and negligible for our 
purposes. The same is approximately true even for the other non-Gaussian mass 
functions. Thus we use the Gaussian ST mass function for our numerical 
predictions.
%===============================================================================
%				PERTURBATION THEORY
%===============================================================================
\section{Perturbation Theory including non-Gaussianity}\label{sec:pt}
%===============================================================================
\subsection{Matter Density Field}
Perturbation theory (PT) aims to solve the cosmological fluid equations
using an expansion about the linear overdensity $\delta_\text{m}^{(1)}(\vec k)$
\cite{Bernardeau2002}
\be
\delta_\text{m}(\vec k)=\delta_\text{m}^{(1)}(\vec k)+\delta_\text{m}^{(2)}(\vec
k)+\delta_\text{m}^{(3)}(\vec k)+\ldots.
\label{eq:matterexpansion}
\ee
In our approach, these modes should be thought of as having wavelength longer that the cutoff $\Lambda^{-1}$ that separates short and long modes.
Let us define a linearly evolved primordial density field ({\it i.e.} evolved without taking into account the gravitational evolution), to be $\delta_\text{m,p}$. 
At first order, $\alpha(k,z)\varphi(\vec
k)=\delta_\text{m,p}^{(1)}(\vec k,z)$, \emph{i.\,e.\ }the primordial Gaussian
potential $\varphi$ is of the same order as the primordial linear overdensity
$\delta^{(1)}_\text{m,p}$. In the following we refer to these two quantities
as first order quantities and count the order of terms by counting the powers of
first order terms. 
\par
The non-Gaussian self-coupling of the potential introduces non-linearities
in the evolved primordial density field, whereas in the Gaussian case we 
would have $\delta_\text{m,p}=\delta_\text{m,p}^{(1)}$. Transforming Eq.~\eqref{eq:ngpot} 
to Fourier space and applying Eq.~\eqref{eq:linpoiss} we obtain for the linearly 
evolved primordial matter density up to third order
\begin{align}
\delta_\text{m,p}(\vec k,z)=&\alpha(k,z) \Phi(\vec k)\\
=&\alpha(k,z) \varphi(\vec k)+\alpha(k,z) \fnl \int \dqc
\varphi(\vec q)\varphi(\vec k-\vec q)\nonumber\\
+&\alpha(k,z) \gnl \int \dqc\int \dqcp
\varphi(\vec q)\varphi(\vec q')\varphi(\vec k-\vec q-\vec q'),\\
=&\delta_\text{m,p}^{(1)}(\vec k,z)+\fnl \delta_\text{m,p}^{(2)}(\vec k,z)+\gnl
\delta_\text{m,p}^{(3)}(\vec k,z)\label{eq:ngexp}.
\end{align}
The resulting field is then subject to
late-time non-linear gravitational clustering, which introduces further
couplings. To take this effect into account we use the evolved primordial distribution
as the source for the late time evolution and insert Eq.~\eqref{eq:ngexp} into
the known PT expressions for $\delta_\text{m}^{(1)}$,
$\delta_\text{m}^{(2)}$ and $\delta_\text{m}^{(3)}$ (see appendix A) 
\begin{align}
\delta^{(1)}_\text{m,nG}(\vec k,z)=&\delta_\text{m,p}^{(1)}(\vec k,z),\\
\delta^{(2)}_\text{m,nG}(\vec k,z)=&\int \dqc \delta^{(1)}_\text{m,p}(\vec
q)\delta^{(1)}_\text{m,p}(\vec k-\vec q)
F_2(\vec q,\vec k-\vec q)\nonumber\\
+&\fnl \delta_\text{m,p}^{(2)}(\vec k,z),\\
\delta^{(3)}_\text{m,nG}(\vec k,z)=&\int \dqc\int \dqcp
\delta^{(1)}_\text{m,p}(\vec q,z)\delta^{(1)}_\text{m,p}(\vec
q',z)\delta^{(1)}_\text{m,p}(\vec k-\vec q-\vec q',z) F_3(\vec q,\vec q',\vec
k-\vec q)\nonumber\\
+& 2\fnl \int \dqc \delta^{(1)}_\text{m,p}(\vec q,z)
\delta^{(2)}_\text{m,p}(\vec k-\vec q,z)F_2(\vec q,\vec k-\vec q)\nonumber\\
+& \gnl \delta_\text{m,p}^{(3)}(\vec k,z),
\end{align}
where $F_2(\vec k_1,\vec k_2)$ and $F_3(\vec k_1,\vec k_2,\vec k_3)$ are the standard second and third order mode coupling kernels.

%===============================================================================
\subsection{Halo Density Field}
The halo density including higher order corrections from biasing,
non-Gaussianity and non-linear clustering can be derived using 
Eq.~\eqref{eq:matterexpansion} in Eq.~\eqref{eq:biasexpk}
\begin{align}
\delta_\text{h}(\vec k)=&b_{10}\left(\delta_\text{m,nG}^{(1)}(\vec k)+\delta_\text{m,nG}^{(2)}(\vec k)\right)+b_{01}\varphi(\vec k)\nonumber\\
+&b_{20}\bigl[\delta_\text{m,nG}^{(1)}*\delta_\text{m,nG}^{(1)}\bigr](\vec k)+b_{11}\bigl[\delta_\text{m,nG}^{(1)}*\varphi\bigr](\vec k)
+b_{02}\bigl[\varphi*\varphi\bigr](\vec k)\\
=&b_{10}\delta_\text{m,p}^{(1)}(\vec
k)+b_{01}\varphi(\vec k)\elnn
+&b_{02}\int \dqc \varphi(\vec q)\varphi(\vec k-\vec q)+b_{20}\int \dqc
\delta_\text{m,p}^{(1)}(\vec q)\delta_\text{m,p}^{(1)}(\vec k-\vec q)\elnn
+&b_{10}\int \dqc \delta_\text{m,p}^{(1)}(\vec q)\delta_\text{m,p}^{(1)}(\vec
k-\vec q)F_2(\vec q,\vec
k-\vec q)+\alpha(k) \fnl b_{10}\int \dqc \varphi(\vec q)\varphi(\vec k-\vec
q)\elnn
+&b_{11}\int \dqc \delta_\text{m,p}^{(1)}(\vec q)\varphi(\vec k-\vec q).
\end{align}
Note that at lowest order we recover the well known result of \cite{Dalal2008}
\be
\delta_\text{h}(\vec k)=\left(b_{10}+\frac{2\fnl\delta_\text{c}
(b_{10}-1)}{\alpha(k)}\right)\delta_\text{m,p}(\vec k).\label{eq:dalalfield}
\ee
At this order the potential can be replaced by the density due to the linearity
of the Poisson equation in $k$-space. Thus one finally obtains an expression
that is proportional to the density, but non-local. The next to leading order, however, is not
proportional to
$\delta^2(\vec k)$ since the density and the potential are convolved with each
other.
It is due to this effect that the bias expansion should be performed both in
$\delta$ and $\varphi$.
\par
We would like to stress an important subtlety concerning the usage of the
bias parameters obtained from the peak background split in perturbation
theory. Depending on the precision required for the calculation, it is possible
that higher order corrections from perturbation theory need to be implemented.
In the 
diagrammatic description, which is explained in the next subsection, there are 
loop diagrams (convolutions of the fields with some kernel) 
involving non-linear bias vertices. Unfortunately, some of these diagrams
are highly dependent on the cutoff $\Lambda$ ({\it i.e.} the
smoothing scale). 
Clearly, physical quantities should not depend on the smoothing scale. This 
spurious dependence can be removed by defining effective or, more 
precisely, {\it renormalized} bias parameters that take into account both the tree 
level and the loop contributions and that are directly connected 
to observable quantities. 
For example, it is possible to 
show that loop corrections to the power spectrum originating for example from 
$b_{20}$ and $b_{30}$ terms effectively renormalize $b_{10}$ and 
$b_{01}$~\cite{McDonald2006,McDonald:2008sc}.\footnote{In fact, it is easy to 
estimate the one loop contribution due to $b_{20}$ combined with an $\fnl$ 
vertex and an $F_2$ vertex (see next section for these definitions). This
diagram induces an effective bias $b_{01}$ numerically equivalent to the one
obtained from the peak-background split if one considers small external $k$'s, a
very high mass scale  and most importantly sets the smoothing scale equal to the
mass scale of the halo as done in \cite{Jeong2009}. This is an effect that
arises from pushing the smoothing scale to very high $k$'s, where it is unclear
whether perturbative calculations can be trusted.
These renormalized parameters do not depend on the 
smoothing scale, while the coupling constants with which we perform perturbation 
theory and the contributions from loop diagrams do depend on the cutoff
$\Lambda$. This is a behavior familiar from quantum field theory. 
Actually, as mentioned in the former section, the bias parameters inferred
from the peak background split method can be interpreted as the renormalized
ones. Therefore, loop diagrams causing the renormalization have been already
accounted for. This guarantees that the tree level
calculation presented here provides the dominant contribution on very large
scales. A
more precise understanding of this point lies beyond the scope of the present
paper, where we concentrate only on the tree-level calculation and do not deal
explicitly with loop orders.}
\par
Finally, it should be stressed that even the remaining perturbation theory 
should be performed in a way that ensures that the remaining contributions are
independent of the smoothing length $\Lambda^{-1}$. This suggests to use a
carefully defined 
perturbation theory, such as for example `renormalized perturbation
theory'~\cite{Crocce2006}, 
or the recently proposed effective fluid description of cosmological 
perturbations~\cite{Baumann2010}.
%===============================================================================
\subsection{Diagrammatic Representation}
In the previous section we derived the perturbative expressions for the matter
and halo density fields. As a consequence of the stochastic nature of
cosmological fluctuations, there is no hope to directly predict the
observed distribution of galaxies and matter in the Universe. Rather, we need to
calculate expectation values of products of the fields and compare them to the
corresponding statistics as measured in the sky. The calculation of these
statistics turns out to be an involved combinatorial task if one goes beyond
second order in the fields.
\par
To facilitate these calculations, we present a diagrammatic representation of the
mode coupling terms that arise from biasing, non-linear clustering and
non-Gaussianity. Similar diagrammatic representations of perturbation theory have
been used in the literature \cite{Goroff1986,Scoccimarro1996, Bernardeau2002,Matsubara2008}
but we are not aware of an intuitive inclusion of all the three effects into one
prescription. These Feynman diagrams show intuitively which coupling terms
arise and can be translated into the corresponding equations by straightforward
application of a set of Feynman rules.
\par
\begin{figure}[h]
	\centering
	\includegraphics{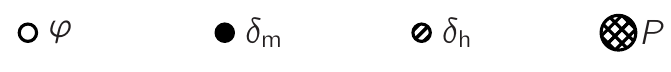}%[width=0.49\textwidth]
	\caption{Symbols use to represent the fields and power spectra.
From left to right, the primordial Gaussian potential $\varphi$, 
the matter density field $\delta_\text{m}$,
the galaxy/halo
density field $\delta_\text{h}$, and
the power spectrum $P(k)$ arising from two linked fields.
}
	\label{fig:fields}
\end{figure}

Let us start representing the fields, as the basic ingredients of the theory.
What we want to calculate in the end are correlators of halo or 
matter density fields, thus we need symbols for the outer points, namely 
$\delta_\text{h}$ and $\delta_\text{m}$. The latter two are represented by the
half filled and filled circles depicted in Fig.~\ref{fig:fields}. Note that
when $\delta_\text{m}$ is used for an outer point it always includes all
the possible non-Gaussian and non-linear contributions up to the considered 
order, whereas the density as a source field is linear.
Next, we consider the primordial potential $\varphi$, represented by an open 
circle. Even if $\alpha(k)\varphi(\vec k)=\delta_\text{m,p}^{(1)}(\vec k)$, we 
introduce symbols for both the density and potential to make the $1/k^2$ 
behavior of the potential terms more obvious and to make sure that the potential
terms arise only directly from the initial conditions. 
However, no difference is made between the evolved primordial and the late time
non-linear matter density field in terms of the symbols, because they can be 
distinguished from the context. For instance, the coupling vertices for 
gravity are sourced by evolved primordial matter fields defined in Eq.~\eqref{eq:ngexp} 
and lead to non-linear fields (see the discussion of the vertices below for more details).
Finally, the initial conditions are known in terms of the power spectra of 
fluctuations. Thus we also introduce the power spectrum symbolized by the
half filled big circle, where subscripts are used to distinguish the density-density,
density-potential and potential-potential power spectra.

%
%>>>>>>>>>>>>>>>>>>>>>>>>>>>>>>>>>>>>>>>>>>>>>>>>>>>>>>>>>>>>>>>>>>>>>>>>>>>>>>>
%	 			Figure 3
%>>>>>>>>>>>>>>>>>>>>>>>>>>>>>>>>>>>>>>>>>>>>>>>>>>>>>>>>>>>>>>>>>>>>>>>>>>>>>>>
%<<<<<<<<<<<<<<<<<<<<<<<<<<<<<<<<<<<<<<<<<<<<<<<<<<<<<<<<<<<<<<<<<<<<<<<<<<<<<<<
%
%>>>>>>>>>>>>>>>>>>>>>>>>>>>>>>>>>>>>>>>>>>>>>>>>>>>>>>>>>>>>>>>>>>>>>>>>>>>>>>>
%	 			Figure 4
%>>>>>>>>>>>>>>>>>>>>>>>>>>>>>>>>>>>>>>>>>>>>>>>>>>>>>>>>>>>>>>>>>>>>>>>>>>>>>>>
\begin{figure}[h]
	\centering
	\includegraphics{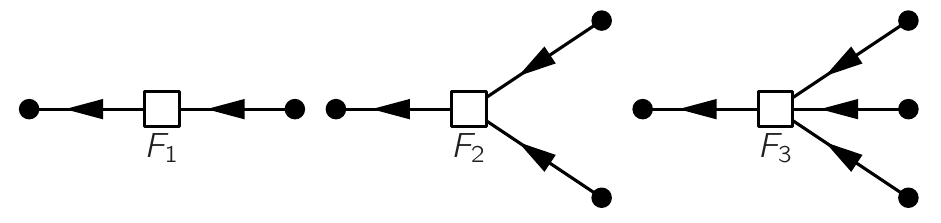}
	\caption{Non-linear gravitational clustering leads to a convolution
integral weighted by a $F_i$ kernel, which we symbolize by a square shaped
vertex. From left to right we show the first, second and third order mode
coupling contributions to the matter density field. Note that $F_1\equiv 1$.}
	\label{fig:clustering}
\end{figure}
%<<<<<<<<<<<<<<<<<<<<<<<<<<<<<<<<<<<<<<<<<<<<<<<<<<<<<<<<<<<<<<<<<<<<<<<<<<<<<<<
\par
\begin{figure}[h]
	\centering
	\includegraphics{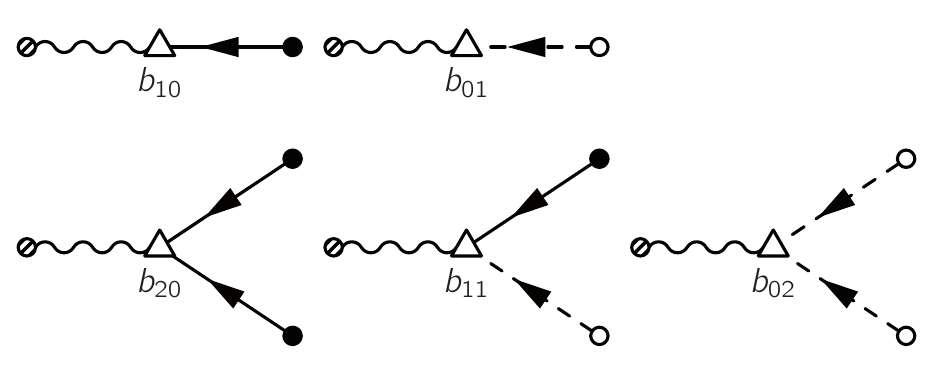}
	\caption{The multivariate bias expansion can be
	expressed by the triangular vertices shown above. The interaction of two
	fields on a vertex corresponds to a convolution integral over
	the ingoing $k$-vectors without any weighting. The vertex is connected to 
	the outer point by the wiggly halo propagator. Ingoing potentials are always
	primordial because the coupling of long and short modes, and thus the 
	enhancement of the short wavelength variance, happens in the early Universe.}
	\label{fig:bias}
\end{figure}
\begin{figure}[h]
	\centering
	\includegraphics{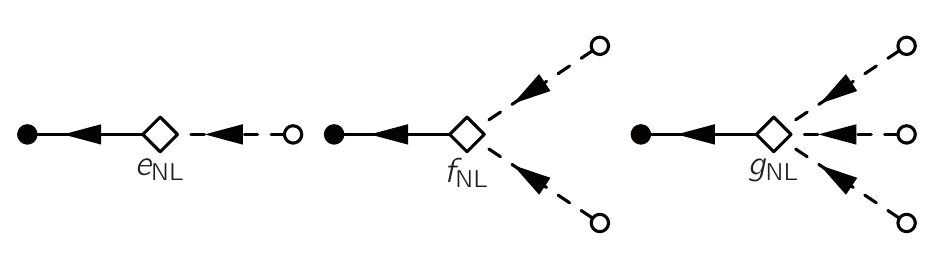}
	\caption{The non-Gaussian vertices correspond to an unweighted
	convolution integral times one factor of $\alpha$ and the corresponding
	$Q_\text{NL}=\{e_\text{NL},f_\text{NL},\ldots\}$. We introduce the
	$e_\text{NL}$ vertex to represent the conversion from $\varphi$ to
	$\delta_\text{m,p}^{(1)}$. Note that the ingoing potentials are always primordial,
	which is highlighted by the dashed line.}
	\label{fig:nongauss}
\end{figure}

The non-linear clustering vertices (we will refer to them also as
gravity vertices) $F_i$, coupling two matter density fields, are
represented by the open squares shown in Fig.~\ref{fig:clustering}. The input
density field on the right hand side is an evolved primordial field
$\delta_\text{m,p}$ defined in Eq.~\eqref{eq:ngexp}, which receives higher order 
corrections from non-Gaussianity only. 
Fig.~\ref{fig:bias} shows the vertices arising from the
multivariate bias expansion, where the input can be a matter density field of
any order or the primordial potential. 
The identification of the higher order bias terms with vertices is possible,
since in $k$-space the products of fields lead to convolution integrals, which
are similar to the non-linear gravity terms, where the $F_i$ kernels are
replaced by the scale independent bias factors $b_{ij}$. Thus
we interpret biasing as an unweighted convolution of source modes.
As noted above, loop diagrams involving higher order bias vertices can thus lead to
large or divergent contributions, effectively renormalising lower order bias
parameters.
Similarly, the non-Gaussian terms are effectively coupling potential modes
to a higher order primordial matter mode $\delta_\text{m,p}^{(n)}$. 
This interaction is represented by the diamond shaped open vertices depicted in 
Fig.~\ref{fig:nongauss}, corresponding to $\alpha(k)Q_\text{NL}$, where 
$Q_\text{NL}=\{e_\text{NL},f_\text{NL},\ldots\}$.
These coupling vertices are a specific property of local
type non-Gaussianity, basically expanding the bispectrum in terms of the
primordial potential
\be
B_{\Phi\Phi\Phi}(\vec k_1,\vec k_2, \vec k_3)=2 \fnl P_{\varphi\varphi}(\vec
k_1)P_{\varphi\varphi}(\vec k_2)+2\cyc.
\ee 
For other shapes of non-Gaussianity such an expansion might take a different
form with a $k$-dependent kernel that can be easily included.
\par
Time integration is trivially performed in standard perturbation theory 
since all the initial fields are linearly evolved, considering only the growing 
mode. Thus the lines correspond to propagators, \emph{i.\,e.\ }linear growth factors. 
However, these linear growth factors can be used to transform the initial fields
to linearly evolved fields such that we can totally ignore the time evolution
as long as we use linearly evolved primordial matter fields as a source.
We only need to ensure causality by following the time evolution of the fields.
Non-Gaussian coupling happens directly after inflation, then non-linear clustering 
and finally biasing. The lines used to connect the primordial potential
with the non-Gaussian vertices and the non-Gaussian bias vertices are dashed to
highlight the fact that the coupling of primordial potentials, and thus the 
imprint of non-Gaussianity, happens directly after inflation.
To facilitate the distinction of the density propagators, we use 
straight and wiggly lines for the matter and halo density field, respectively.
\par
The $i$-th order contribution to an evolved matter or halo field can be obtained 
following the time evolution step by step starting from the initial conditions 
and going all the way to the final field
\begin{enumerate}
\item Draw $i$ initial fields $\delta_\text{m,p}^{(1)}$ or $\varphi$.
\item Draw the non-Gaussian $Q_\text{NL}$ vertices and connect them to the 
primordial potential using dashed lines.
\item Draw the gravity vertices and connect them to the non-Gaussian vertices or 
initial density fields by solid lines.
\item For biased tracers, draw the bias vertices
and connect them to either initial density fields, gravity vertices,
non-Gaussian vertices or primordial potentials. Use a wiggly line to connect them
with the outer point.
\end{enumerate}
\par
So far we focused our attention on the fields. To compute the $i$-th order
contribution\footnote{The index $i$ has to be even since the
correlator of an odd number of Gaussian fields vanishes due to  the
Wick-theorem} to the $n$-spectra we need to glue $n$ diagrams with $i$ source
fields and $n$ outer points in all possible ways and then pair the source fields
in all possible ways. Two linked source fields lead to a power spectrum and a
momentum conserving delta function $\la\delta(\vec k)\delta(\vec
k')\ra=(2\pi)^3\delta^{(D)}(\vec k +\vec k')P_{\delta\delta}(k)$. 
The potential-potential and density-potential power spectra are related to the 
density-density power spectra by the appropriate Poisson factors $\alpha(k,z)$. 
For the translation of the above diagrams into mathematical expressions we
assign $k$-vectors to each of the outer fields. The different ways of
performing this assignment are accounted for by the cyclic permutations of
the $k$-vectors. 
%>>>>>>>>>>>>>>>>>>>>>>>>>>>>>>>>>>>>>>>>>>>>>>>>>>>>>>>>>>>>>>>>>>>>>>>>>>>>>>>
%	 			Figure 5
%>>>>>>>>>>>>>>>>>>>>>>>>>>>>>>>>>>>>>>>>>>>>>>>>>>>>>>>>>>>>>>>>>>>>>>>>>>>>>>>
%<<<<<<<<<<<<<<<<<<<<<<<<<<<<<<<<<<<<<<<<<<<<<<<<<<<<<<<<<<<<<<<<<<<<<<<<<<<<<<<
%
%>>>>>>>>>>>>>>>>>>>>>>>>>>>>>>>>>>>>>>>>>>>>>>>>>>>>>>>>>>>>>>>>>>>>>>>>>>>>>>>
%	 			Figure 6
%>>>>>>>>>>>>>>>>>>>>>>>>>>>>>>>>>>>>>>>>>>>>>>>>>>>>>>>>>>>>>>>>>>>>>>>>>>>>>>>
%<<<<<<<<<<<<<<<<<<<<<<<<<<<<<<<<<<<<<<<<<<<<<<<<<<<<<<<<<<<<<<<<<<<<<<<<<<<<<<<
%

%===============================================================================
\subsection{Feynman Rules for the $n$-Spectra}
\renewcommand{\labelenumii}{\roman{enumii}.)}
Even though the diagrams can be straightforwardly translated into the
corresponding mathematical expressions we write down the Feynman rules
explicitly for the sake of definiteness.
For the calculation of the $i$-th order contribution to the $n$-spectrum do the 
following
\begin{enumerate}
\item Draw all distinct connected diagrams with $n$ external lines up to the 
desired order $i$ in $\varphi$ 
\begin{enumerate}
\item For each vertex with ingoing momenta $\vec q_i$ and outgoing momenta $\vec
p_j$ write a delta function $\delta^\text{(D)}\left(\sum_i \vec q_i-\sum_j \vec
p_j\right)$
\item Assign a linear power spectrum $(2\pi)^3\delta^{(D)}(\vec q +\vec q')P_\text{lin}(q)$ 
to each of the big circles with outgoing momenta $\vec q$ and $\vec q'$. Divide 
by $\alpha(q)$ or $\alpha^2(q)$ for $P_{\delta\varphi}$ and $P_{\varphi\varphi}$,
respectively.
\item For the outer fields with momenta $\vec k_i$ write a delta function 
$\delta^\text{(D)}\left(\sum_i \vec k_i\right)$
\item For each square shaped vertex $F_n$ with ingoing momenta $\vec q_i$ write
a mode coupling kernel $F_n(\vec q_1,\ldots,\vec q_n)$
\item For each triangular shaped vertex write a bias factor $b_{ij}$
\item For each diamond shaped vertex 
$Q_\text{NL}=\left\{e_\text{NL},\fnl,\gnl,\ldots\right\}$ with outgoing 
momentum $\vec q$ write $\alpha(\vec q) Q_\text{NL}$
\item Integrate over all inner momenta $\int d^3q_i/(2\pi)^3$
\item Multiply with the symmetry factor
\item Sum over all distinct labelings of the external lines
\end{enumerate}
\item Add up the resulting expressions from all diagrams
\end{enumerate}

%===============================================================================
%				MATTER POWER SPECTRUM
%===============================================================================
\subsection{Matter Power Spectrum}
\begin{figure}[h]
	\centering
	\includegraphics[width=0.8\textwidth]{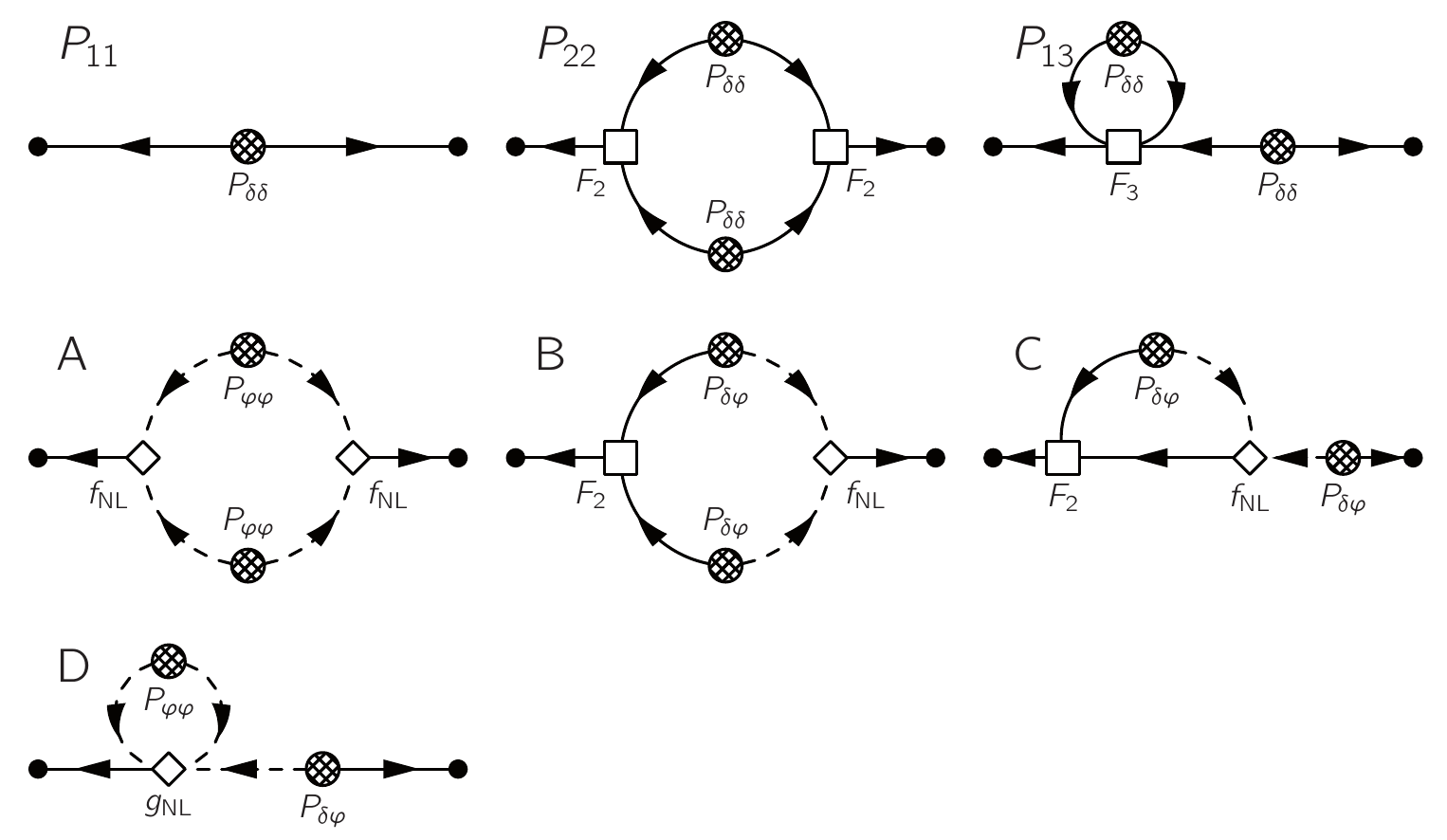}
	\caption{Diagrams contributing to the matter power spectrum as calculated 
	in Eq.~\eqref{eq:matpow}. The first line
	shows the standard PT terms, whereas the terms in the second and third line 
	are purely non-Gaussian. 
	}
	\label{fig:pmm}
\end{figure}

As a first application of our diagrammatic approach we write down the terms
contributing to the non-Gaussian matter power spectrum and show the 
corresponding diagrams in Fig.~\ref{fig:pmm}
\begin{align}
P_\text{mm}(k)=&P_{11}(k)+P_{22}(k)+P_{13}(k)\elnn
+&\left(2\alpha^2(k)\fnlsq\int\dqc\frac{P(q)}{\alpha^2(q)}\frac{P(\vec k-\vec q)}{
\alpha^2(\vec k-\vec q) }\right)_\text{A}\elnn
+&\left(4\alpha(k)\fnl\int\dqc
\frac{P(q)}{\alpha(q)}\frac{P(\vec k-\vec q)}{\alpha(\vec k-\vec q)}F_2(\vec
q,\vec k-\vec q)\right)_\text{B}\label{eq:matpow}\\
+&\left(8\fnl\frac{P(k)}{\alpha(k)} \int\dqc \frac{P(q)}{\alpha(q)}F_2(\vec q,\vec
k-\vec q)\alpha(\vec k-\vec q)\right)_\text{C}\elnn
+&\left(6\alpha(k)\gnl\frac{P(k)}{\alpha(k)}\int\dqc\frac{P(q)}{\alpha^2(q)}\right)_\text{D},\nonumber
\end{align}
where $P_{13}$ and $P_{22}$ are the standard one-loop corrections to the power
spectrum (see \cite{Bernardeau2002} and Appendix A). The subscripts on the 
brackets in the above equation can be used to identify the corresponding terms in
Fig.~\ref{fig:pmm}.
The functional form agrees with previous results as presented in 
\cite{Taruya2008}, who also performed a numerical evaluation which is thus not 
repeated here. The corrections arising from the $\fnl$ terms are generally small
and most prominent for high $k$, where the validity of perturbation theory has
to be doubted. 
%>>>>>>>>>>>>>>>>>>>>>>>>>>>>>>>>>>>>>>>>>>>>>>>>>>>>>>>>>>>>>>>>>>>>>>>>>>>>>>>
%	 			Figure 7
%>>>>>>>>>>>>>>>>>>>>>>>>>>>>>>>>>>>>>>>>>>>>>>>>>>>>>>>>>>>>>>>>>>>>>>>>>>>>>>>
%<<<<<<<<<<<<<<<<<<<<<<<<<<<<<<<<<<<<<<<<<<<<<<<<<<<<<<<<<<<<<<<<<<<<<<<<<<<<<<<
%===============================================================================
%				MATTER BISPECTRUM
%===============================================================================
\subsection{Matter Bispectrum}
\begin{figure}[h]
	\centering
	\includegraphics{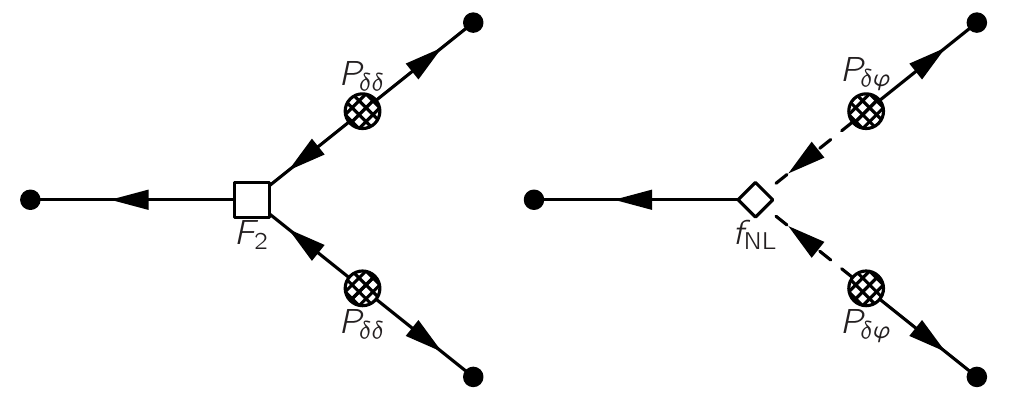}
	\caption{The two diagrams contributing to the tree level matter
	bispectrum $B_\text{mmm}$ in Eq.~\eqref{eq:bmmm}.}
	\label{fig:bmmm}
\end{figure}

A non-vanishing matter bispectrum beyond the non-linear gravitational
contribution would be a direct sign of non-Gaussian initial conditions. From
the diagrams depicted in Fig.~\ref{fig:bmmm} we can derive the following
expression for the tree-level matter bispectrum
\begin{align}
B_\text{mmm}(\vec k_1,\vec k_2,\vec k_3)
%=&\alpha(k_1)\fnl\int
%\dqc\la\phi(\vec q)\phi(\vec k_1-\vec q)\delta(\vec k_2)\delta(\vec
%k_3)\ra\nonumber\\
%+&\int \dqc F_2(\vec q,\vec k_3-\vec q) \la\delta(\vec k_1) \delta(\vec k_2)
%\delta(\vec q)\delta(\vec k_3-\vec q)\ra\\
=&\biggl(2P(k_1)P(k_2)F_2(\vec k_1,\vec k_2)+2\cyc\biggr)
+\biggl(2\fnl\frac{P(k_1)P(k_2)\alpha(k_3)}{\alpha(k_1)\alpha(k_2)}
+2\cyc\biggr)\\
=&B_{F_2}(\vec k_1,\vec k_2,\vec k_3)+B_{\fnl}(\vec k_1,\vec k_2,\vec
k_3)\label{eq:bmmm}
\end{align}
Here $\cyc$ is used to symbolise that the function arguments in the
preceding terms have to be cyclically permuted as
$\left\{(k_1,k_2,k_3),(k_2,k_3,k_1),(k_3,k_1,k_2)\right\}$. An evaluation of
the latter term and comparison to $n$-body simulations is provided in a recent
study by \cite{Sefusatti2010}. They find that the inclusion of one-loop terms
leads to a considerable improvement of the agreement between theory and 
simulation on scales exceeding $k=0.1 \ihMpc$. 
\par
Estimating the bispectrum of the dark matter distribution is an involved task
even for the upcoming lensing surveys. Thus we will focus our attention on the
bispectra of biased tracers such as galaxies in the next section.
%
%>>>>>>>>>>>>>>>>>>>>>>>>>>>>>>>>>>>>>>>>>>>>>>>>>>>>>>>>>>>>>>>>>>>>>>>>>>>>>>>
%	 			Figure 8
%>>>>>>>>>>>>>>>>>>>>>>>>>>>>>>>>>>>>>>>>>>>>>>>>>>>>>>>>>>>>>>>>>>>>>>>>>>>>>>>
%<<<<<<<<<<<<<<<<<<<<<<<<<<<<<<<<<<<<<<<<<<<<<<<<<<<<<<<<<<<<<<<<<<<<<<<<<<<<<<<
%
%===============================================================================
%				Bispectra
%===============================================================================
\section{Bispectra of Biased Tracers}\label{sec:bispect}
In the following section we present the main result of this paper, the 
derivation of the halo bispectra. The bispectrum measures the correlation 
between three fields and thus halo and matter fields lead to four possible 
combinations. The matter auto-bispectrum was already discussed in the previous
section such that we can focus our attention in this section to bispectra 
involving at least one biased tracer.
%===============================================================================
\subsection{Halo Bispectrum}
\begin{figure}[h!]
	\centering
	\includegraphics{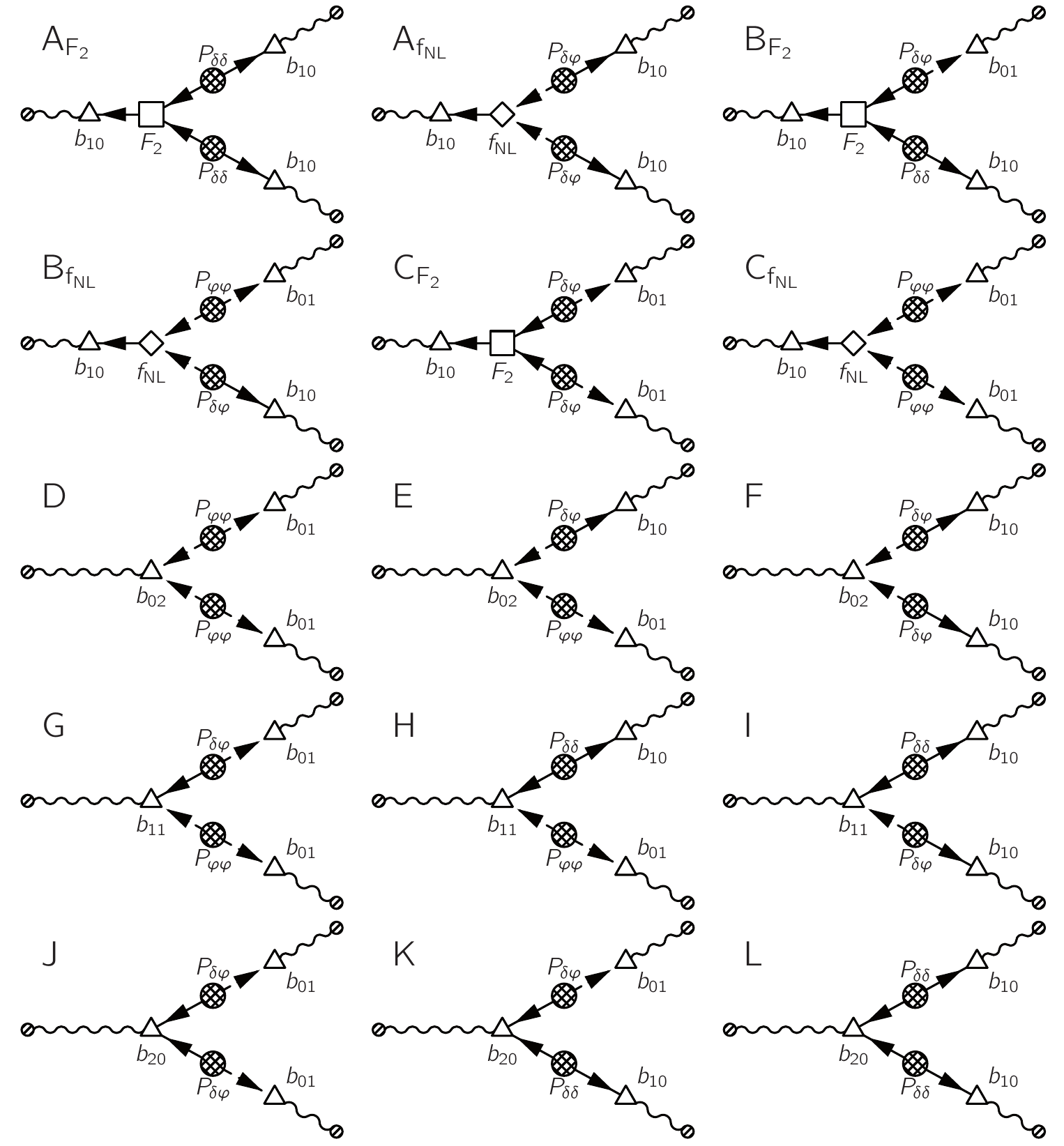}
	\caption{Tree level bispectrum diagrams considered for
Eq.~\eqref{eq:bhhh}. The scaling of the different terms with $\fnl$ as well as
their $k$ dependence are described in Table \ref{tab:bhhh}.}
	\label{fig:bhhh}
\end{figure}

Summing up the diagrammatic expressions shown in Fig.~\ref{fig:bhhh} we can
write down the tree-level expression for the halo auto-bispectrum
\begin{align}
B_\text{hhh}(\vec k_1,\vec k_2,\vec
k_3)=&b_{10}^3\left(2P(k_1)P(k_2)F_2(\vec k_1,\vec k_2)
+2\fnl\frac{P(k_1)P(k_2)\alpha(k_3)}{\alpha(k_1)\alpha(k_2)}
+2\cyc\right)_\text{A}\nonumber\\
+&b_{10}^2b_{01}\left(2P(k_1)P(k_2)\left(\frac{1}{\alpha(k_1)}+\frac{1}{
\alpha(k_2)}\right)F_2(\vec k_1,\vec k_2)\right.\nonumber\\
+&\left.2\fnl\frac{P(k_1)P(k_2)\alpha(k_3)}{\alpha(k_1)\alpha(k_2)}\left(\frac{1
}{\alpha(k_1)}+\frac{1}{\alpha(k_2)}\right)+2\cyc\right)_\text{B}\nonumber\\
+&b_{10}b_{01}^2\left(2\frac{P(k_1)P(k_2)}{\alpha(k_1)\alpha(k_2)}F_2(\vec k_1,
\vec k_2)+2\fnl\frac{P(k_1)P(k_2)\alpha(k_3)}{\alpha^2(k_1)\alpha^2(k_2)}
+2\cyc\right)_\text{C}\nonumber\\
+&b_{01}^2b_{02}\left(2\frac{P(k_1)P(k_2)}{\alpha^2(k_1)\alpha^2(k_2)}
+2\cyc\right)_\text{D}\nonumber\\
+&b_{01}b_{10}b_{02}\left(2\frac{P(k_1)P(k_2)}{\alpha(k_1)\alpha(k_2)}
\left(\frac{1}{\alpha(k_1)}+\frac{1}{\alpha(k_2)}\right)+2\cyc\right)_\text{E}\nonumber\\
+&b_{10}^2b_{02}\left(2\frac{P(k_1)P(k_2)}{\alpha(k_1)\alpha(k_2)}
+2\cyc\right)_\text{F}\label{eq:bhhh}\\
+&b_{01}^2b_{11}\left(\frac{P(k_1)P(k_2)}{\alpha(k_1)\alpha(k_2)}\left(\frac{1}{
\alpha(k_1)}+\frac{1}{\alpha(k_2)}\right)+2\cyc\right)_\text{G}\nonumber\\
+&b_{01}b_{10}b_{11}\left(P(k_1)P(k_2)\left(\frac{1}{\alpha^2(k_1)}+\frac{2}{
\alpha(k_1)\alpha(k_2)}+\frac{1}{\alpha^2(k_2)}\right)+2\cyc\right)_\text{H}\nonumber\\
+&b_{10}^2b_{11}\left(P(k_1)P(k_2)\left(\frac{1}{\alpha(k_1)}+\frac{1}{
\alpha(k_2)}\right)+2\cyc\right)_\text{I}\nonumber\\
+&b_{01}^2b_{20}\left(2\frac{P(k_1)P(k_2)}{\alpha(k_1)\alpha(k_2)}
+2\cyc\right)_\text{J}\nonumber\\
+&b_{01}b_{10}b_{20}\left(2P(k_1)P(k_2)\left(\frac{1}{\alpha(k_1)}+\frac{1}{
\alpha(k_2)}\right)+2\cyc\right)_\text{K}\nonumber\\
+&b_{10}^2b_{20}\left(2 P(k_1)P(k_2)+2\cyc\right)_\text{L}\nonumber
\end{align}
The subscripts on the brackets can be used to identify the terms with the
corresponding diagrams in Fig.~\ref{fig:bhhh}. In the diagrams and the
above equation there are no explicit $\gnl$ terms, since the $\gnl$ vertices are
not explicitly contributing to the tree level halo bispectrum. However, there
is an implicit dependence via the second order bias parameter $b_{02}$. We will
not further examine this dependence since it changes only a parameter of the
model.
Note that for $\gnl=0$ the bias factors scale approximately as $b_{01}\propto
\fnl$,
$b_{02}\propto\fnlsq$ and $b_{11}\propto\fnl$. Thus the exponent of $\fnl$ in
the terms is the same as the exponent of $\alpha(k)$ in the denominator.
Hence, it is the ratio of $\fnl/\alpha(k)$, where
$k=\min{\left\{k_1,k_2,k_3\right\}}$ that is dominating the overall
amplitude on large scales. The highest order contribution is $f_\text{NL}^4$ in
the
D term. This term is negative and gains importance
for extremely small $k$ and large $\fnl$.
\par
The wave vectors are related by $\vec k_1+\vec k_2+\vec k_3=0$ and thus the
configuration is fully determined by the magnitude of one vector $k_1$, one
angle $\mu=\vec k_1\cdot \vec k_2/(k_1k_2)$ and the ratio of two vectors
$x_2=k_2/k_1$. The magnitude of the third vector is then given by
\be
x_3=\frac{k_3}{k_1}=\sqrt{1+2\mu x_2 +x_2^2}\label{eq:zeta},
\ee
which for the isosceles configuration $k_1=k_2$ simplifies to
$x_3=\sqrt{2(1+\mu)}$. 
The transfer function is unity for large scale modes entering horizon after
matter radiation equality and is damped on small scales, leading to an
asymptotic slope of $\approx-1.75$. Starting from a primordial power spectrum
$P_0(k)=A k^n$ with $n\approx 1$ we see that the primordial Gaussian potential
is scale invariant with $P_{\varphi\varphi}\propto k^{-3}$. The matter power
spectrum in contrast is given by $P_{\delta\delta}=T^2(k) P_0(k)$ and thus it
scales as $k^1$ at low $k$'s and approximately as $k^{-2.5}$ at high $k$'s. For the
isosceles configuration and for low $k=k_1=k_2$ the dominant contribution to 
terms including second order
bias (terms $D$-$L$ of the above equation) scales as 
\be
\frac{P(k_1)P(k_3)}{\alpha^i(k_1)\alpha^j(k_3)}\propto \frac{k^2 x_3}{k^{
2i+2j}x_3^{2j}}=k^{2-2i-2j}x_3^{1-2j},
\label{eq:bhhhkdep}
\ee
while the dominant contribution to the $A,B,C$ terms scales as
\be
\frac{P(k_1)P(k_3)}{\alpha^{i-1}(k_1)\alpha^j(k_3)}\propto \frac{k^2 x_3}{k^{
2i+2j-2}x_3^{2j}}=k^{4-2i-2j}x_3^{1-2j}.
\label{eq:bhhhkdep2}
\ee
The latter equation considers only the $\fnl$ contribution, 
dominating at small $x_3$'s. In Table \ref{tab:bhhh} we quote the exponents 
$i, j$ and the corresponding power of $\fnl$. 
The combination $2(i+j-1)$ is the exponent of the dominating
short mode in the squeezed limit and can go up to $k^6$. The estimation of the
importance of the terms is further complicated by the different bias prefactors.
We thus evaluate the expression numerically and 
discuss the results in Sec.~\ref{sec:numresults} below.
\par
Recent studies of the tree-level bispectrum \cite{Sefusatti2007} are based on
the univariate bias parameters $b_{10}$ and $b_{20}$ only (univariate or
Gaussian biasing). At tree level they thus consider only a subset of the above terms (see
their Eq.~(18)) leading to
\be
B_\text{hhh}(\vec k_1,\vec k_2,\vec k_3)=b_{10}^3
\left[B_{F_2}(\vec k_1,\vec k_2,\vec k_3)+B_{\fnl}(\vec k_1,\vec k_2 ,
\vec k_3)\right]+b_{10}^2b_{20}\left[2 P(k_1)P(k_2)+2\cyc\right],
\label{eq:sef2007}
\ee
where $B_{F_2}$ and $B_{\fnl}$ are implicitly defined in Eq.~\eqref{eq:bmmm}.
This approach neglects the influence of non-Gaussianity on the bias
parameters, whereas it has been explicitly shown in simulations
\cite{Desjacques2009a,Grossi2008,Giannantonio2010} that local non-Gaussianity
introduces a scale dependent bias.
%
%>>>>>>>>>>>>>>>>>>>>>>>>>>>>>>>>>>>>>>>>>>>>>>>>>>>>>>>>>>>>>>>>>>>>>>>>>>>>>>>
%	 			Figure 9
%>>>>>>>>>>>>>>>>>>>>>>>>>>>>>>>>>>>>>>>>>>>>>>>>>>>>>>>>>>>>>>>>>>>>>>>>>>>>>>>
\clearpage
%<<<<<<<<<<<<<<<<<<<<<<<<<<<<<<<<<<<<<<<<<<<<<<<<<<<<<<<<<<<<<<<<<<<<<<<<<<<<<<<
%
\begin{table}[t]\footnotesize
\begin{center}
\begin{tabular}{lrrrrr}
\hline
\hline
 & $i$ & $j$ & $2(i+j-1)$ & $2j-1$ & {$f_\text{NL}$} \\ 
\hline
A$_{F_2}$ & 0 & 0 & -2 & -1 & 0 \\ 
A$_{f_\text{NL}}$ & 1 & 1 & 2 & 1 & 1 \\ 
B$_{F_2}$ & 0 & 1 & 0 & 1 & 1 \\ 
B$_{f_\text{NL}}$ & 1 & 2 & 2 & 3 & 2 \\ 
C$_{F_2}$ & 1 & 1 & 2 & 1 & 2 \\ 
C$_{f_\text{NL}}$ & 2 & 2 & 6 & 3 & 3 \\ 
D & 2 & 2 & 6 & 3 & 4 \\ 
E & 2 & 1 & 4 & 1 & 3 \\ 
 & 1 & 2 & 4 & 3 & 3 \\ 
F & 1 & 1 & 2 & 1 & 2 \\ 
G & 2 & 1 & 4 & 1 & 3 \\ 
 & 1 & 2 & 4 & 3 & 3 \\ 
H & 2 & 0 & 2 & -1 & 2 \\ 
 & 1 & 1 & 2 & 1 & 2 \\ 
 & 0 & 2 & 2 & 3 & 2 \\ 
I & 1 & 0 & 0 & -1 & 1 \\ 
 & 0 & 1 & 0 & 1 & 1 \\ 
J & 1 & 1 & 2 & 1 & 2 \\ 
K & 1 & 0 & 0 & -1 & 1 \\ 
 & 0 & 1 & 0 & 1 & 1 \\ 
L & 0 & 0 & -2 & -1 & 0 \\ 
\hline
\hline
\end{tabular}
\end{center}
\caption{Order of the terms contributing to the bispectrum $B_\text{hhh}$ in
Eq.~\eqref{eq:bhhh} as defined via Eq.~\eqref{eq:bhhhkdep}. From left to right
we quote the exponents of the $k$-vectors, the exponent of the long $k$-mode,
the exponent of the long-short ratio and the exponent of $\fnl$. More than one
line per diagram can arise if there are different possibilities for combining
the components of the diagram.}
\label{tab:bhhh}
\end{table}
%===============================================================================
\subsection{Cross-Bispectra}
The measurement of the halo auto-bispectrum is limited by the finite number 
of haloes and shotnoise. Thus it is interesting to consider also the cross
spectra between matter and
haloes. Weak gravitational lensing in cross-correlation with galaxies could be a 
possible observational probe providing
measurements of these effects. Without showing the diagrams we write down
the expressions for the halo-halo-matter bispectrum:
\begin{align}
B_\text{hhm}(\vec k_1,\vec k_2,\vec k_3)=&b_{10}^2 \biggl(6 F_2(\vec k_1,\vec
k_2)P(k_1)P(k_2)+6\fnl
\alpha(k_3) \frac{P(k_1)}{\alpha(k_1)}\frac{P(k_2)}{\alpha(k_2)}
+2\cyc\biggr)\elnn
+&  b_{01} b_{10} \biggl( 4 F_2(\vec k_1,\vec
k_2)P(k_1)P(k_2)
\left(\frac{1}{\alpha(k_1)}+\frac{1}{\alpha(k_2)}\right)\elnn
+& 4  \fnl \alpha(k_3)P(k_1)P(k_2) 
\left(\frac{1}{\alpha^2(k_1)\alpha(k_2)}+\frac{1}{\alpha(k_1)\alpha^2(k_2)}
\right)+2\cyc\biggr)\elnn
+&b_{01}^2\biggl(2 F_2(\vec k_1,\vec
k_2)\frac{P(k_1)P(k_2)}{\alpha(k_1)\alpha(k_2)}
+2
\fnl\alpha(k_3)\frac{P(k_1)P(k_2)}{\alpha^2(k_1)\alpha^2(k_2)}+2\cyc\biggr)\elnn
+&b_{10}b_{20}\biggl(4P(k_1)P(k_2)+2\cyc\biggr)\label{eq:bhhm}\\
+&b_{20}b_{01}\biggl(2P(k_1)P(k_2)\left(\frac{1}{\alpha(k_1)}+\frac{1}{
\alpha(k_2)}
\right)+2\cyc\biggr)\elnn
+&b_{11}b_{10}\biggl(2P(k_1)P(k_2)\left(\frac{1}{\alpha(k_1)}+\frac{1}{
\alpha(k_2)}
\right)+2\cyc \biggr)\elnn
+&b_{11}b_{01}\biggl(P(k_1)P(k_2)\left(\frac{1}{\alpha^2(k_1)}+\frac{2}{
\alpha(k_1)\alpha(k_2)}+\frac{1}{\alpha^2(k_2)} \right) +2\cyc \biggr)\elnn
+&b_{10}b_{02}\biggl(4\frac{P(k_1)P(k_2)}{\alpha^2(k_1)\alpha^2(k_2)}
+2\cyc\biggr)\elnn
+&b_{02}b_{01}\biggl(2P(k_1)P(k_2)\left(\frac{1}{\alpha^2(k_1)\alpha(k_2)}+\frac
{1}{\alpha(k_1)\alpha^2(k_2)}
\right)+2\cyc\biggr)\nonumber
\end{align}
The corresponding prediction of a purely univariate bias model is given by
\be
B_\text{hhm}(\vec k_1,\vec k_2,\vec k_3)=3 b_{10}^2
\left[B_{F_2}(\vec k_1,\vec k_2,\vec k_3)+B_{\fnl}(\vec k_1,\vec k_2 ,
\vec k_3)\right]+2b_{10}b_{20}\left[2 P(k_1)P(k_2)+2\cyc\right].
\label{eq:sef2007hhm}
\ee
Finally, one can also correlate two matter and one halo density field: 
\begin{align}
B_\text{hmm}(\vec k_1,\vec k_2,\vec k_3)=& b_{10}\biggl(6F_2(\vec k_1,\vec
k_2)P(k_1)P(k_2)+6\fnl
\alpha(k_3)\frac{P(k_1)P(k_2)}{\alpha(k_1)\alpha(k_2)}+2\cyc\biggr)\elnn
+&b_{01}\biggl(2F_2(\vec k_1,\vec
k_2)P(k_1)P(k_2)\left(\frac{1}{\alpha(k_1)}+\frac{1}{\alpha(k_2)}
\right)\elnn
+&2\fnl \alpha(k_3) P(k_1)P(k_2)\left(\frac{1}{\alpha^2(k_1)\alpha(k_2)}
+\frac{1}{\alpha(k_1)\alpha^2(k_2)}\right)+2\cyc\biggr)\elnn
+&b_{20}\biggl(2P(k_1)P(k_2)+2\cyc\biggr)\label{eq:bhmm}\\
+&b_{11}\biggl(P(k_1)P(k_2)\left(\frac{1}{\alpha(k_1)}+\frac{1}{\alpha(k_2)}
\right)+2\cyc\biggr)\elnn
+&b_{02}\biggl(2\frac{P(k_1)P(k_2)}{\alpha(k_1)\alpha(k_2)}
+2\cyc\biggr)\nonumber,
\end{align}
where the corresponding prediction of the univariate bias model reads as
\be
B_\text{hmm}(\vec k_1,\vec k_2,\vec k_3)=3 b_{10}
\left[B_{F_2}(\vec k_1,\vec k_2,\vec k_3)+B_{\fnl}(\vec k_1,\vec k_2 ,
\vec k_3)\right]+b_{20}\left[2 P(k_1)P(k_2)+2\cyc\right].
\label{eq:sef2007hmm}
\ee
As for the halo auto-bispectrum, we show the results of a numerical
evaluation of the above results in the next subsection.
%===============================================================================
\subsection{Discussion of the Results}
\label{sec:numresults}
%>>>>>>>>>>>>>>>>>>>>>>>>>>>>>>>>>>>>>>>>>>>>>>>>>>>>>>>>>>>>>>>>>>>>>>>>>>>>>>>
%	 			Figure 10
%>>>>>>>>>>>>>>>>>>>>>>>>>>>>>>>>>>>>>>>>>>>>>>>>>>>>>>>>>>>>>>>>>>>>>>>>>>>>>>>
\begin{figure}[h]
	\centering
	\includegraphics[width=0.49\textwidth]{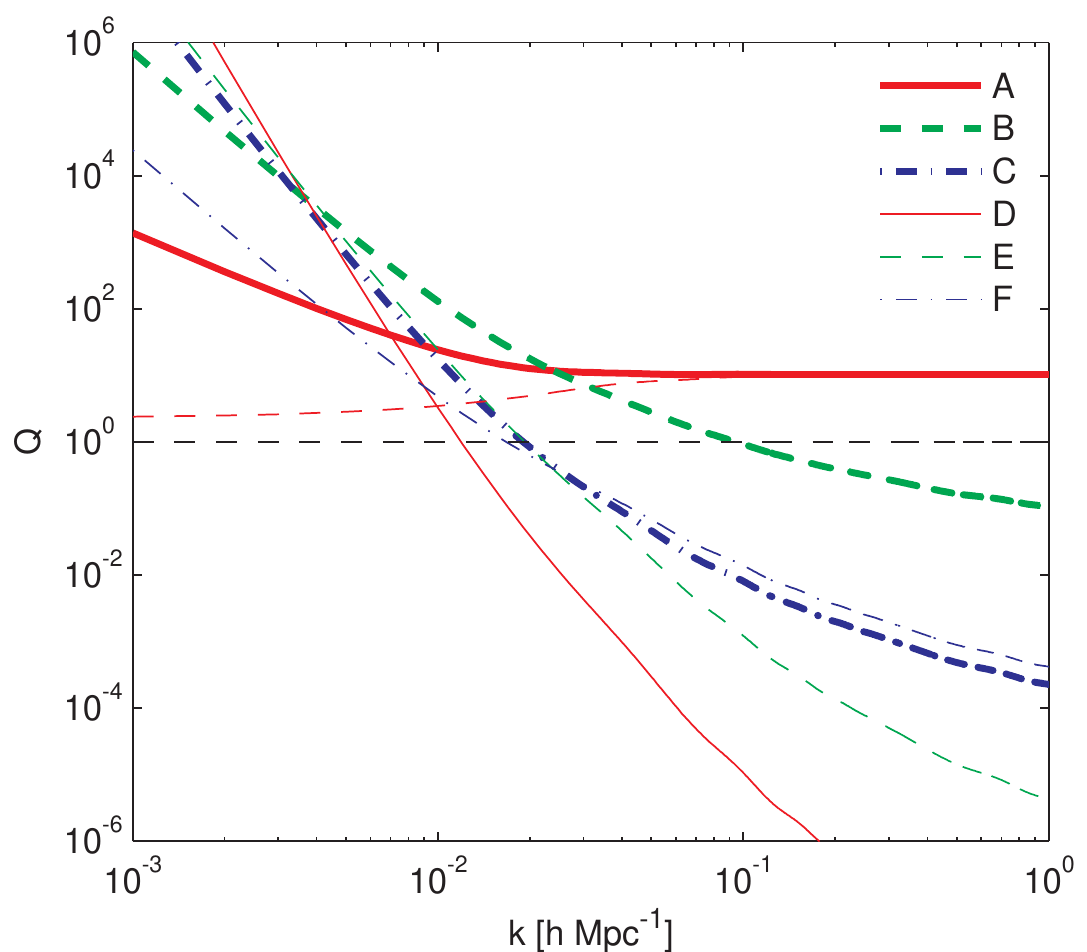}
	\includegraphics[width=0.49\textwidth]{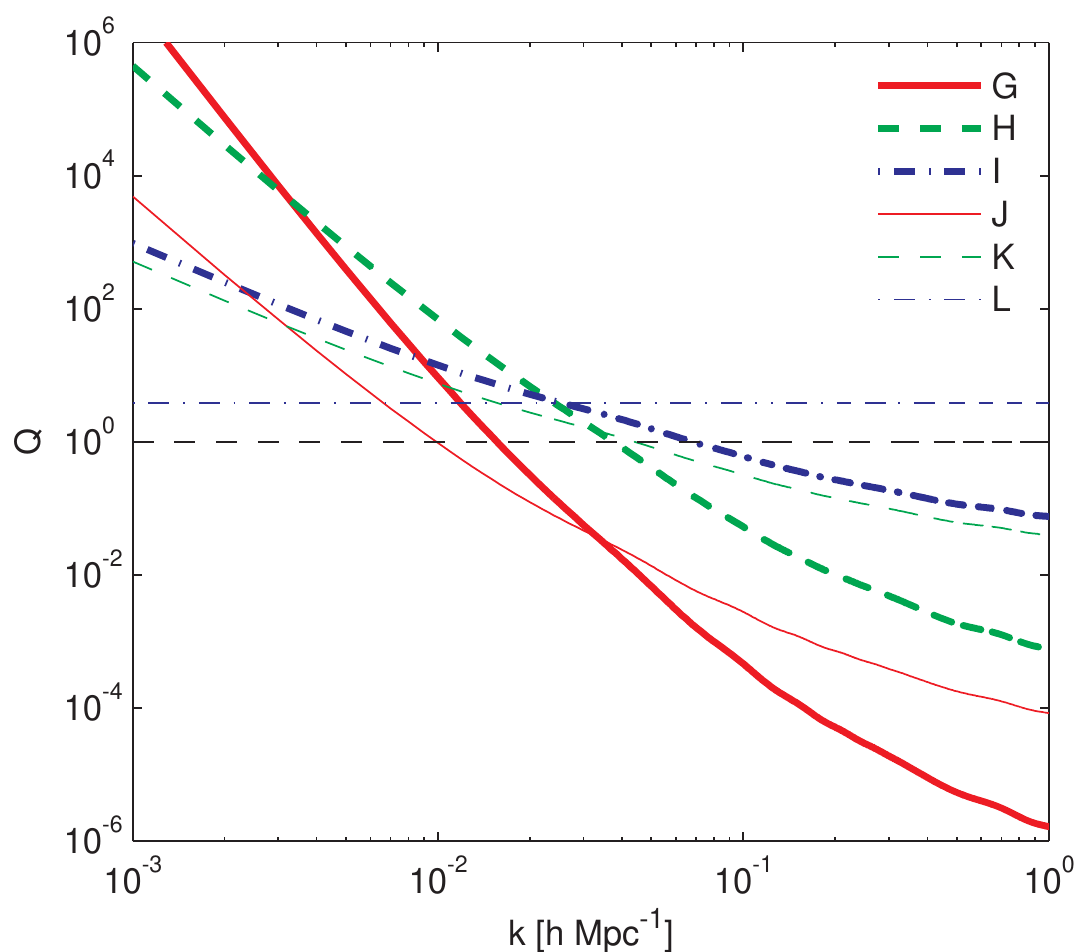}
	\caption{Scale dependence of the terms contributing to the halo
	bispectrum $B_\text{hhh}$ in Eq.~\eqref{eq:bhhh} for $\fnl=100$ and $\nu=4$
    in an isosceles configuration $k=k_1=k_2$, $\mu=-0.99$.
	The lines are labeled according to the corresponding diagrams in
	Fig.~\ref{fig:bhhh} and Table \ref{tab:bhhh}. The thin red dashed line
	in the left panel shows the bispectrum in the Gaussian case, which
	vanishes on large scales due to the shape dependence of the $F_2$
	kernel. We show the reduced bispectrum defined in Eq.~\eqref{eq:redbi}.
	The amplitude of the terms is determined both by the bias
	prefactors and	the amplitude of the $k$-dependent factors. We see
	that the B and H terms	are dominating on intermediate scales $k\approx
	1\tim{-2} \ihMpc$, whereas D,E and G take over on smaller $k$'s.}
	\label{fig:compareterms}
\end{figure}
%<<<<<<<<<<<<<<<<<<<<<<<<<<<<<<<<<<<<<<<<<<<<<<<<<<<<<<<<<<<<<<<<<<<<<<<<<<<<<<<
%
%>>>>>>>>>>>>>>>>>>>>>>>>>>>>>>>>>>>>>>>>>>>>>>>>>>>>>>>>>>>>>>>>>>>>>>>>>>>>>>>
%	 			Figure 11
%>>>>>>>>>>>>>>>>>>>>>>>>>>>>>>>>>>>>>>>>>>>>>>>>>>>>>>>>>>>>>>>>>>>>>>>>>>>>>>>
\begin{figure}[h]
	\centering
	\includegraphics[width=0.49\textwidth]{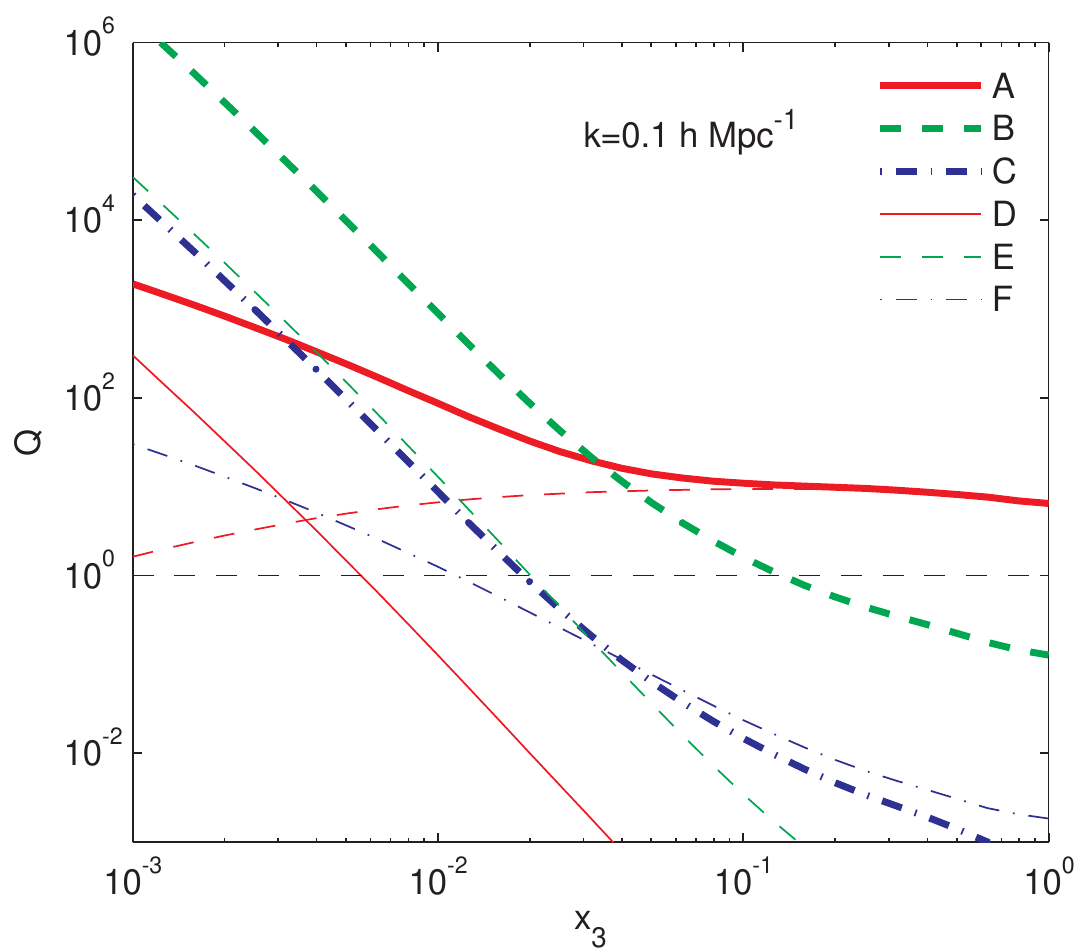}
	\includegraphics[width=0.49\textwidth]{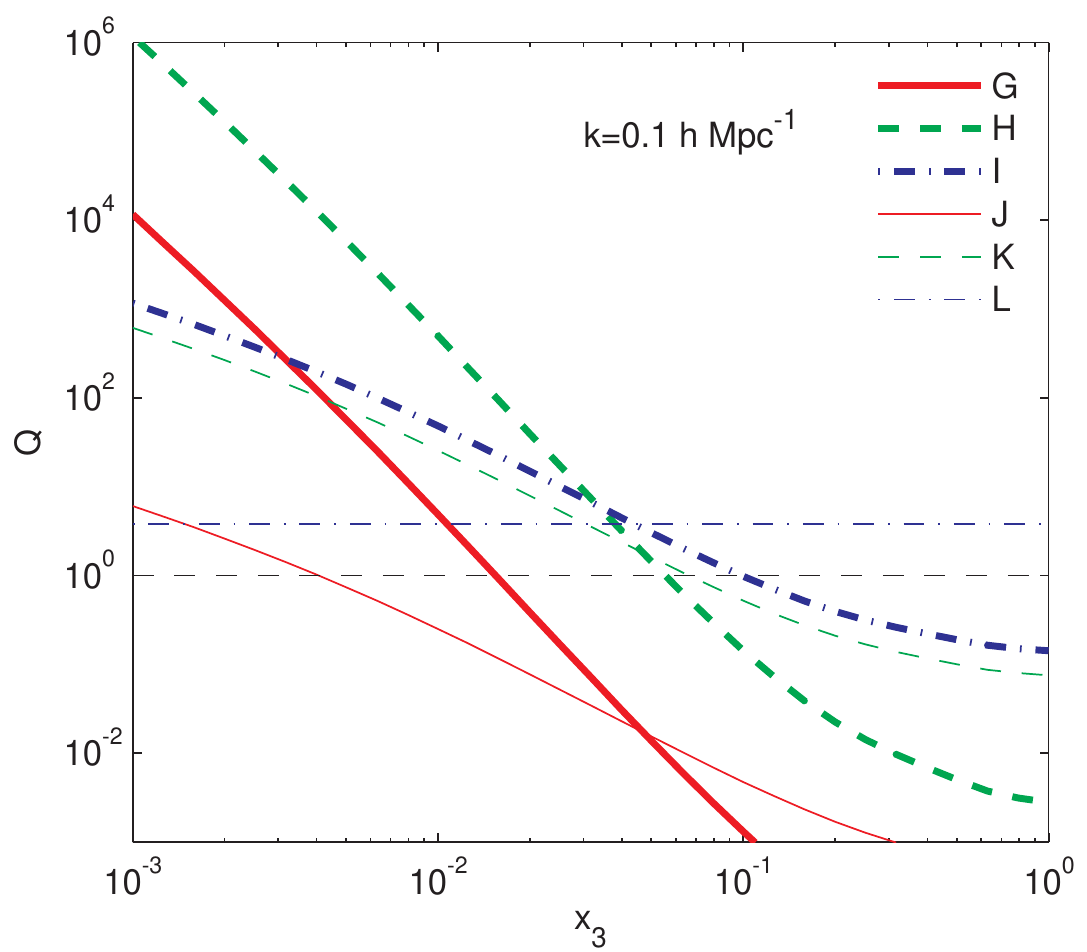}\\
	\includegraphics[width=0.49\textwidth]{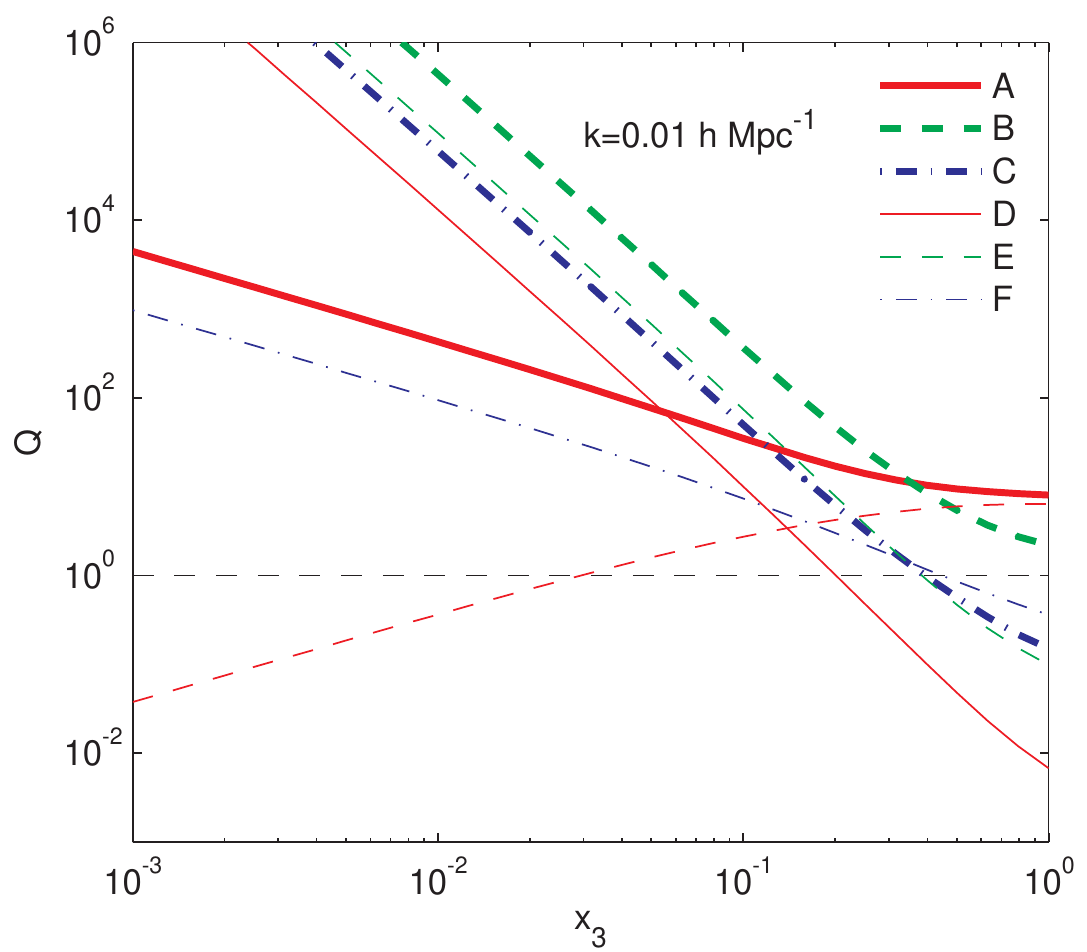}
	\includegraphics[width=0.49\textwidth]{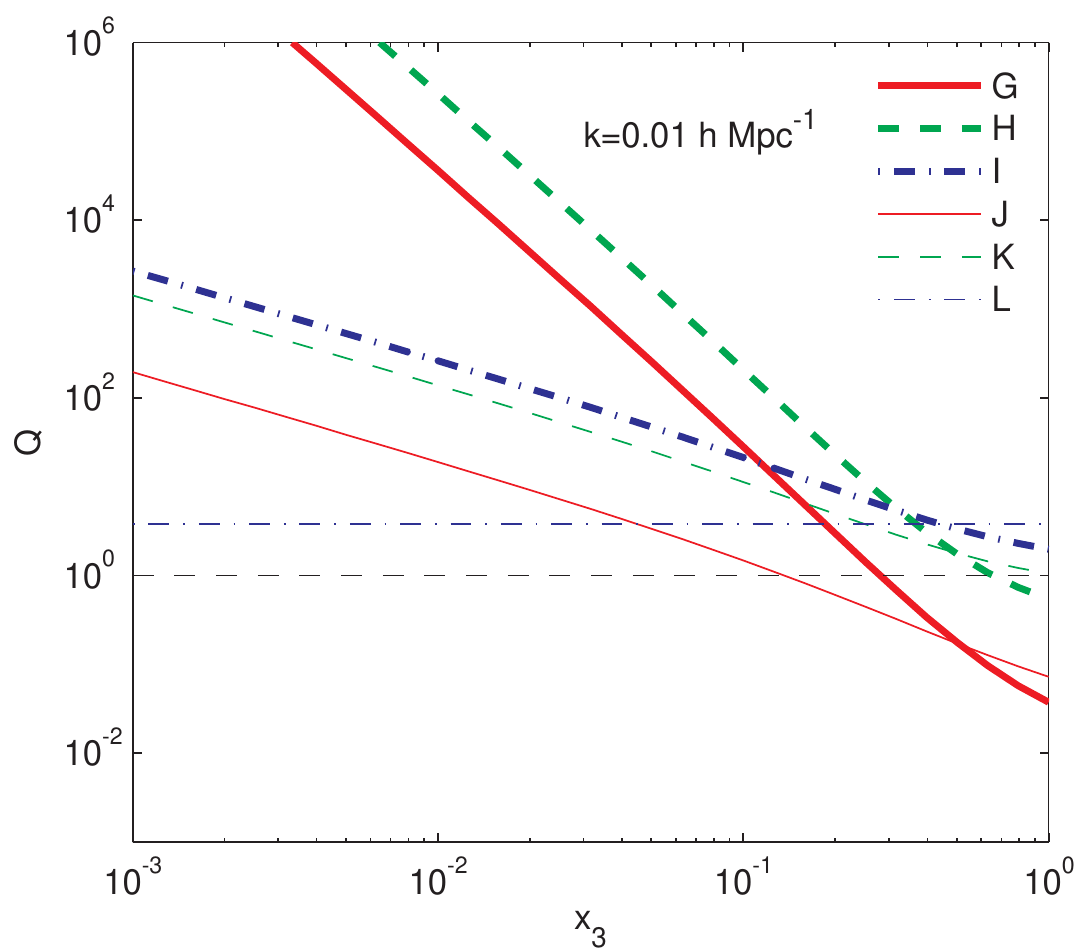}
	\caption{Dependence of the terms contributing to the halo
	bispectrum $B_\text{hhh}$ in Eq.~\eqref{eq:bhhh} on the long-short axis ratio
	$x_3$ for an isosceles configuration with $k=k_1=k_2$.
	The overall scale of the triangle is $k=0.1 \ihMpc$ for the upper panels
	and $0.01 \ihMpc$ for the lower panels. This plot clearly shows the effect of 
	the $\varphi$-bias corrections in the squeezed limit $x_3\to 0$. As in the above
	plot the thin red dashed line in the left panels shows the $A_{F_2}$
	contribution.
	}
	\label{fig:comparetermsqueezedb}
\end{figure}
%<<<<<<<<<<<<<<<<<<<<<<<<<<<<<<<<<<<<<<<<<<<<<<<<<<<<<<<<<<<<<<<<<<<<<<<<<<<<<<<
%
%>>>>>>>>>>>>>>>>>>>>>>>>>>>>>>>>>>>>>>>>>>>>>>>>>>>>>>>>>>>>>>>>>>>>>>>>>>>>>>>
% 				Figure 12
%>>>>>>>>>>>>>>>>>>>>>>>>>>>>>>>>>>>>>>>>>>>>>>>>>>>>>>>>>>>>>>>>>>>>>>>>>>>>>>>
\begin{figure}
	\centering
	\includegraphics[width=0.49\textwidth]{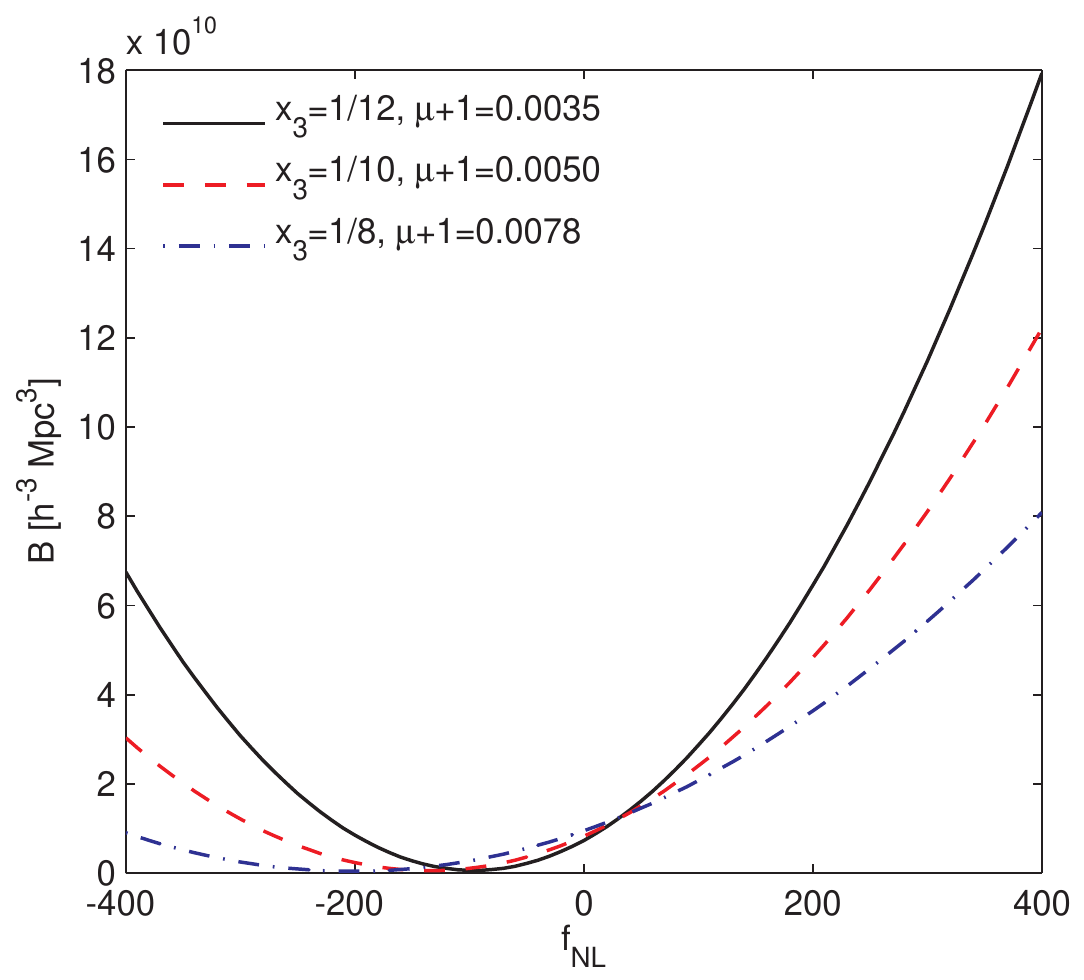}
	\includegraphics[width=0.49\textwidth]{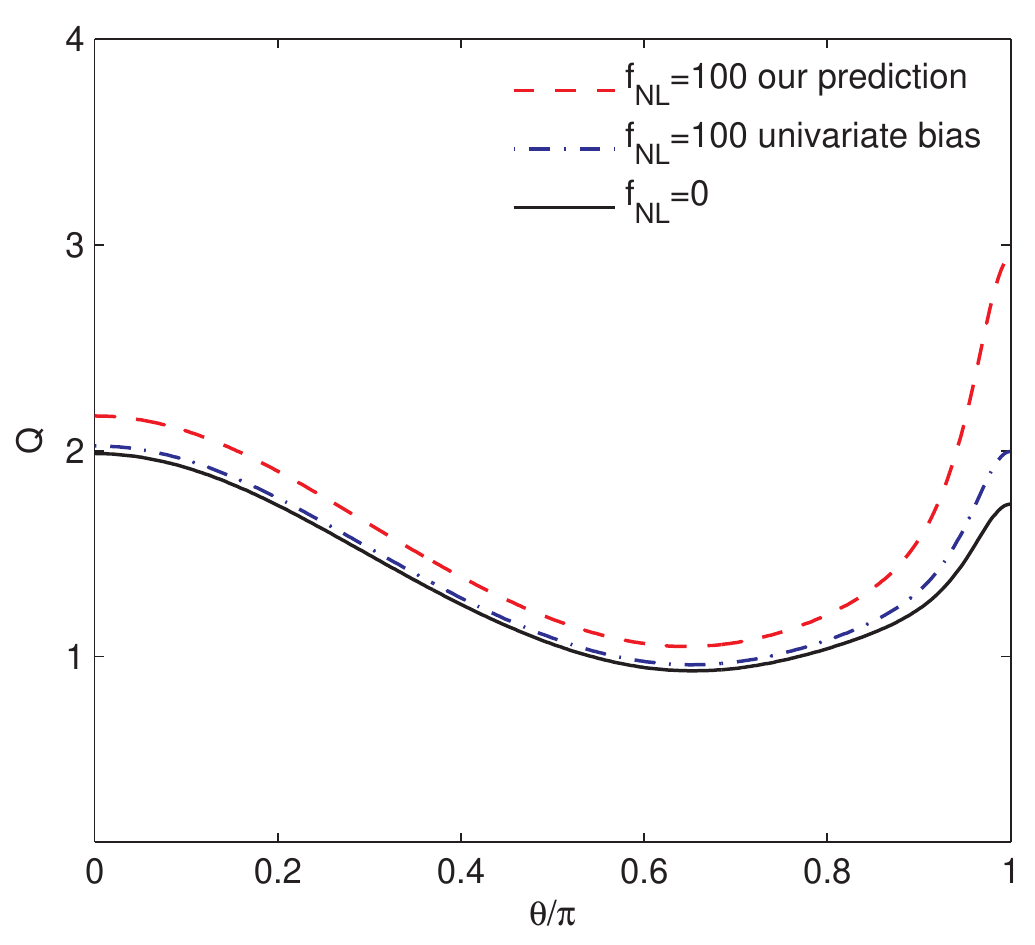}
	\caption{
	\emph{Left panel: }Dependence of the amplitude of the halo bispectrum
	$B_\text{hhh}$ on the non-Gaussianity parameter $\fnl$ for $\gnl=0$ and
	$k_1=k_2=0.042 \ihMpc$. We show the bispectrum amplitude for three
	different values of the short-long ratio $x_3$ defined in
	Eq.~\eqref{eq:zeta}. This plot agrees qualitatively with Fig.~4 in \cite{Nishimichi2009}. It is obvious from this plot that the
	bispectrum decreases with respect to the Gaussian non-linear clustering
	case for weakly negative $\fnl$.
	\emph{Right panel: }Angular dependence of the non-Gaussian and Gaussian
	reduced bispectra for $\fnl=100$ as a function of $\theta=\arccos{\left(\vec
	k_1, \vec k_2\right)}=\arccos{(\mu)}$ for a halo sample with $\nu=4$ and a nearly
	isosceles configuration with	$k_1=k_2/1.2=0.03 \ihMpc$. We show our
	result \eqref{eq:bhhh} (red dashed) and the univariate biasing result
	\eqref{eq:sef2007} (blue dash-dotted). Besides the overall enhancement
	of the multivariate bias result with respect to the univariate bias
	result there is a remarkable upturn for $\theta\to \pi$.
	}
	\label{fig:fnldep}
	\label{fig:angularratio}
\end{figure}
%<<<<<<<<<<<<<<<<<<<<<<<<<<<<<<<<<<<<<<<<<<<<<<<<<<<<<<<<<<<<<<<<<<<<<<<<<<<<<<<
%
In this section we focus on the evaluation and discussion of the 
bispectrum expressions derived in Section \ref{sec:bispect} above. For the 
evaluation no smoothing is required since at tree level there are no 
integrations. The effect of smoothing at tree level breaks down to a
multiplication with a smoothing function and thus suppresses the result on
small scales keeping the large scales unaffected. The bias parameters are evaluated for a peak height of $\nu=4$ corresponding to haloes of $M\approx 1\tim{14} \hMs$ and our fiducial cosmology introduced in Section \ref{sec:basics}.
\par
% Singleterms
In Fig.~\ref{fig:compareterms} we evaluate the terms in Eq.~\eqref{eq:bhhh}
 separately to asses their importance. To make the comparison of the terms
easier
we plot the reduced bispectrum
\be
Q(k_1,k_2,k_3)=\frac{B(k_1,k_2,k_3)}{P(k_1)P(k_2)+P(k_2)P(k_3)+P(k_3)P(k_1)}.
\label{eq:redbi}
\ee

\begin{itemize}
\item We see that diagram $A$'s contribution is constant at high-$k$'s and grows 
at low-$k$'s. This is due to the fact that at high-$k$'s the signal from the 
subdiagram $A_{F_2}$ grows (a fact that by the way $Q$ is designed means that 
$Q$ is constant), while on large scales the signal from the $A_{\fnl}$ grows. 
The two subdiagrams have a different scale dependence.
\item It is easy to see that diagrams $B$ are obtained by substituting 
$\delta$-bias with the one in $\varphi$. Since $\varphi$ is suppressed 
with respect to $\delta$ for high-$k$'s, we have that the signal in $B$ is peaked 
at smaller $k$'s. The value of $\fnl$ used for this plot is rather large, 
thus the $B$ contribution is larger than the one of $A$ at small $k$'s.
\item In order to pass to the $C$ diagrams, we exchange one $\delta$ leg with 
a $\varphi$ leg and exchange a $b_{10}$ factor with a $b_{01}$. Since in the $B$ 
diagram one of the $\varphi$ legs is already associated to the low $k$ mode, we need 
to associate the remaining leg to a high $k$ mode. This means that the $C$ diagram 
scales with respect to the $B$ diagram by a factor of $b_{01}/(b_{10}\alpha(k_{\rm high}))$ 
that, for the values of $\fnl$ and $k_{\rm high}$, that we are using is about a 
factor of 10 at the smallest $k$'s we plot. Notice that since the terms here 
have additional factors of $1/k^2$, it raises more steeply at low $k$'s. Even 
more than for the $B$ diagram, this term becomes irrelevant at high $k$'s 
because of the many $\varphi$ factors that it involves.
\item The remaining diagrams $D - L$ are constructed using the non-linear 
biases in all possible contractions. This means that the steepest at low 
$k$'s will be the $D$ diagram that involves $b_{02}$ and $b_{01}^2$, while the 
most important at high $k$'s will be the $L$ diagram, that does not involve any 
$\varphi$. By construction the $L$ diagram is scale independent when plotted in 
terms of its contribution to the reduced bispectrum $Q$.
\end{itemize}
In Fig.~\ref{fig:comparetermsqueezedb}, we show how the 
various diagrams scale with the shape parameter $x_3$ in the isosceles configuration.
For to overall size we consider $k=0.1 \ihMpc$ and $k=0.01 \ihMpc$. 
The $x_3$ scaling is also given in Table~\ref{tab:bhhh} and can be easily 
reconstructed from Eqs.~\eqref{eq:bhhhkdep} 
and~\eqref{eq:bhhhkdep2}. There is a clear correction at small $k$'s from the 
$\varphi$ terms. Since the signal is not scale invariant, as we move to higher 
$k$'s, the contribution from terms involving $\delta_\text{m}$ and its non-linearities, 
such as the $A_{F_2}$ diagram, gain importance.
\par
% Bispectra
% fnl dependence and angular dependence
The left panel of Fig.~\ref{fig:fnldep} shows the dependence of the bispectrum
amplitude on the variation of the non-Gaussianity parameter $\fnl$. 
This plot shows that higher than linear powers in $\fnl$ have to be considered
to describe the full bispectrum amplitude. The plot also shows that the sensitivity to $\fnl$
increases as one considers more squeezed configurations. While evaluated for a different
halo sample and redshift, the plot shows qualitative agreement with the
simulation measurements of \cite{Nishimichi2009}.
In the right panel of Fig.~\ref{fig:fnldep} we plot the angular dependence of
the bispectrum for an almost isosceles configuration $k_1=k_2/1.2$ since the
exact isosceles configuration is divergent for $\theta \to \pi$. 
Here $\theta=\arccos{(\mu)}$ refers to the angle between $\vec k_1$ and 
$\vec k_2$.\footnote{The divergence of the isosceles case means nothing 
but the fact that as $\theta\rightarrow\pi$ one of the sides of the triangle 
goes to zero, and so the bispectrum amplitude goes formally to infinity.}
Besides an overall enhancement the most remarkable feature is the upturn of the
multivariate bias prediction in the squeezed limit.
\par
\begin{figure}[p]
	\centering
	\includegraphics[width=0.49\textwidth]{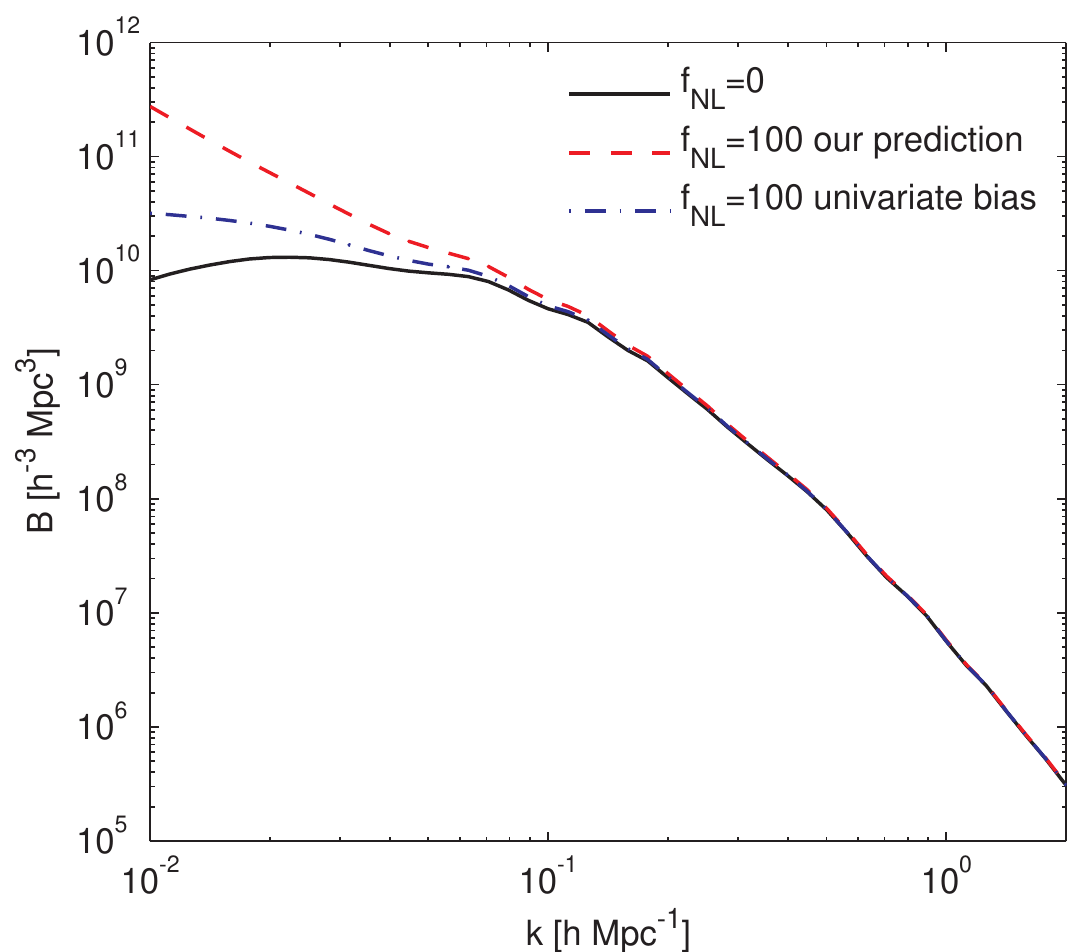}
	\includegraphics[width=0.49\textwidth]{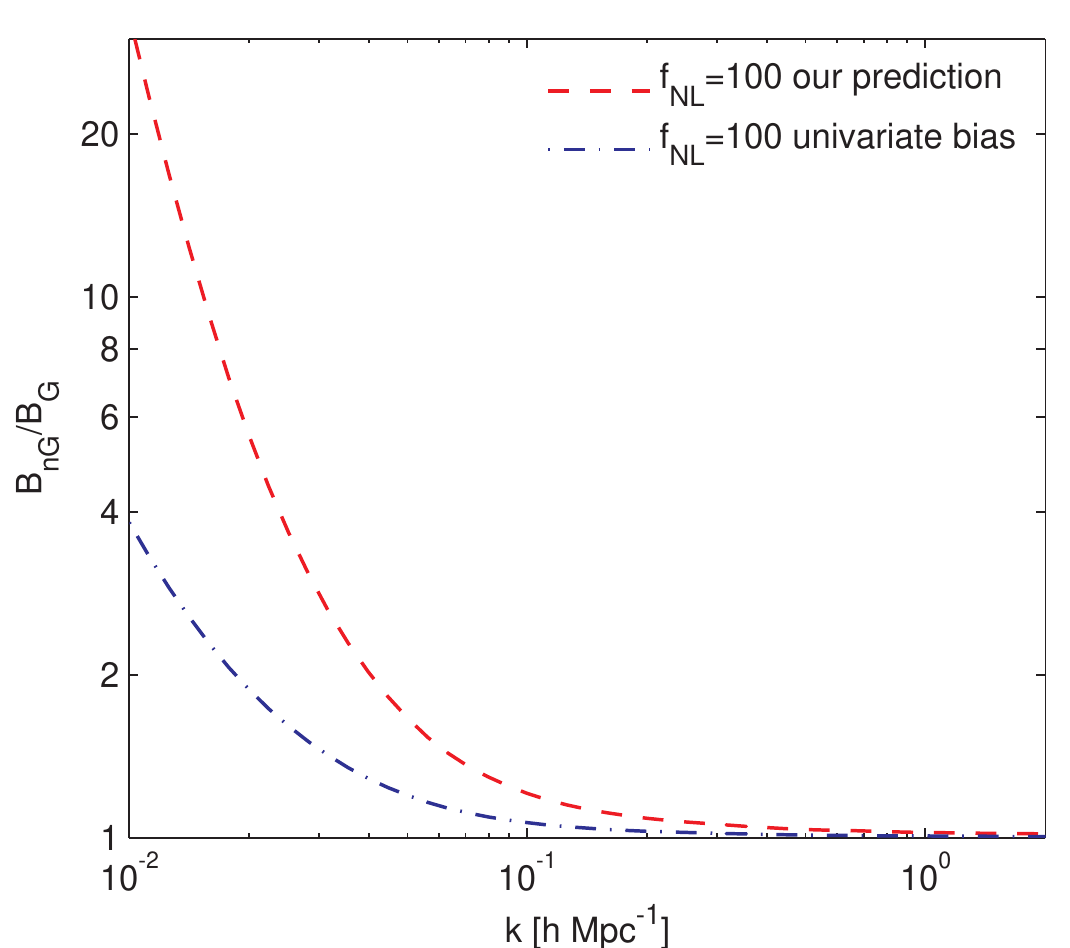}\\
	\includegraphics[width=0.49\textwidth]{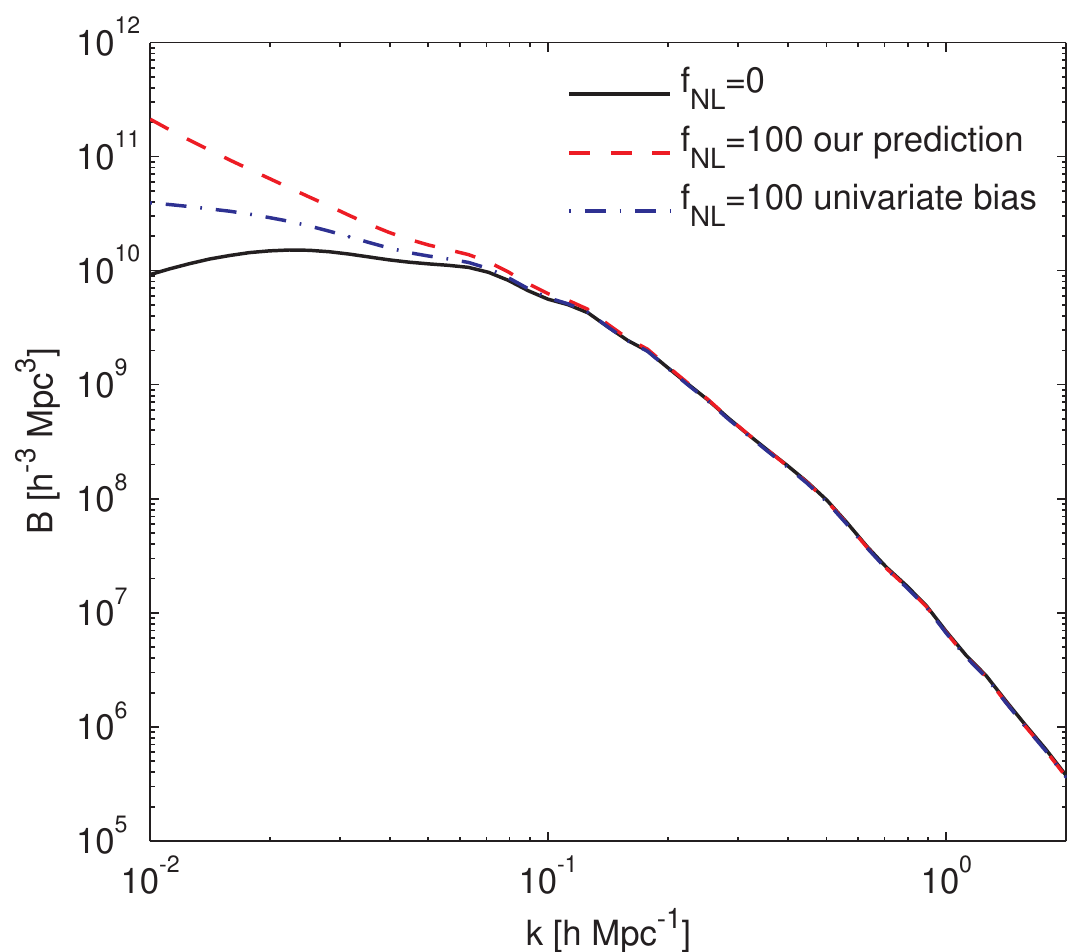}
	\includegraphics[width=0.49\textwidth]{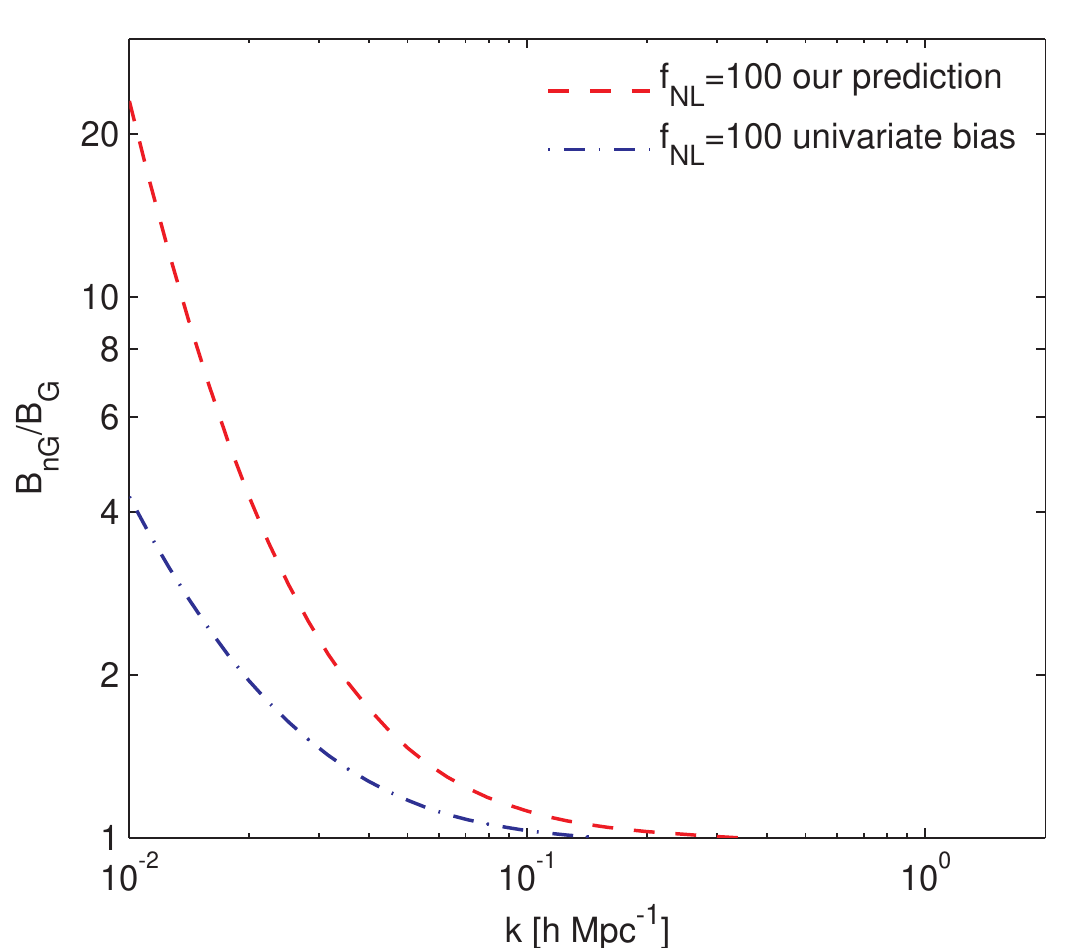}\\
	\includegraphics[width=0.49\textwidth]{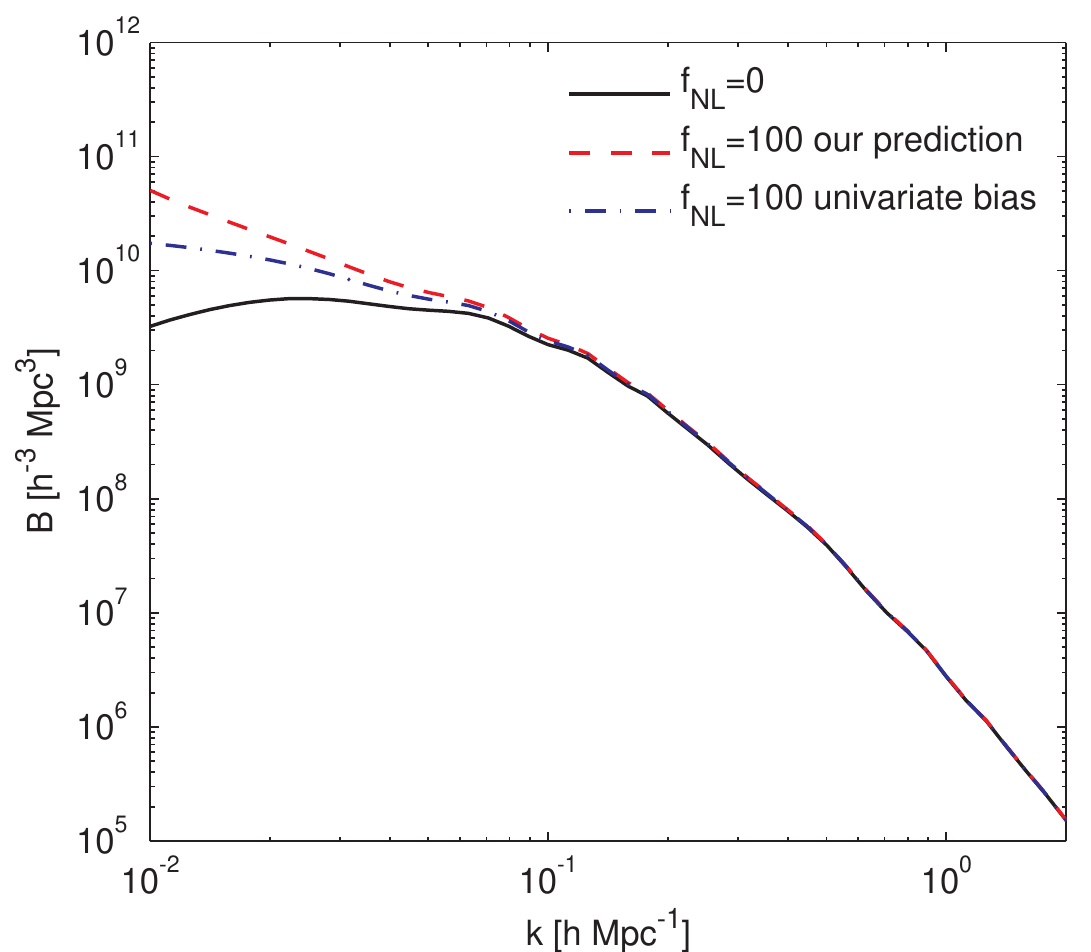}
	\includegraphics[width=0.49\textwidth]{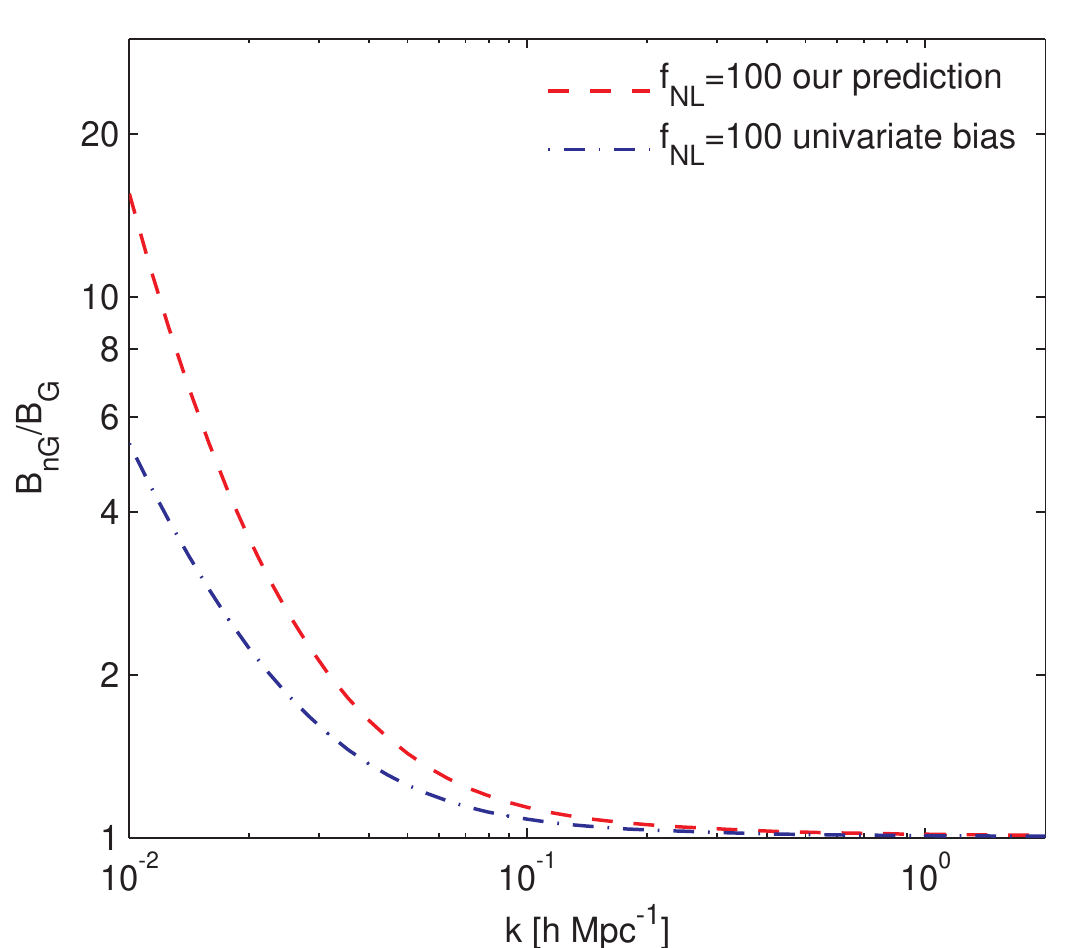}
    \caption{\emph{Upper Left Panel: }Halo auto bispectrum $B_\text{hhh}$
	according to Eq.~\eqref{eq:bhhh} for $\fnl=100$ and $\mu=-0.99$
in the isosceles configuration $k_1=k_2=k$. We
	show the Gaussian bispectrum (black solid line), the
	univariate bias prediction of Eq.~\eqref{eq:sef2007} (blue
	dash dotted) and our predictions (red dashed). \emph{Upper Right panel:
}
	Ratio of the non-Gaussian bispectra and the Gaussian bispectrum shown
	in the left panel.   
	\emph{Middle Panels:} Same as above, but for the halo-matter cross
bispectrum
	$B_\text{hhm}$ according to Eq.~\eqref{eq:bhhm}.
	\emph{Lower Panels:} Same as above, but for the halo-matter cross
bispectrum
	$B_\text{hmm}$ according to Eq.~\eqref{eq:bhmm}.
}
	\label{fig:scaledepbhhh}
	\label{fig:scaledepbhhm}
	\label{fig:scaledepbhmm}
\end{figure}

Fig.~\ref{fig:scaledepbhhh} %,  \ref{fig:scaledepbhhm} and \ref{fig:scaledepbhmm}
shows the halo auto- and cross-bispectra. We show both their 
scale dependence as well as the ratio to the Gaussian expression for a
squeezed isosceles configuration $k_1=k_2, \mu=-0.99, x_3\approx 1/7$ and 
compare to the predictions of the univariate bias model. These figures show that 
the non-Gaussian bispectrum asymptotes to the Gaussian one for high $k$'s 
(small scales). On larger scales both the univariate bias model as well as our 
prediction are increased with respect to the Gaussian case due to the coupling 
between long and short modes. Our expression for $B_\text{hhh}$ predicts an 
enhancement by a factor of 2 with respect to the existing univariate biasing 
calculations on scales of $k \approx 0.03 \ihMpc$. The enhancement is less 
pronounced for the mixed halo-matter bispectra.
From the scaling with $k$ and $x_3$ in Eq.~\eqref{eq:bhhhkdep} and Table~\ref{tab:bhhh}
it is clear that the bispectrum amplitude is largest for squeezed configurations $\mu \to -1,\ \theta\to \pi$
and low overall scale $k \to 0$. However, the shape and scale of the triangle
are limited by the fundamental mode and the sampling variance in exactly this
limit. We will quantify the signal-to-noise ratio in the next section.

%===============================================================================
\subsection{Comparison to Simulations}
\begin{figure}[t]
	\centering
	\includegraphics[width=0.49\textwidth]{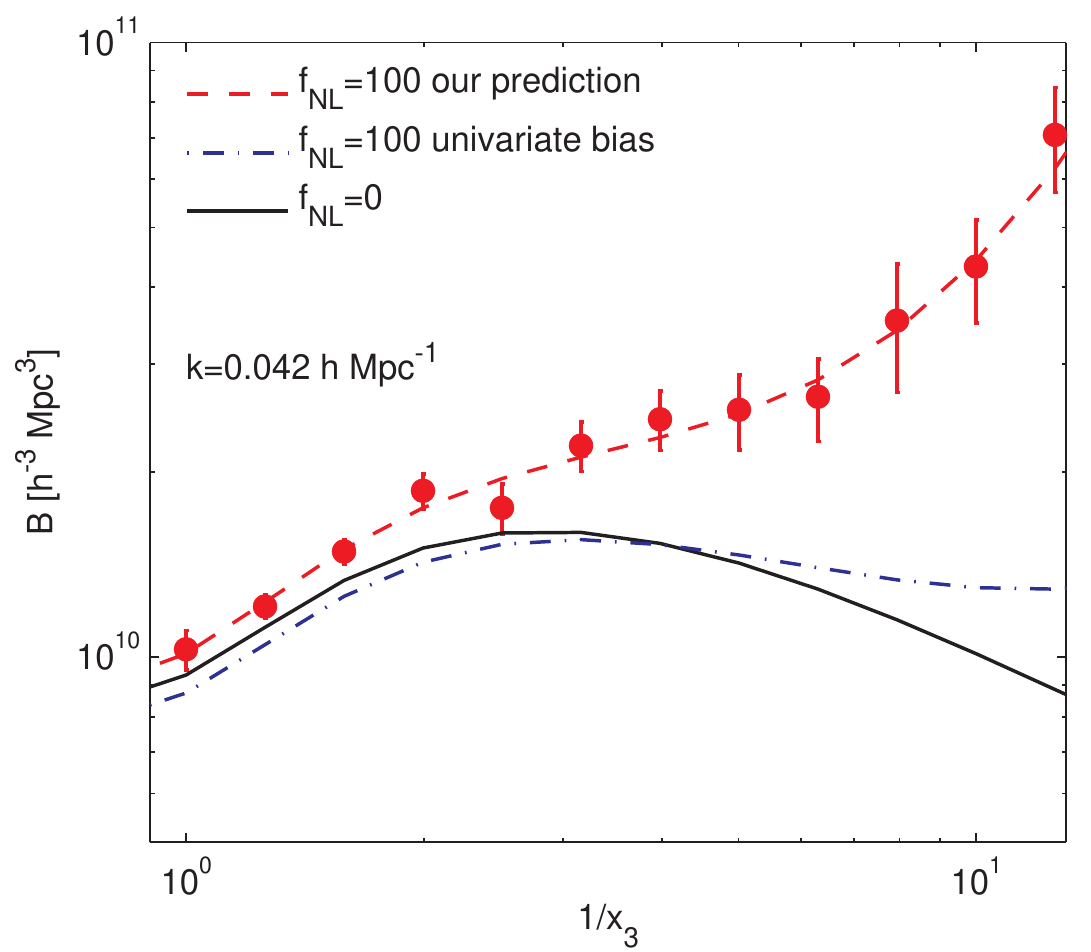}
	\includegraphics[width=0.49\textwidth]{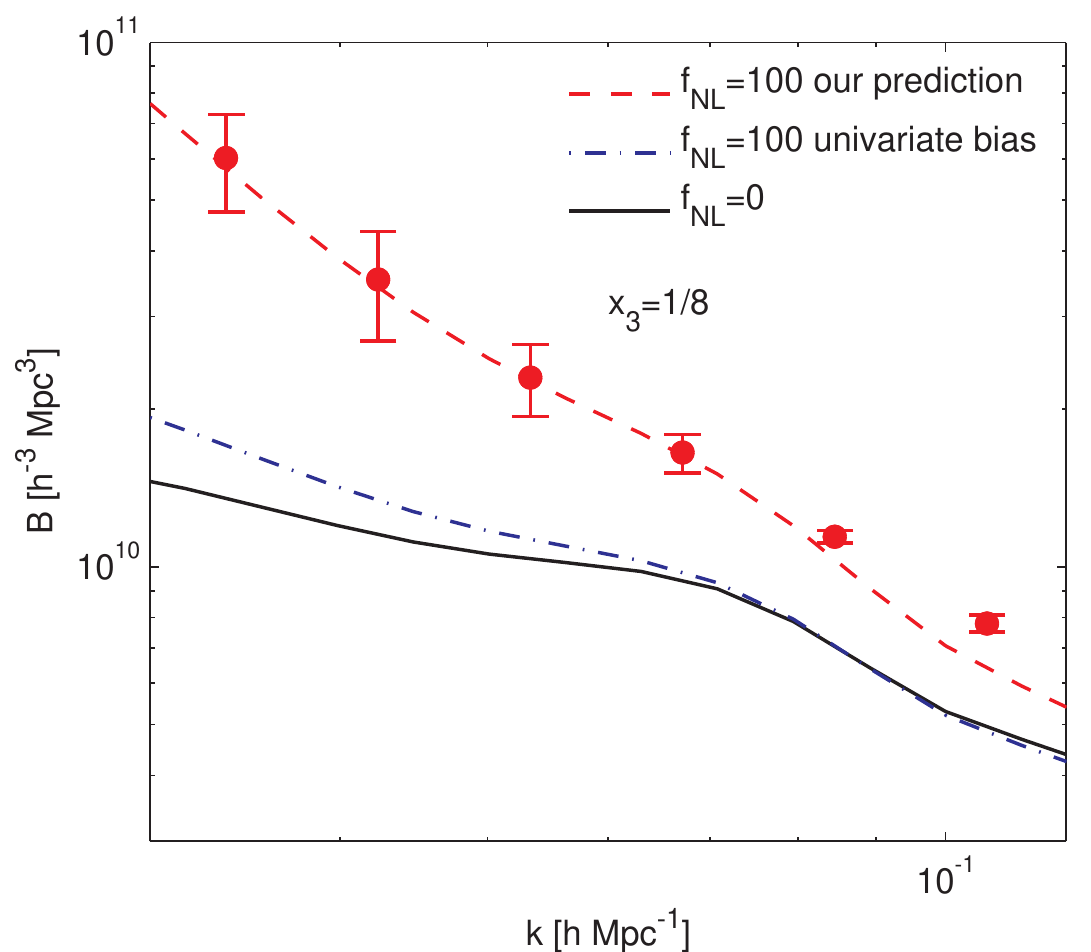}
	\caption{Halo bispectrum in the isosceles configuration
	$k_1=k_2$ for $z=0.5$ and $M>4.6 \tim{13} \hMs$ in 
	comparison to the data points in Fig.~(6) of \cite{Nishimichi2009}. 
	The lines show the multivariate bias (red dashed) and univariate bias (blue dash-dotted)
	non-Gaussian bispectrum as well as the Gaussian bispectrum (black solid).
	We adjust the low mass cutoff to bring the bias parameters derived from
	the mass function in agreement with the simulation measurement $b_{10}=2.9$
	and rescale $b_{01}\to 0.75\, b_{01}$.
	\emph{Left panel: }Bispectrum as a function of the high-low-$k$ ratio $1/x_3=k/k_3$
	and $k=0.042 \ihMpc$. 
	\emph{Right panel: }Bispectrum as a function of scale $k$ for $1/x_3=7.94$.
	}
	\label{fig:nishimishi}
\end{figure}

While a full simulation based analysis of the halo bispectrum goes beyond the 
scope of this paper, we can nevertheless compare to published simulation 
measurements. The only published simulation measurement of the halo bispectrum
we are aware of, was performed on a $L=2000 \hMpc$ cosmological simulation 
by \cite{Nishimichi2009}.
They consider a cosmology with $\sigma_8=0.816$, $\Omega_\text{m}=0.28$,
$\Omega_\Lambda=0.72$ and present their results for $z=0.5$.
In Fig.~\ref{fig:nishimishi} we show the comparison between our theoretical
predictions and their simulation measurement for a halo sample
with $M>4.6 \tim{13} \hMs$. The bias value inferred from the simulations 
$b_\text{10,sim}=2.9$ \cite{Nishimichi2010}
points towards a halo sample with $M>7\tim{13}\hMs$. This deviation to
the low mass cutoff quoted by the authors might be due to an incomplete sampling
of haloes around the cutoff. We adopt $M>7\tim{13}\hMs$ as the lower boundary of the
halo sample and calculate the average bias parameters using the LV mass function.
Consistent with \cite{Nishimichi2009} we rescale $b_{01}$ as $b_{01}\to 0.75\, b_{01}$, 
a modification motivated by ellipsoidal collapse (although this seems to apply only to 
Friends-of-Friends identified halos and not to spherical overdensity halos,
\cite{Desjacques2010}).
Fig.~\ref{fig:nishimishi} shows that the scale and shape dependence is
quite well described by the theory, especially the upturn of
the bispectrum for squeezed configurations. The terms dominating the
non-Gaussian signal for $x_3 \approx 0.1$ in this
comparison are A,B,H and I in Eq.~\eqref{eq:bhhh}. Concentrating on the
dominating terms for the squeezed isosceles configuration $k_1=k_2=k$,
$x_3 \approx 0$ the bispectrum reads as
\begin{align}
B_\text{hhh}(\vec k_1,\vec k_2,\vec k_3)=&\left(4b_{10}^3\fnl+2b_{10}^2
b_{11}\right)\frac{P(k)P(x_3 k)}{\alpha(x_3
k)}+\left(4 \fnl b_{10}^2 b_{01}+2 b_{10} b_{11} b_{01}\right)\frac{P(k)P(x_3
k)}{\alpha^2(x_3 k)}\elnn
+&\left(4 \fnl b_{10}^2 b_{01}+4 b_{10} b_{11} b_{01}\right)\frac{P(k)P(x_3
k)}{\alpha(x_3 k)\alpha(k)}
+b_{10}^2b_{01} \frac{P(k)P(x_3 k)}{\alpha(x_3k)}
\frac{13-5x_3^2}{7}
\end{align}
%
%>>>>>>>>>>>>>>>>>>>>>>>>>>>>>>>>>>>>>>>>>>>>>>>>>>>>>>>>>>>>>>>>>>>>>>>>>>>>>>>
% 				Figure 13
%>>>>>>>>>>>>>>>>>>>>>>>>>>>>>>>>>>>>>>>>>>>>>>>>>>>>>>>>>>>>>>>>>>>>>>>>>>>>>>>
\clearpage
%<<<<<<<<<<<<<<<<<<<<<<<<<<<<<<<<<<<<<<<<<<<<<<<<<<<<<<<<<<<<<<<<<<<<<<<<<<<<<<<
%
We see that for the scales and shapes considered, the bispectrum amplitude
is dominated by $\fnl$ and $\fnlsq$ terms. The disagreement between our
predictions and the simulation measurement on scales of $k\approx 0.1 \ihMpc$ is
probably due to loop corrections in the matter bispectrum \cite{Sefusatti2010}
or to the inaccuracy of the non-Gaussian mass function that we have used to
extract the bias parameters.\footnote{It is indeed possible that other
non-Gaussian mass functions might give improved results, but a more careful
treatment of these effects goes beyond the scope of our paper.}
%>>>>>>>>>>>>>>>>>>>>>>>>>>>>>>>>>>>>>>>>>>>>>>>>>>>>>>>>>>>>>>>>>>>>>>>>>>>>>>>
% 				Figure 14
%>>>>>>>>>>>>>>>>>>>>>>>>>>>>>>>>>>>>>>>>>>>>>>>>>>>>>>>>>>>>>>>>>>>>>>>>>>>>>>>
%\clearpage
%<<<<<<<<<<<<<<<<<<<<<<<<<<<<<<<<<<<<<<<<<<<<<<<<<<<<<<<<<<<<<<<<<<<<<<<<<<<<<<<
%===============================================================================
%				SIGNAL TO NOISE
%===============================================================================
\section{Signal-to-Noise}\label{sec:snr}

One goal of this work is to show the viability of the bispectrum to put
constraints on primordial non-Gaussianity. The question is whether it is
worth the additional effort of measuring the halo bispectrum given
that the halo power spectrum can be used to put constraints on non-Gaussianity
as well. 
%
%>>>>>>>>>>>>>>>>>>>>>>>>>>>>>>>>>>>>>>>>>>>>>>>>>>>>>>>>>>>>>>>>>>>>>>>>>>>>>>>
% 				Figure 15
%>>>>>>>>>>>>>>>>>>>>>>>>>>>>>>>>>>>>>>>>>>>>>>>>>>>>>>>>>>>>>>>>>>>>>>>>>>>>>>>
\begin{figure}[htb]
	\centering
	\includegraphics[width=0.49\textwidth]
	{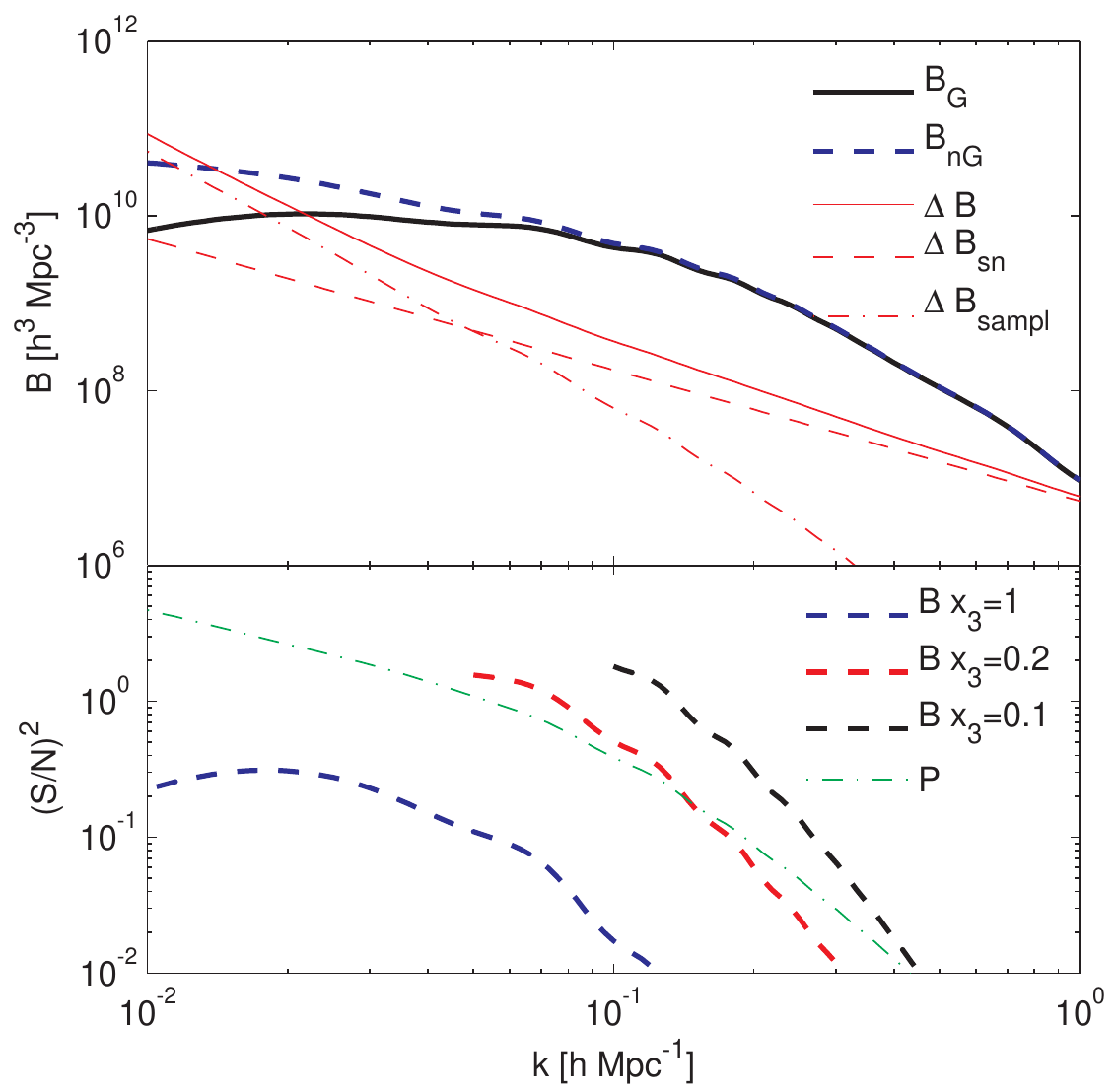}
	\includegraphics[width=0.49\textwidth]
	{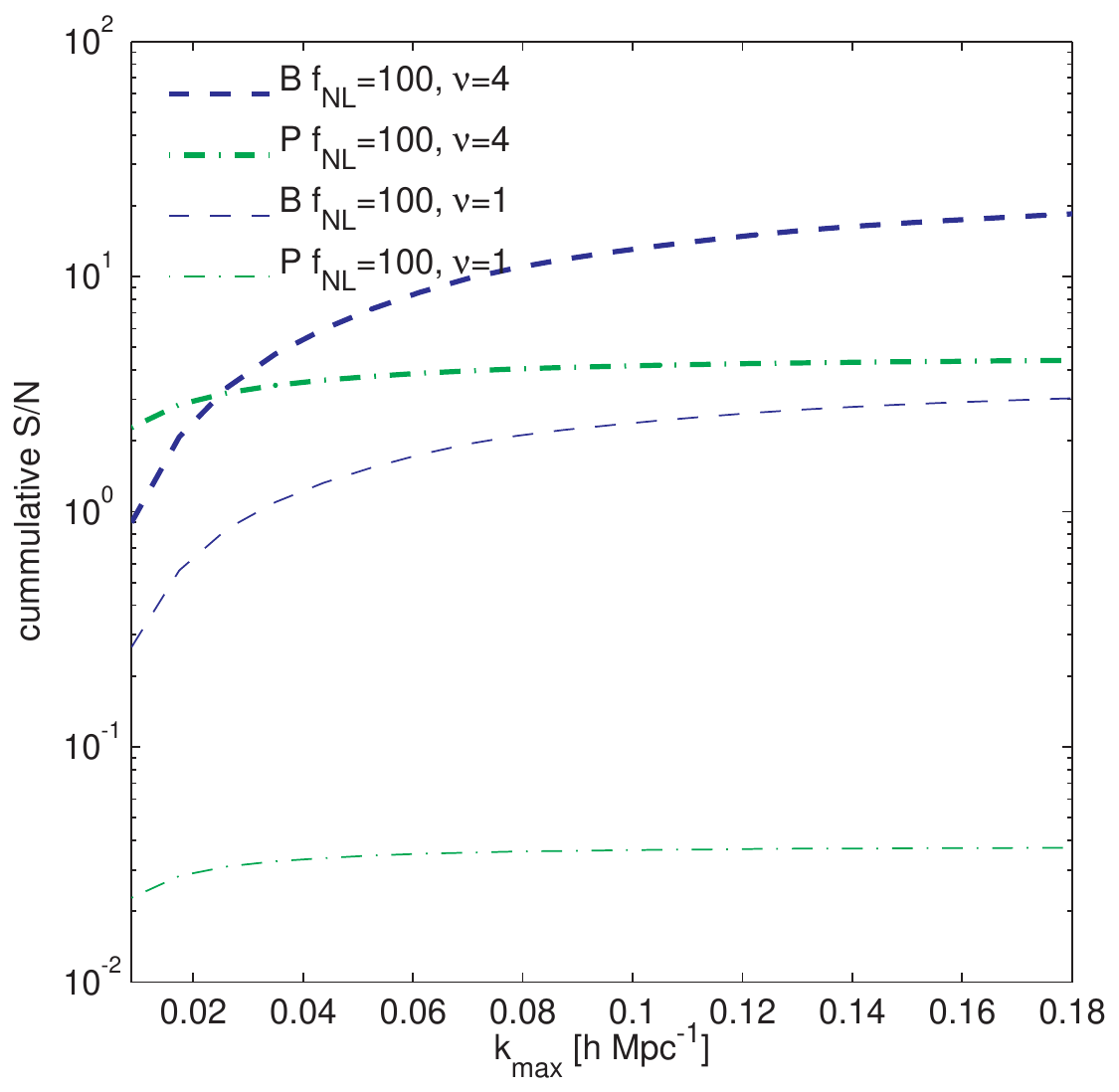}
	\caption{
	\emph{Left top panel:} 
	Variance of the halo bispectrum $B_\text{hhh}$ for $\fnl=100,\ x_3=0.1$.
	Gaussian (thick black) and non-Gaussian (thick blue dashed) bispectrum 
	amplitude and error on the bispectrum for
	$\bar{n}=1\tim{-5} h^3\unit{Mpc}^{-3}$, $V=10 \hGpcc$
	$k_\text{f}=2.9	\tim{-3}\ihMpc$ and $\delta k=3k_\text{f}=0.008
	\ihMpc$. For the biased tracer we assume a halo sample with $\nu=4$,
	corresponding to $M\approx 1 \tim{14} \hMs$ ($b_{10}=2.23,\ b_{20}=0.38$).
	We plot the full error
	$\Delta B(k)$ (red) as well as the sampling variance (red dash-dotted)
	and shotnoise (red dashed) contributions. 
	\emph{Left bottom panel:} SNR for the halo
	bispectrum $B_\text{hhh}$ (blue, red and green dashed for $x_3=1,\ 0.2$ and $0.1$ from bottom to top) 
	and the halo power spectrum
	$P_\text{hh}$ (green dash-dotted). The signal is defined as the
	difference between the non-Gaussian and Gaussian prediction. As we go to higher
	$k$ more squeezed configurations gain importance.
	\emph{Right panel: }Cumulative signal-to-noise for the bispectrum
	(blue dashed) and power spectrum (green dash-dotted). We evaluate the
	SNR for	$\nu=4$ (thick) and $\nu=1$ (thin).
	}
	\label{fig:snratio}
\end{figure}
%<<<<<<<<<<<<<<<<<<<<<<<<<<<<<<<<<<<<<<<<<<<<<<<<<<<<<<<<<<<<<<<<<<<<<<<<<<<<<<<
The bispectrum 
analysis is naively more sensitive than the power spectrum to non-Gaussianities
but it remains to show that this can
overcome the enhanced errors. To estimate the errors we assume a survey of 
volume $V$ from which the halo density field $\delta_\text{h}(\vec k)$
is estimated. The bispectrum and power spectrum estimators are constructed from
a decomposition of $k$-space into spherical shells of width $\delta k$. Then the
$k$-modes within the shell are spherically averaged to obtain the estimators
\cite{Scoccimarro2004}
\be
\hat{P}(k)=\frac{V_\text{f}}{V_{12}}
\int_k \frac{\derivd^3q_1}{(2\pi)^3}
\int_k \frac{\derivd^3q_2}{(2\pi)^3}
\delta(\vec q_1)\delta(\vec q_2)\delta^\text{(D)}(\vec q_1 +\vec q_2)
\ee
for the power spectrum and
\be
\hat{B}(\vec k_1,\vec k_2,\vec k_3)=\frac{V_\text{f}}{V_{123}}
\int_{k_1}\frac{\derivd^3q_1}{(2\pi)^3}
\int_{k_2}\frac{\derivd^3q_2}{(2\pi)^3}
\int_{k_3}\frac{\derivd^3q_3}{(2\pi)^3} 
\delta(\vec q_1)\delta(\vec q_2)\delta(\vec q_3)
\delta^\text{(D)}(\vec q_1 +\vec q_2 +\vec q_3)
\ee
for the bispectrum, where the integrals are over spherical shells
$q_i\in[k_i-\delta k/2,k_i+\delta k/2]$.
As shown in \cite{Scoccimarro2004} the covariance of this bispectrum estimator
is given by
\be
\left(\Delta
B\right)^2=s_{123}\frac{V_\text{f}}{V_{123}}\left(P_\text{h}(k_1)+\frac{1}{
\bar{n}}
\right)\left(P_\text{h}(k_2)+\frac{1}{\bar{n}}\right)\left(P_\text{h}(k_3)+\frac
{1}{\bar{n}}\right)
\ee
and the one for the power spectrum \cite{Scoccimarro2004,Smith2009} by 
\be
\left(\Delta
P\right)^2=2\frac{V_\text{f}}{V_{12}}\left(P_\text{h}(k_1)+\frac{1}{\bar{n}
}
\right)\left(P_\text{h}(k_2)+\frac{1}{\bar{n}}\right).
\ee
Here $s_{123}$ is a symmetry factor, with $s_{123}=6,2,1$ for
equilateral, isosceles and general configurations, respectively.
In the noise estimators we approximate the halo power spectrum by the leading
contribution $P_\text{h}(k)=\left(b_{10}+b_{01}/\alpha(k)\right)P_\text{lin}(k)$
for simplicity. The $k$-space volume of the fundamental cell is
$V_\text{f}=k_\text{f}^3=(2\pi)^3/L^3$ and the norm volumes are given by
\be
V_{123}=
\int_{k_1}\frac{\derivd^3q_1}{(2\pi)^3}
\int_{k_2}\frac{\derivd^3q_2}{(2\pi)^3}
\int_{k_3}\frac{\derivd^3q_3}{(2\pi)^3} 
\delta^\text{(D)}(\vec q_{123})\approx 8 \pi^2 k_1 k_2 k_3 \delta k^3
\ee
and
\be
V_{12}=
\int_{k_1}\frac{\derivd^3q_1}{(2\pi)^3}
\int_{k_2}\frac{\derivd^3q_2}{(2\pi)^3}
\delta^\text{(D)}(\vec q_{12})\approx 4 \pi k_1 k_2 \delta k.
\ee
If $P\gg 1/\bar{n}$ the cosmic variance dominates whereas shotnoise dominates when $P\ll 1/\bar{n}$.

As our fiducial case we consider a survey of $V=10 \hGpcc$ and $\delta k=3
k_\text{f}=9 \tim{-3} \ihMpc$ and a halo sample with overdensity $\nu=4$ 
corresponding to $M\approx 1 \tim{14} \hMs$ ($b_{10}=2.23,\ b_{20}=0.38$). The
number density of the halo sample is assumed to be $\bar{n}=1\tim{-5} \ihMpc$.
The left panel of Fig.~\ref{fig:snratio} shows the errors and the
signal-to-noise ratio (SNR) for the halo bispectrum $B_\text{hhh}$ in a
squeezed isosceles configuration $k=k_1=k_2$, $x_3=0.1$. Here we define the
signal as the signal-to-noise weighted difference between the total non-Gaussian and the fiducial Gaussian
bispectrum amplitude
\be
\snr_B^2(k_i,k_j,k_l)=\frac{\left(B_\text{nG}(k_i,k_j,k_l)-B_\text{G}(k_i,k_j,
k_l)\right)^2}{(\Delta B)^2(k_i,k_j,k_l)}
\ee
with an equivalent expression for the power spectrum. We see that the $\fnl$ 
contribution enhances the signal on a wide range of scales and most relevantly 
on the smallest $k$'s. The left bottom panel of Fig.~\ref{fig:snratio} shows the 
SNR for a given bin of $k$ modes, for isosceles triangles with three different 
values of $x_3$. We see that the bispectrum signal is enhanced as we make the 
triangle more and more squeezed and as we take the highest $k$'s of the triangle 
closer to the non-linear scale. We should stress that our tree-level calculation 
does not apply for $k$'s close or larger than the non-linear scale. We finally 
also plot the binned SNR from the power spectrum, which is peaked at the smallest 
$k$'s. So far we restricted our discussion to a particular bispectrum 
configuration. For the extraction of a non-Gaussian signal from a survey one 
would rather add up all the possible information, \emph{i.\,e.\ }sum over all 
possible configurations up to a maximum wavenumber $k_\text{max}$.
The total signal-to-noise is given by a sum over all the possible combinations
of shells up to the maximum wavenumber $k_\text{max}$
\be
\left(\frac{S}{N}\right)_{\text{tot},B}^2=\sum_{i=1}^{i_\text{max}}\sum_{j=1}^{i}\sum_{l=i-j}^{i+j}\ \snr_B^2(k_i,k_j,k_l)
\ee
where we used $k_i=i \delta k$.
In the right panel of Fig.~\ref{fig:snratio} we plot the cumulative
signal-to-noise for the bispectrum and power spectrum for $\fnl=100$ and tracers with
peak height $\nu=4$ and $\nu=1$. For the high bias tracer we see 
that for the considered survey and going up to 
$k_\text{max}=0.18 \ihMpc$ we could constrain $\sigma_{f_\text{NL}}\approx 5$ 
from the bispectrum analysis, whereas the power spectrum leads to constraints
of $\sigma_{f_\text{NL}}\approx 25$. Note that we are plotting the total non-Gaussian
signal. To obtain the SNR on $\fnl$, the lines in the figure have to be divided by $f_\text{NL}$.
For $k$ higher than about $0.1\ihMpc$ non-linear corrections should 
be implemented. The bispectrum wins over the power spectrum even if the maximum 
wave number is restricted to $k_\text{max}\approx 0.03 \ihMpc$. Furthermore the 
plot shows that extraction of $\fnl=\mathcal{O}(10)$ is possible only with the
bispectrum analysis but not with the power spectrum analysis for these tracers.
Note that the choice of bias, volume and number density 
is similar to values for luminous red galaxies expected in SDSS-III (BOSS) redshift survey \cite{Schlegel2009}. 
The bispectrum of the low bias $\nu=1$ tracers shows an even more remarkable improvement
in SNR compared to the corresponding power spectrum.

%===============================================================================
%				DISCUSSION
%===============================================================================
\section{Conclusions}\label{sec:discussion}
In this paper we present a diagrammatic prescription for the calculation of
multipoint statistics of biased tracers of the cosmological density field 
accounting for the effect of non-Gaussian initial fluctuations. The diagrammatic 
approach combines biasing, non-Gaussianity and non-linear 
clustering into one consistent and intuitive picture. While we focused on the 
bispectrum, the generalisation to higher order correlators should be 
straightforward. Also, non-local shapes can be implemented using their
primordial bispectrum instead of the mode coupling vertices in 
Fig.~\ref{fig:nongauss}.
\par
We use this diagrammatic prescription to derive the halo bispectrum 
accounting for all tree level/zero loop contributions. Unsurprisingly, the 
bispectrum is largest in the squeezed limit $x_3\to 0$, where our 
prediction exceeds the results obtained using univariate bias only, by about a 
factor of $2$ on scales of $k\approx 0.03 \ihMpc$. 
Given the fact, that we see these corrections compared to the univariate bias 
approach already at tree level, we caution the use of the univariate bias expansion
as presented in \cite{Sefusatti2009} and \cite{Jeong2009}.
While the qualitative results are quite general, the quantitative results presented in Section
\ref{sec:numresults} are somewhat dependent on the choice of the halo sample. We
considered a $\nu=4$ tracer, corresponding to haloes of mass
$M\approx 1 \tim{14} \hMs$ ($b_{10}=2.23,\ b_{20}=0.38$).
\par
Comparing our results to the ones obtained in \cite{Jeong2009} we see that
their one-loop terms proportional to $b_2$ are comparable to some of our tree
level terms if i) the smoothing scale equals the halo mass scale, ii) high
mass haloes are considered and iii) large scales are considered. But, these
terms can not be seen as a replacement to the non-Gaussian terms obtained in
our approach because all the one loop terms arising in the univariate bias
approach are also present in our approach. The final comparison of the terms
would require a careful resummation of the bias parameters, which was not
considered in \cite{Jeong2009}.
\par
Probably the most important result of this paper is that the halo bispectrum 
analysis offers an alternative to the power spectrum for detecting
 local non-Gaussianities with an even higher constraining power. For the $\nu=4$ tracers
in a $V=10 \hGpcc$ survey, our signal-to-noise analysis predicts a factor of 
five improvement in the constraints on $\fnl$ compared to the power spectrum.
For lower bias tracers at the same number density the total signal-to-noise
is a bit lower, but it is in fact much higher relative to the power spectrum 
analysis, which contains no signal for unbiased tracers. 
\par
Our results suggest that the bispectrum should be the statistic of choice for detecting primordial 
non-Gaussianity in current and future survey data.
However, some additional work has to be done before this method can be applied to the real data. 
One extension is to make predictions for photometric surveys, where 
only projected density fluctuations are observed. Alternatively, if a redshift survey 
is used in the analysis then the predictions here should be generalized
to include redshift space distortions \cite{Smith2008}. 
In the light of ever increasing surveys and simulations one might also be concerned 
about general relativistic corrections on horizon scales \cite{Yoo2009}. 
\par
Convergence is probably one of the most important problems for perturbative 
calculations. For Gaussian initial conditions, comparisons of the matter bispectrum in
simulations to the theoretical predictions as presented in \cite{Sefusatti2010}
and \cite{Smith2007} conclude that 1-loop calculations are required to achieve a
reasonable agreement on scales of $k\approx 0.1 \hMpc$. 
The bispectrum in 
presence of local non-Gaussianity is the perfect statistic to apply perturbation 
theory at tree level combined with the bias from the peak-background split since 
the non-Gaussian effects are most prominent for low $k$ or large smoothing scales. 
Still, we focus on a tree 
level calculation and find that the signal receives relevant contributions from 
scales close to the non-linear scale. 
For this reason, loop-corrections should be examined to fully assess the 
detailed amplitude of the signal. As we discuss, this will probably require to 
study the cutoff dependence both of the loop corrections and of the bias 
coefficients, such that final observables do not depend on the cutoff.\footnote{
Previous studies \cite{Jeong2009,Sefusatti2009} did not stress the conceptual 
separation between the smoothing scale, the cutoff of the loops and the mass 
scale of the halos.}
Some of these 1-loop terms will be renormalized and absorbed into the 
tree level terms discussed here. Still, a consistent calculation of the halo 
bispectrum at the next order would
require consideration of non-linear couplings up to $F_4$ and biasing up to
fourth order, both in $\delta$ and $\varphi$. This increases the number of terms
by a large amount and goes beyond the scope of the current paper.
The final decision about the validity of the perturbative calculation presented in
this paper has to be based on a detailed comparison to the halo bispectrum measured
in simulations. A first comparison of our results to the measurements published by
\cite{Nishimichi2009} shows an encouraging level of agreement.

%===============================================================================
%				ACKNOWEDGEMENTS
%===============================================================================
\acknowledgments
We acknowledge M.\ Zaldarriaga for initial collaboration on this project and for 
many discussions. Furthermore, we would like to thank N.~Hamaus, P.~McDonald,
C.~Porciani, F.~Schmidt, R.~Scoccimarro and in particular E.~Sefusatti for helpful
discussions. Special thanks to V.\, Desjacques both for fruitful discussions and
comments on the manuscript. 
We would also like to thank T.~Nishimichi for providing the data points of the 
simulation bispectrum measurement and the Asian Pacific
Centre for Theoretical Physics in Pohang, Korea, for their kind hospitality during the workshop on
"Cosmology and Fundamental Physics". This work is supported by DOE, 
the Swiss National Foundation under contract 200021-116696/1 and WCU
grant R32-2009-000-10130-0.

\bibliographystyle{JHEP}
\bibliography{fnl}
\appendix
\section{Standard Perturbation Theory}
This appendix reviews the essence of cosmological perturbation theory and
serves as a source for the most important equations. For a more detailed
treatment we refer the reader to the comprehensive review on the subject by
\cite{Bernardeau2002}.
The evolution equations for the cosmic fluid in an expanding Friedmann-Robertson-Walker
Universe can be formulated in terms of the overdensity $\delta$ and the velocity
divergence $\theta=\vec \nabla \cdot \vec v$ as
\begin{align}
\frac{\partial \delta(\vec x,\tau)}{\partial \tau}+\theta(\vec
x,\tau)=&0\\
\frac{\partial \theta(\vec x,\tau)}{\partial \tau}+\mathcal{H}(\tau)\theta(\vec
x,\tau)+\frac{3}{2}\Omega_\text{m}\mathcal{H}^2(\tau)\delta(\vec x,\tau)=&0.
\end{align}
A direct solution to this coupled differential equations does not exist.
Therefore one has to restrain to a perturbative solution, expanding the density
field and velocity divergence in a power series
\begin{align}
\delta_\text{m}(\vec k)=\sum_{n=1}^\infty \delta_\text{m,p}^{(n)}(\vec k), &&
\theta(\vec k)=-\mathcal{H} \sum_{n=1}^\infty \theta^{(n)}(\vec k),
\end{align}
where $\delta^{(n)}_\text{m,p}$ and $\theta^{(n)}$ are $\mathcal{O}(\delta^n_\text{m,p})$.
The solutions for the $n$-th order contribution to the fields are given by
\be
\delta_\text{m}^{(n)}(\vec k)=\int \frac{d^3q_1}{(2\pi)^3}\ldots \int
\frac{d^3q_n}{(2\pi)^3} \delta_\text{m,p}(\vec q_1)\ldots \delta_\text{m,p}(\vec
q_n)F_n(\vec q_1,\ldots,\vec q_n)\delta^\text{(D)}(\vec q_1+\ldots+\vec q_n-\vec k)
\ee
\be
\theta^{(n)}(\vec k)=\int \frac{d^3q_1}{(2\pi)^3}\ldots \int
\frac{d^3q_n}{(2\pi)^3} \delta_\text{m,p}(\vec q_1)\ldots \delta_\text{m,p}(\vec
q_n)G_n(\vec q_1,\ldots,\vec q_n)\delta^\text{(D)}(\vec q_1+\ldots+\vec q_n-\vec k),
\ee
where the coupling kernels $F_n$ and $G_n$ can be obtained using recursion
relations. For the results presented in this paper we need $F_1=G_1=1$, and
\be
F_2(\vec k_1,\vec k_2)=\frac{5}{7}+\frac{1}{2}\frac{\vec k_1 \cdot \vec k_2}{k_1
k_2}\left(\frac{k_1}{k_2}+\frac{k_2}{k_1}\right)+\frac{2}{7}\left(\frac{\vec k_1
\cdot \vec k_2}{k_1 k_2}\right)^2.
\ee
Using the above results, the one-loop corrections to the matter power spectrum
read as
\be
P_{22}=2\int \dqc P(q)P(\left|\vec k-\vec q\right|)\left|F_2(\vec q,\vec k-\vec
q)\right|^2,
\ee
and
\be
P_{13}=6 P(k)\int \dqc P(q) F_3(\vec k,\vec q,-\vec q).
\ee

\section{Explicit Derivation of the Bias Parameters}
\label{app:B}
As shown in Section \ref{sec:bias}, long wavelength modes in presence of
primordial non-Gaussianity change the effective collapse threshold and
variance of density fluctuations. In the mass function, the presence of a long
wavelength mode can be accounted for by rescaling the peak height as
\be
\nu=\left(\frac{\delta_\text{c}}{\sigma_\text{G}}\right)^2\to\tilde{\nu}
=\left(\frac { \delta_\text{c} -\delta_\text{l} } { \sigma_\text{G}
(1+2\fnl\varphi_\text{l}+3\gnl \varphi_\text{l}^2+2 f_\text{NL}^2
\varphi_\text{l}^2) }\right)^2.\label{eq:conditionalpeak}
\ee
As we will need them for the explicit calculation of the bias parameters, we
write down the partial derivatives of the peak height with respect to
primordial potential and long wavelength density fluctuation
\begin{align}
\frac{\partial \tilde{\nu}}{\partial
\delta_\text{l}}\Big\rvert_{\delta_\text{l}=0,\varphi_\text{l}=0}=-2\frac{\nu}{
\delta_\text{c} }
&&
\frac{\partial \tilde{\nu}}{\partial
\varphi_\text{l}}\Big\rvert_{\delta_\text{l}=0,\varphi_\text{l}=0}=-4\fnl\nu\\
%\end{align}
%\begin{align}
\frac{\partial^2 \tilde{\nu}}{\partial
\delta_\text{l}^2}\Big\rvert_{\delta_\text{l}=0,\varphi_\text{l}=0}=2\frac{\nu}{
\delta_\text{c}^2}
&&
\frac{\partial^2 \tilde{\nu}}{\partial \varphi_\text{l} \partial
\delta_\text{l}}\Big\rvert_{\delta_\text{l}=0,\varphi_\text{l}=0}=8 \fnl
\frac{\nu}{\delta_\text{c}}
\end{align}
\be
\frac{\partial^2 \tilde{\nu}}{\partial
\varphi_\text{l}^2}\Big\rvert_{\delta_\text{l}=0,\varphi_\text{l}=0}=(24
\fnlsq-12 \gnl)\nu
\ee
Using these expressions and the results of Section \ref{sec:bias} we can first
derive the Gaussian bias parameters arising from the rescaling of the collapse
threshold
\begin{align}
b_{10}^\text{L}=&\frac{1}{\bar{n}}\partfrac{n}{\delta_\text{l}}
=\frac{1}{\bar n}\partfrac{n}{\nu}\partfrac{\tilde{\nu}}{\delta_\text{l}}
=-\frac{1}{\bar n}\frac{2\nu}{\delta_c }\partfrac{n}{\nu}\\
%%%%%%
b_{20}^\text{L}=&\frac{1}{\bar{n}}\partsqfrac{n}{\delta_l}=
\frac{1}{\bar
n}\partsqfrac{n}{\nu}\left(\partfrac{\tilde{\nu}}{\delta_\text{l}}
\right)^2+\frac { 1 } { \bar
n}\partfrac{n}{\nu}\partsqfrac{\tilde{\nu}}{\delta_\text{l}}\elnn
=&\frac{4}{\bar n}
\frac{\nu^2}{\delta_c^2}\partsqfrac{n}{\nu}+\frac{2}{\bar{n}}\frac{\nu}{
\delta_c^2}\partfrac{n}{
\nu}
\end{align}
The non-Gaussian bias parameters are now arising from the rescaling of the
variance in the denominator of Eq.~\eqref{eq:conditionalpeak}
\begin{align}
b_{01}^\text{L}=&\frac{1}{\bar{n}}\partfrac{n}{\varphi_\text{l}}
=\frac{1}{\bar n}\partfrac{n}{\nu}\partfrac{\tilde{\nu}}{\varphi_\text{l}}
=-\frac{4\fnl\nu}{\bar n}\partfrac { n } { \nu}
=2\fnl\delta_cb_{10}^\text{L}\\
%%%%%%
b_{11}^\text{L}=&\frac{1}{\bar{n}}\frac{\partial^2 n}{\partial
\varphi_\text{l}\partial\delta_\text{l}}
=\frac{1}{\bar n}\partsqfrac{n}{\nu}\partfrac{\tilde{\nu}}{\varphi_\text{l}}
\partfrac{\nu}{\delta_\text{l}}+\frac{1}{
\bar n}\partfrac{n}{\nu}\frac{\partial^2 \tilde{\nu}}{\partial \varphi_\text{l}
\partial \delta_\text{l}}\elnn
=&\frac{8\fnl}{\bar n}\left(\frac{\nu^2}{\delta_\text{c}}
\partsqfrac { n } { \nu }
+\frac{\nu}{\delta_c}\partfrac{n}{\nu}\right)\elnn
=&2\fnl\left( b_{20}^\text{L}\delta_\text{c}-b_{10}^\text{L}\right)\\
%%%%%%
b_{02}^\text{L}=&\frac{1}{\bar{n}}\partsqfrac{n}{\varphi_\text{l}}
=\frac{1}{\bar
n}\partsqfrac{n}{\nu}\left(\partfrac{\tilde{\nu}}{\varphi_\text{l}}
\right)^2+\frac{1}{
\bar n}\partfrac{n}{\nu}\partsqfrac{\tilde{\nu}}{\varphi_\text{l}}\elnn
=&\frac{8\fnlsq}{\bar n}\left(2\nu^2\partsqfrac{n}{\nu}+3\nu\partfrac{n}{\nu}
\right)-\frac{12\nu\gnl}{\bar n}\partfrac{n}{\nu}\elnn
=&4\fnlsq\delta_\text{c}\left(b_{20}^\text{L}
\delta_\text{c}-2b_{10}^\text{L}\right)+6\delta_c\gnl b_{10}^\text{L}
\end{align}
where we used the results for $b_{10}^\text{L}$ and $b_{20}^\text{L}$ obtained
above to replace the partial derivatives of the massfunction with respect to
the peak height.

\end{document}